\documentclass[aps,pra,graphicx,9pt,onecolumn,superscriptaddress,twocolumn]{revtex4-2}
\usepackage[colorlinks,linkcolor=blue,urlcolor=blue,anchorcolor=blue,citecolor=blue]{hyperref}
\usepackage{amsmath}
\usepackage{graphicx}

\usepackage{dcolumn}
\usepackage{mathrsfs}
\usepackage{amssymb}
\usepackage{amsfonts,multirow}
\usepackage{bm,changes}
\usepackage{color}
\usepackage{ulem} 

\begin{document}
\draft

\title{Control of Non-Markovianity via Channel Exceptional Points in a Driven-Dissipative Kerr-Cat Qubit}

\author{Tian-Le Yang}
\affiliation{Fujian Key Laboratory of Quantum Information and Quantum Optics, Fuzhou University, Fuzhou 350108, China}
\affiliation{Department of Physics, Fuzhou University, Fuzhou 350108, China}
\author{Hao-Long Zhang}
\affiliation{Fujian Key Laboratory of Quantum Information and Quantum Optics, Fuzhou University, Fuzhou 350108, China}
\affiliation{Department of Physics, Fuzhou University, Fuzhou 350108, China}
\author{Pei-Rong Han} 
\affiliation{School of Physics and Mechanical and Electrical Engineering, Longyan University, Longyan 364012, China}
\author{Zhen-Biao Yang} \email{zbyang@fzu.edu.cn}
\affiliation{Fujian Key Laboratory of Quantum Information and Quantum Optics, Fuzhou University, Fuzhou 350108, China}
\affiliation{Department of Physics, Fuzhou University, Fuzhou 350108, China}
\affiliation{Hefei National Laboratory, Hefei 230088, China}
\author{Shi-Biao Zheng} \email{t96034@fzu.edu.cn}
\affiliation{Fujian Key Laboratory of Quantum Information and Quantum Optics, Fuzhou University, Fuzhou 350108, China}
\affiliation{Department of Physics, Fuzhou University, Fuzhou 350108, China}
\affiliation{Hefei National Laboratory, Hefei 230088, China}
\date{\today}

\begin{abstract}
Exceptional points (EPs) in open quantum systems are conventionally defined at the Liouvillian level, making them representation-dependent and often requiring auxiliary degrees of freedom.
We study EPs directly at the quantum-channel level in a driven-dissipative Kerr-cat qubit, in which dissipation drives bidirectional jumps between its two basis states.
Interpolating two Markovian channels yields single EPs when the endpoints straddle a spectral-phase boundary and, remarkably, double EPs within the same phase.
The interpolated EPs map out a Markovian--non-Markovian phase diagram: the parameter plane separates certified non-Markovian and Markovian domains, with domain boundaries switching the channel between the two characters.
Three spectral criteria certify every domain. 
As the evolution time advances, the domains drift, split, and partially shrink, making the switching increasingly sensitive to parameters and time.
This establishes a representation-independent paradigm for classifying and switching quantum memory effects, applicable to continuous-variable, discrete-variable, and hybrid systems.
\end{abstract}

\maketitle

\textit{Introduction}---Open quantum system dynamics is governed by the system-environment interaction, which typically induces decoherence and information loss~\cite{Gorini1978,Breuer2007}.
The reduced evolution is described by a completely positive trace-preserving (CPTP) map $\mathcal{E}$, known as a quantum channel~\cite{Ruskai2002}.
In the Markovian regime, $\mathcal{E}$ is CP-divisible and generated either by a time-independent Gorini-Kossakowski-Sudarshan-Lindblad (GKSL) Liouvillian $\mathcal{L}$,  yielding the semigroup $e^{(t-t_{0})\mathcal{L}}$~\cite{Gorini1976,Lindblad1976}, or by a time-dependent GKSL Liouvillian $\mathcal{L}(t)$ with non-negative decay rates, yielding $\mathcal{E}(t,t_0) = \mathcal{T}\exp\!\left(\int_{t_0}^{t}\!\mathcal{L}(\tau)\,d\tau\right)$, with $\mathcal{T}$ the time-ordering operator~\cite{Chruscinski2010,Breuer2016}.
When memory effects become non-negligible, the dynamics enters the non-Markovian regime~\cite{Breuer2007,Breuer2016}, captured either by a time-local master equation $d\rho(t)/dt=\mathcal{L}_{t}\rho(t)$ with temporarily negative decay rates~\cite{Breuer2009,Breuer2016,Hall2014}, or by a time-nonlocal integral-differential equation with a memory kernel~\cite{Breuer2007,Chruscinski2010,Lin2025}.

Numerous operational strategies have been developed to quantify non-Markovianity of quantum channels, including trace-distance backflow~\cite{Breuer2009,Laine2010}, violation of CP divisibility~\cite{Chruscinski2011,Wolf2008CMP}, and entanglement revival~\cite{Rivas2010,Liu2011,Han2024APL}.
These approaches, however, predominantly rely on input-state optimization or intermediate-map verification.
It remains an open question whether intrinsic spectral features of the quantum channel itself can serve as direct structural witnesses of non-Markovianity, independent of external probes or specific dynamical representations.

Exceptional points (EPs), non-Hermitian singularities at which eigenvalues and eigenvectors simultaneously coalesce, have emerged as powerful probes of open quantum systems~\cite{Ganainy2018,Miri2019,Ozdemir2019,Xue2026PRL}. 
In the Markovian regime, EPs have been studied in effective non-Hermitian Hamiltonians~\cite{Han2024NC,Zhang2025,Abbasi2022,Han2023PRL,Wu2024,Yang2026SCPMA,Yang2026SCPMA-Hyper,Yan2026} and full GKSL Liouvillians~\cite{Minganti2020,Chen2022,Zhang2022,Sun2024,Khandelwal2021,Bu2023,Han2026SCPMA,Yang2026}.
Beyond the Markovian limit, however, EPs at the Liouvillian-superoperator level face fundamental obstacles: the time-local Liouvillian $\mathcal{L}_{t}$ is generically nonstationary~\cite{Breuer2007,Breuer2009,Breuer2016,Hall2014}, so its spectral degeneracies typically occur only at isolated instants rather than persisting over finite time intervals; meanwhile, the Nakajima-Zwanzig formulation is inherently nonlocal in  time and admits no instantaneous superoperator~\cite{Nakajima1958,Breuer2016,Zwanzig1960,Chruscinski2010}.
Embedding strategies can convert these transient features into stationary EPs of extended Liouvillians~\cite{Lin2025,Zhang2025PRL}, reaction-coordinate dynamics~\cite{Holst2026}, or quantum-state-diffusion coefficient equations~\cite{Chen2026NMQSD}, but at the cost of introducing auxiliary degrees of freedom that mask the native temporal structure.
Furthermore, channel EPs were first introduced and demonstrated in a discrete-variable experiment~\cite{Wong2026}.
Studies so far remain confined to time-independent channels, whereas channels whose EP locations drift with the evolution time remain uncharted.

We introduce channel-level EPs, defined intrinsically by the coalescence of eigenvalues and eigenvectors of the CPTP map $\mathcal{E}$ itself. These representation-independent objects have locations that drift continuously in time.
We investigate such EPs in a driven-dissipative Kerr-cat qubit, a representative platform for continuous-variable encoded qubits~\cite{Wielinga1993,Goto2016,Minganti2016,Puri2017,Puri2019,Grimm2020,Lin2024}. Dissipation in this cat qubit drives bidirectional jumps between its two basis states \cite{Han2026SCPMA,Yang2026}, in contrast to discrete-variable encoded qubits. 
By interpolating between two Markovian channels (MCs) away from EPs, we show that the interpolated channel hosts single interpolated channel EPs (ICEPs) when the endpoints straddle a spectral-phase boundary, consistent with the general expectation of Ref.~\cite{Wong2026}, and, more strikingly, double ICEPs within the same spectral phase, beyond the conventional phase-boundary picture.
Most importantly, the ICEP landscape forms a Markovian--non-Markovian phase diagram in the control-parameter space.
Interpolating between two MCs divides the parameter plane into extended domains of non-Markovian and Markovian ICEPs that alternate along parameter sweeps, so that the channel can be switched repeatedly between different domains by crossing these boundaries.
As the evolution time advances, the domains drift, split, and partially shrink.
In the double-EP region, the non-Markovian domains fragment and shrink, whereas in the single-EP region all domains persist and split into smaller and more numerous pieces.
The switching thus becomes increasingly sensitive to parameters and time, as the multiplying boundaries introduce ever more switch points, and complements the dynamical control of Ref.~\cite{Liu2011} by providing a spectral, snapshot-level certification. This certification is formulated here for a continuous-variable platform and applies equally to discrete-variable and hybrid continuous-discrete-variable platforms.
We provide three unexplored yet spectral criteria that certify every domain of phase diagram, thereby ensuring that no ICEP remains undecided.
These results establish interpolation between MCs as a controllable route to engineer and switch Markovian and non-Markovian dynamics at EPs, revealing a certified phase structure of ICEPs inaccessible at the Liouvillian level.

\begin{figure}
\centering
\includegraphics[width =0.48\textwidth]{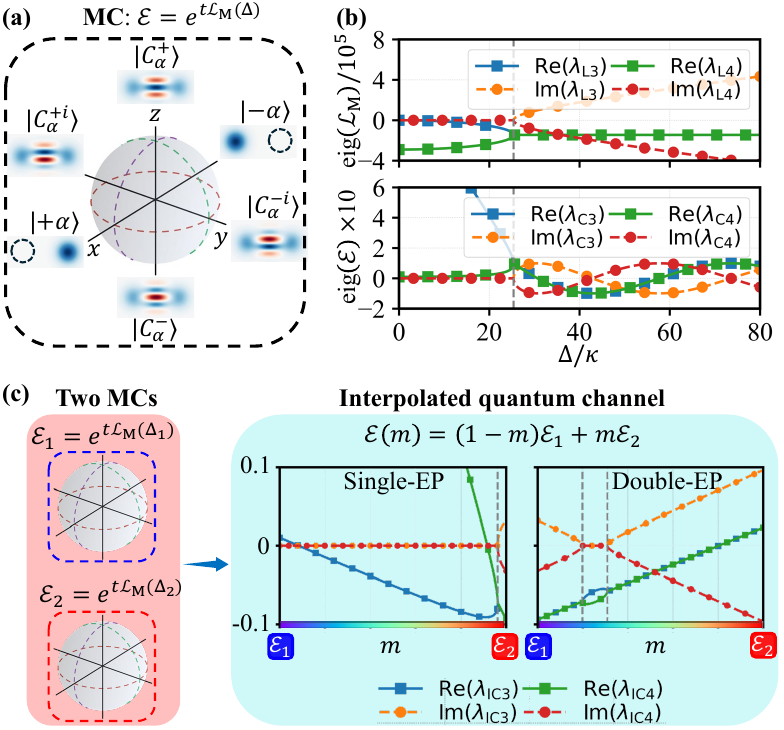}
\caption{\label{Fig1}(a) (Left) Bloch-sphere representation of the protected Kerr-cat qubit, which forms an MC $\mathcal{E}$.
(b) Third and fourth eigenvalues of $\mathcal{L}_{\text{M}}$ and $\mathcal{E}$ at $\kappa t=1$; the gray dashed lines mark the Liouvillian EP and CEP, respectively.
(c)
Interpolating between two MCs, the interpolated channel selectively exhibits single-EP or double-EP behavior.
Parameters: $K/\kappa=0.65\times10^3$ and $\varepsilon_{2}/\kappa=1.50\times10^3$.
}
\end{figure}

\textit{Markovian channel}---We consider a driven-dissipative Kerr-cat qubit~\cite{Wielinga1993,Goto2016,Minganti2016,Puri2017,Puri2019,Wang2019,Grimm2020,Lin2024}, governed by the Lindblad master equation ($\hbar=1$):
\begin{align}
\dot{\rho} &= \mathcal{L}[\rho]= -i[H,\,\rho] + \kappa \, \mathcal{D}[a] \rho,
\label{eq:master} \\
H &= \Delta a^\dagger a -Ka^{\dagger2}a^{2} + (\varepsilon_2 a^{\dagger 2} + \varepsilon_2^* a^2),
\label{eq:H_new}\\
\mathcal{D}[a]\rho &= a\rho a^\dagger - \frac{1}{2}\{a^\dagger a, \rho\}.
\end{align}
Here, $\varepsilon_2$ is the two-photon drive amplitude, $a^{\dagger}$ ($a$) is the annihilation (creation) operator, $\Delta$ is drive-resonator detuning, $K$ is the Kerr coefficient, and $\kappa$ is the single-photon loss rate.
For $\Delta,\,\kappa\ll4K|\alpha|^{2}$ with $\alpha=\sqrt{\varepsilon_2/K}$, the dynamics is restricted to the subspace spanned by the even and odd cat states $|\mathcal{C}^\pm_\alpha\rangle=\mathcal{N}_{\alpha}^{\pm}(|\alpha\rangle\pm|-\alpha\rangle)$ [Fig.~\ref{Fig1}(a)], with $\mathcal{N}_{\alpha}^{\pm}=1/\sqrt{2(1\pm e^{-2|\alpha|^{2}})}$. Single-photon loss induces bidirectional jumps between the logical states \cite{Han2026SCPMA,Yang2026}, whereas a finite $\Delta$ lifts the cat-state degeneracy, and their interplay shapes the Liouvillian spectrum underlying the Markovian character of the dynamics.

Vectorizing the density matrix, $\rho \mapsto \mathbf{V} = \mathrm{vec}(\rho)$, maps Eq.~\eqref{eq:master} to a linear ordinary differential equation $\dot{\mathbf{V}}(t) = \mathcal{L} \mathbf{V}(t)$.
Projecting onto the logical subspace $\{ |\mathcal{C}^+_\alpha\rangle, |\mathcal{C}^-_\alpha\rangle \}$ and using
$a |\mathcal{C}^\pm_\alpha\rangle = \alpha p^{\mp 1} |\mathcal{C}^\mp_\alpha\rangle$ and $a^\dagger |\mathcal{C}^\pm_\alpha\rangle = \alpha^{*} p^{\pm 1} |\mathcal{C}^\mp_\alpha\rangle$, with $p=\mathcal{N}_{\alpha}^+/\mathcal{N}_{\alpha}^-$,
the projected Liouvillian reduces to a $4\times4$ matrix $\mathcal{L}_{\mathrm{M}}$ (see Supplemental Material~\cite{supplement}), whose eigenvalues are
\begin{align}
\lambda_{\text{L1}} &= 0,\,
&\lambda_{\text{L2}} &= -\alpha^2\kappa p^{+}_{2}, \nonumber\\
\lambda_{\text{L3}} &= \alpha^2(\sqrt{\Lambda}-\frac{\kappa p^{+}_2}{2}),\,
&\lambda_{\text{L4}} &= \alpha^2(-\sqrt{\Lambda}-\frac{\kappa p^{+}_2}{2}),
\end{align}
with $p^{\pm}_j = p^{-j} \pm p^{j}$ and $\Lambda=\kappa^2-(\Delta p^{-}_{2})^2$.
A Liouvillian EP occurs when $\lambda_{\text{L3}}=\lambda_{\text{L4}}$, as shown in the upper panel in Fig.~\ref{Fig1}(b), where the gray dashed line marks the EP condition $\Delta/\kappa=1/p^{-}_{2}$ ($\Lambda=0$)~\cite{Han2026}.
The simultaneous eigenvector coalescence is verified in the Supplemental Material~\cite{supplement}.
The spectrum exhibits two phases: it is fully real for $\Delta/\kappa<1/p^{-}_{2}$ (the overdamped phase with $\Lambda>0$), i.e., $\{0,\,\lambda_{\text{L}2},\,\lambda_{\text{L}3},\,\lambda_{\text{L}4}\}\in\mathbb{R}$, and contains one complex-conjugate pair for $\Delta/\kappa>1/p^{-}_{2}$ (the underdamped phase with $\Lambda<0$), i.e., $\{0,\,\lambda_{\text{L}2}\}\in\mathbb{R}$ while $\{\lambda_{\text{L}3},\,\lambda_{\text{L}4}=\lambda^{*}_{\text{L}3}\}\in\mathbb{C}$.
We use dimensionless units normalized by $\kappa$, with experimentally feasible ratios~\cite{Grimm2020} $K/\kappa=0.65\times10^{3}$ and $\varepsilon_{2}/\kappa=1.50\times10^{3}$ (giving $\alpha\approx1.52$).
The spectral features reported below depend on these dimensionless ratios, not on absolute frequency scales.

Within the Markovian assumption, the time evolution is $\mathbf{V}(t)=e^{\mathcal{L}_{\mathrm{M}}t}\mathbf{V}(0)$. We define the Markovian quantum channel as $\mathcal{E}=e^{\mathcal{L}_{\mathrm{M}}t}$.
Because $\mathcal{E}$ and $\mathcal{L}_{\mathrm{M}}$ share the same eigenvectors, the CEP of $\mathcal{E}$ inherits the time-independent Liouvillian EP structure of $\mathcal{L}_{\mathrm{M}}$, with channel eigenvalues $\lambda_{\text{C}k}=e^{\lambda_{\text{L}k}t}$; see the bottom panel in Fig.~\ref{Fig1}(b).

\textit{Interpolated channel}---Interpolating two MCs that are away from CEPs can selectively generate non-Markovian channels (NMCs) at ICEPs.
Analyzing EPs at the channel level thus gives access a broader landscape of degeneracies than Liouvillian analysis alone.
Consider the linear interpolation
\begin{equation}
\mathcal{E}(m)=(1-m)\mathcal{E}_1+m\mathcal{E}_2,\qquad m\in[0,1],
\end{equation}
between the two endpoint MCs $\mathcal{E}_{i}=e^{\mathcal{L}_{\mathrm{M}}(\Delta_{i})t}$ ($i=1,2$), whose detunings $\Delta_i$ lie in the same or different spectral phases.
Physically, this describes a state undergoing $\mathcal{E}_{1}$ with probability $1-m$ and $\mathcal{E}_{2}$ with probability $m$, as depicted in Fig.~\ref{Fig1}(c).
The eigenvalues $\lambda_{{\text{IC}}k}$ and eigenvectors $\mathbf{V}_{{\text{IC}}k}$ of $\mathcal{E}(m)$ depend on $m$ (see Supplemental Material~\cite{supplement}).
The ICEP locations are given by the two solutions for $m$:
\begin{align}
m^{\pm}_{\text{EP}} =&\left[\Lambda_1\Lambda_2 \mathcal{C}_1 -2\sqrt{\Lambda_1}\sqrt{\Lambda_2}\Lambda_{12}\mathcal{S}_{12}\right.\nonumber\\
&\left.\pm\sqrt{4\Lambda_1\Lambda_2\Lambda^2_{12}\mathcal{S}^2_{12}-\Lambda^2_1\Lambda^2_2\mathcal{C}_1\mathcal{C}_2}\right]\nonumber\\
&/\left[\Lambda_1\Lambda_2(\mathcal{C}_1+\mathcal{C}_2)-4\sqrt{\Lambda_1}\sqrt{\Lambda_2}\Lambda_{12}\mathcal{S}_{12}\right],
\end{align}
with 
\begin{align}
\Lambda_j=&\kappa^2-(\Delta_{j}p^{-}_{2})^2,\, &\Lambda_{12}&=\kappa^2-\Delta_1\Delta_2 (p^{-}_{2})^2, \nonumber\\
\mathcal{C}_{j}=&\cosh(2x_{j})-1, \,&\mathcal{S}_{12}&=\prod_{j=1}^{2} \sinh(x_{j}),
\end{align}
where $x_{j}=t\alpha^2\sqrt{\Lambda_j}$.
An ICEP emerges whenever $0<m^{\pm}_{\text{EP}}<1$.
Conversely, the boundary solutions $m^{-}_{\text{EP}}=1$ ($m^{+}_{\text{EP}}=0$) or $m^{+}_{\text{EP}}=1$ ($m^{-}_{\text{EP}}=0$) recover of the original MC $\mathcal{E}_1$ or $\mathcal{E}_2$.
See Supplemental
Material~\cite{supplement} for the time evolution of $m^{\pm}_{\text{EP}}$ as functions of $\Delta_{1}/\kappa$ and $\Delta_{2}/\kappa$.
At an ICEP, the coalesced eigenvalue is
\begin{equation}
\lambda_{\mathrm{EP}}
=e^{-\frac{p^{+}_{2}}{2}\alpha^{2}\kappa t}
\Big[(1-m^{\pm}_{\mathrm{EP}})\cosh x_{1}+m^{\pm}_{\mathrm{EP}}\cosh x_{2}\Big].
\label{eq:lamEP}
\end{equation}
Thus, the interpolated channel can exhibit either a single or a double ICEP, with the coalesced eigenvalue $\lambda_{\mathrm{EP}}$.

\textit{Markovian--Non-Markovian phase diagram}---A snapshot channel is Markovian if it is an element of a time-independent Markovian semigroup, $\mathcal{E}=e^{\mathcal{L}t}$, and this holds if and only if some branch of $\log\mathcal{E}$ is Hermiticity preserving (HP), trace annihilating, and conditionally completely positive (cCP)~\cite{Wolf2008}.
Trace annihilating is automatically satisfied if we start from a trace preserving map.
Whether a snapshot at an ICEP meets these conditions is certified by the three criteria proved below~\cite{supplement}.

First, suppose a snapshot channel $\mathcal{E}$ at an EP is Markovian, $\mathcal{E}=e^{t\mathcal{L}}$ with $\mathcal{L}$ HP. Then $\lambda_k=e^{t\nu_k}$, where $\lambda_k$ and $\nu_k$ are the eigenvalues of $\mathcal{E}$ and $\mathcal{L}$, respectively, and HP requires a conjugate-symmetric spectrum $\{\nu_k\}=\{\nu_k^{*}\}$~\cite{Wolf2008,Minganti2018}.
An EP with a coalesced, defective eigenvalue $\lambda_{\mathrm{EP}}<0$ in a single Jordan block forces $\nu=t^{-1}[\ln|\lambda_{\mathrm{EP}}|+i(2k+1)\pi]$ ($k\in\mathbb{Z}$). This is non-real for every $k$ and appears twice without a conjugate partner, so the conjugate symmetry fails:
\begin{align}
\lambda_{\mathrm{EP}}<0
\ \Rightarrow\
\text{no HP logarithm} 
\ \Rightarrow\
\text{NMC}.
\label{eq:criterion-non-HP}
\end{align}
We refer to this as the non-HP criterion.

Second, when $\lambda_{\mathrm{EP}}>0$, the principal branch $\nu=t^{-1}\ln\lambda_{\mathrm{EP}}$ ($k=0$) is real and degenerate, hence conjugate symmetric, so the unique logarithm $\mathcal{L}_{0}=t^{-1}\log\mathcal{E}$ is HP. It remains to determine whether $\mathcal{L}_{0}$ generates a valid evolution, i.e., whether it is cCP.
Writing the Choi map as $J(\mathcal{A})=\sum_{ij}\mathcal{A}(|i\rangle\langle j|) \otimes|i\rangle\langle j|$ ~\cite{Wolf2008,Wolf2008CMP}, the smallest eigenvalue of $J(e^{s\mathcal{L}_{0}})$ reads
\begin{align}
\lambda_{\min}(s)&=\lambda_{\min}\big[J\big(e^{s\mathcal{L}_{0}}\big)\big] =s\,\mu+O(s^{2}),\\
\mu&=\lambda_{\min}[M],
\qquad
M_{lr}=\langle\mathbf{e}_{l}|J(\mathcal{L}_{0})|\mathbf{e}_{r}\rangle,
\label{eq:mu}
\end{align}
with $\lambda_{\min}[\mathcal{M}]$ the minimal eigenvalue of $\mathcal{M}$, $|\Omega\rangle=\sum_i|ii\rangle$, and $\{|\Omega\rangle/\sqrt{\langle\Omega|\Omega\rangle},|\mathbf{e}_1\rangle, |\mathbf{e}_2\rangle,|\mathbf{e}_3\rangle\}$ is an orthonormal basis with $|\mathbf{e}_1\rangle=(|00\rangle-|11\rangle)/\sqrt{2}$, $|\mathbf{e}_2\rangle=|01\rangle$, and $|\mathbf{e}_3\rangle=|10\rangle$.
Hence
\begin{align}
\lambda_{\mathrm{EP}}>0,\ 
\mu<0
\ &\Rightarrow\
\exists\, s>0:\ \lambda_{\min}\!\big[J(e^{s\mathcal{L}_{0}})\big]<0\nonumber\\
\ &\Rightarrow\
\text{no cCP logarithm}
\ \Rightarrow\
\text{NMC},
\label{eq:criterion-non-cCP}
\end{align}
is the non-cCP criterion. Third, a snapshot with $\lambda_{\mathrm{EP}}>0$ and $\mu\geq0$ satisfies the HP-cCP criterion and is an MC.

\begin{figure*}
\centering
\includegraphics[width =\textwidth]{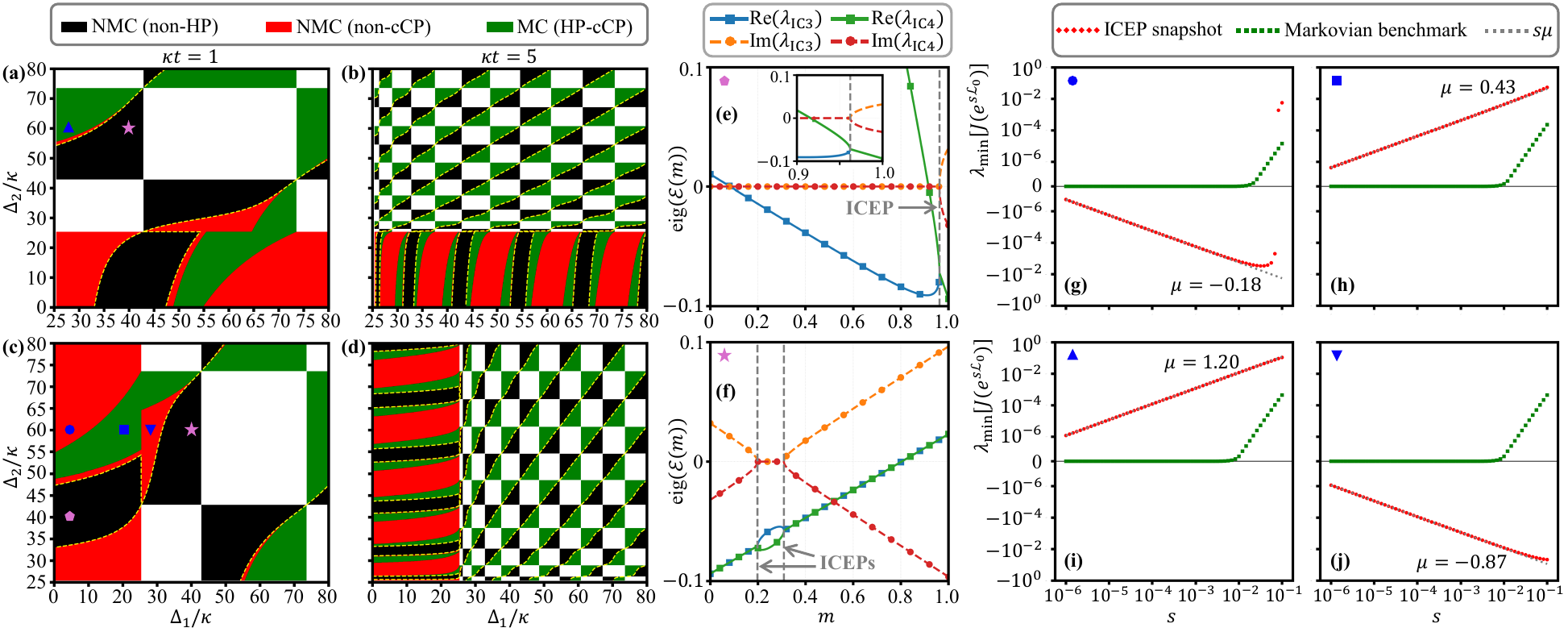}
\caption{\label{Fig2} Certified Markovian--non-Markovian phase diagram
of the ICEP snapshots.
(a)--(d): Phase diagrams in the $(\Delta_1/\kappa,\Delta_2/\kappa)$ plane at $\kappa t$ $=$ $1$ and $5$ (left to right), for the ICEP solutions $m^{-}_{\mathrm{EP}}(\Delta_{1},\Delta_{2})$ (upper row) and $m^{+}_{\mathrm{EP}}(\Delta_{1},\Delta_{2})$ (lower row).
$\lambda_{\mathrm{EP}}$ is real throughout the colored regions, with the yellow dashed line marking $\lambda_{\mathrm{EP}}=0$; white indicates that no valid ICEP exists.
Black: $\lambda_{\mathrm{EP}}<0$, certified as an NMC by the non-HP criterion [Eq.~\eqref{eq:criterion-non-HP}].
Red: $\lambda_{\mathrm{EP}}>0$ with $\mu<0$, certified as an NMC by the non-cCP criterion [Eq.~\eqref{eq:criterion-non-cCP}].
Green: $\lambda_{\mathrm{EP}}>0$ with $\mu\geq0$, certified as an MC by passing the HP-cCP test.
(e), (f): Eigenvalues $\lambda_{\mathrm{IC}3,4}$ versus $m$ at points $(5,40)$ in (a) and $(40,60)$ in (c) of the $(\Delta_1/\kappa,\Delta_2/\kappa)$ plane, confirming ICEPs with negative coalesced eigenvalues.
(g)--(j): Smallest eigenvalue of $J[e^{s\mathcal{L}_{0}}]$ versus $s$ for the marked ICEPs: (g) $(5,60)$ and (h) $(20,60)$ with $m=m^{+}_{\mathrm{EP}}$; (i) and (j) $(28,60)$ with $m=m^{-}_{\mathrm{EP}}$ and $m=m^{+}_{\mathrm{EP}}$, respectively.
Green dashed curve: Markovian benchmark; gray dotted lines: small-$s$ asymptote $s\mu$.
Parameters: $K/\kappa=0.65\times10^{3}$ and $\varepsilon_{2}/\kappa=1.50\times10^{3}$.}
\end{figure*}

With these three certification criteria established, we examine how certification is distributed in the $(\Delta_1/\kappa,\Delta_2/\kappa)$ plane and how it evolves with time. 
Two structures stand out in Figs.~\ref{Fig2}(a)--\ref{Fig2}(d).
First, at each time, the lower panel mirrors the upper one about $\Delta_2=\Delta_1$: $m^{+}_{\mathrm{EP}}(\Delta_1,\Delta_2)=m^{-}_{\mathrm{EP}}(\Delta_2,\Delta_1)$ between rows (not within a panel; see Supplemental Material~\cite{supplement}), and the certified domains inherit this symmetry.
Second, the three colors form extended certified domains.

At $\kappa t=1$, we first consider the single-EP region, where one endpoint is overdamped and the other underdamped, i.e.\ $\Lambda_j>0>\Lambda_{j^{\prime}}$ ($j,j^{\prime}\in\{1,2\}$, $j\neq j^{\prime}$, with $j$ and $j^{\prime}$ labeling the overdamped and underdamped endpoints, respectively).
For fixed $\Delta_j/\kappa$, sweeping $\Delta_{j'}/\kappa$ yields alternating certified domains, with the red domain being the largest and the green and black domains comparable in size.
Physically, this alternation follows from Eq.~\eqref{eq:lamEP}: $\cosh x_j\geq1$ gives a positive, exponentially growing contribution, whereas the purely imaginary $x_{j'}$ makes $\cos|x_{j^\prime}|$ oscillate with $|x_{j^\prime}|$ increasing in $\Delta_{j^\prime}$. The oscillatory term therefore periodically dominates, changing the sign of the weighted sum.
Thus, along the sweep the certified domains form one single connected region, with their spacing set by the single oscillatory phase, and every point is a certified working point, with detuning acting as a continuous switch between MC and NMC.

In the double-EP region, both endpoints are underdamped ($\Lambda_{1,2}<0$): $x_1$ and $x_2$ are both purely imaginary, and $\lambda_{\mathrm{EP}}$ is a weighted sum of two oscillations $x_{1(2)}=\cos|x_{1(2)}|$, rather than an exponential background plus an oscillation.
The double-EP region is not a single connected region but four disconnected colored blocks separated by white no-EP regions.
A sweep therefore alternates between certified domains within each block and crosses no-EP regions in which EP-based switching is suspended.
This offers a second switching mode on the same platform: block-to-block switching across no-EP regions, in contrast to the continuous alternation in the single-EP region.
This distinction underlies different time evolutions of the two regions.

As $\kappa t$ increases, the two regions evolve distinctly.
In the single-EP region the total certified area remains unchanged and only the internal domain boundaries shift, since the alternation is driven by the single oscillation phase $x_{j'}\propto\kappa t$, whose number of sign changing over a fixed $\Delta_{j'}$ interval grows linearly with time.
In the double-EP region, by contrast, the fragmentation is structural: the rectangular blocks multiply and shrink because both oscillation phases scale as $x_{1,2}\propto\kappa t$, and the $\lambda_{\mathrm{EP}}=0$ boundaries accumulate faster than for a single phase; meanwhile the red domains shrink, split, and disappear, so that each small block ends up with comparably sized green and black domains separated by a $\lambda_{\mathrm{EP}}=0$ boundary.
The switching between Markovian and non-Markovian behavior thereby becomes progressively more sensitive to parameters and time in both regions: the multiplying domain boundaries offer a growing number of switch points along a single parameter sweep, so memory effects can be switched on and off repeatedly and at ever finer parameter intervals.
This complements the control demonstrated in Ref.~\cite{Liu2011}, where the transition is monitored by following the trace distance between two states in time; our criteria instead certify from the channel spectrum of a single snapshot.
Formulated here on a continuous-variable platform, such criteria apply equally to discrete-variable and hybrid continuous-discrete-variable platforms. 
In short, Figs.~\ref{Fig2}(a)--\ref{Fig2}(d) map out a fully certified Markovian--non-Markovian phase diagram whose character is controllable through the detunings, with a nontrivial evolution under increasing time.

To verify the criteria across regions, Figs.~\ref{Fig2}(e)--\ref{Fig2}(j) examine the marked points of Figs.~\ref{Fig2}(a) and~\ref{Fig2}(c).
We first verify the non-HP criterion in Eq.~\eqref{eq:criterion-non-HP} for $\lambda_{\mathrm{EP}}<0$ by plotting $\lambda_{\mathrm{IC3,4}}$ versus $m$ in Figs.~\ref{Fig2}(e)--\ref{Fig2}(f).
At $(\Delta_1/\kappa,\Delta_2/\kappa)=(5,40)$ (purple pentagon in the single-EP region of Fig.~\ref{Fig2}(c)), a single ICEP appear at $m=m^{+}_{\mathrm{EP}}\simeq0.9623$ in Fig.~\ref{Fig2}(e).
In contrast, at $(\Delta_1/\kappa,\Delta_2/\kappa)=(40,60)$ (purple stars in the double-EP regions of Figs.~\ref{Fig2}(a) and \ref{Fig2}(c)), double ICEPs appear in Fig.~\ref{Fig2}(f) at $m=m^{-}_{\mathrm{EP}}\simeq0.3092$ and $m=m^{+}_{\mathrm{EP}}\simeq0.1979$.
At all these ICEPs the imaginary parts vanish and the coalesced real eigenvalue is negative, so the non-HP criterion certifies non-Markovianity.

We next verify the non-cCP and HP-cCP criteria in Eq.~\eqref{eq:criterion-non-cCP} for $\lambda_{\mathrm{EP}}>0$ by plotting the smallest eigenvalue of $J[e^{s\mathcal{L}_{0}}]$ versus $s$ in Figs.~\ref{Fig2}(g)--\ref{Fig2}(j).
Two logarithms are examined in each panel: for the ICEP snapshot (red), $\mathcal{L}_{0}=t^{-1}\log\mathcal{E}(m^{\pm}_{\mathrm{EP}})$; for a Markovian benchmark (green), $\mathcal{L}_{0}=t^{-1}\log\mathcal{E}_{1} =\mathcal{L}_{\mathrm{M}}(\Delta_{1})$.
A genuine GKSL generator that must give $\lambda_{\min}(s)\geq0$ for all $s$.
Panels (g) and (h) examine the single-EP points $(5,60)$ (blue circle, red domain) and $(20,60)$ (blue square, green domain), both evaluated with $m=m^{+}_{\mathrm{EP}}$.
In (g), $\lambda_{\min}(s)$ remains negative (with $\mu\simeq-0.18$) up to $s\sim10^{-2}$ and turns positive only near $s\sim10^{-1}$, tracking $s\mu$ at small $s$ as Eq.~\eqref{eq:mu} predicts: the unique HP logarithm fails cCP, and the snapshot is certified non-Markovian by the non-cCP criterion.
The overlap with $s\mu$ at small $s$ reflects the linear term of Eq.~\eqref{eq:mu}, whose validity requires $s\|\mathcal{L}_{0}\|\ll1$; the bending away from the straight line beyond $s\sim10^{-2}$ marks the
breakdown of the first-order approximation, while the negativity at small $s$ remains the decisive evidence that no infinitesimal generator exists.
In (h), $\lambda_{\min}(s)$ stays positive throughout (with $\mu\simeq0.43$) and tracks $s\mu$: the snapshot passes the cCP test and is Markovian.
Panels (i) and (j) examine the double-EP point $(28,60)$ in Figs.~\ref{Fig2}(a) and~\ref{Fig2}(c) (blue upward and downward triangles, green and red domains), evaluated with $m=m^{-}_{\mathrm{EP}}$ and $m=m^{+}_{\mathrm{EP}}$, respectively.
Both behave as their colors dictate: in (i), $\lambda_{\min}(s)>0$ throughout with $\mu\simeq1.20$, whereas in (j) it stays negative with $\mu\simeq-0.87$, each tracking $s\mu$ at small $s$; the former passes and the latter fails the cCP test, certifying a Markovian and a non-Markovian ICEP, respectively.
In all panels the Markovian benchmark remains non-negative for every $s$, in contrast to the failing red curves.

\textit{Discussions and conclusions}---We have established ICEPs at the quantum-channel level as intrinsic, representation-independent witnesses whose spectra certify whether the underlying channel is Markovian.
Using a driven-dissipative cat qubit, we have shown that interpolating between two Markovian channels generates single and double EPs organized into a Markovian--non-Markovian phase diagram: domains certified non-Markovian by the non-HP and non-cCP criteria and Markovian domains identified by the HP-cCP test, with the character switched by crossing the domain boundaries.
All three criteria act directly on the channel map; no Liouvillian is constructed for the interpolated channel, and the master equation serves only to define the two Markovian endpoints. 
As the evolution time advances, the domains drift, split, and partially shrink, making the switching progressively more sensitive to parameters and time.
By shifting the focus from Liouvillians to quantum channels~\cite{Wong2026,Brandao2026}, our work establishes the intrinsic spectral structure of the channel as a faithful, representation-independent characterization of non-Markovian dynamics, with certified and controllable handles at EPs that are accessible across a broad range of quantum platforms.

Although we focus on the Kerr-cat qubit, a platform already implemented in several experiments~\cite{Leghtas2015,Grimm2020,He2023,Xu2025}, our framework requires only the channel map and applies equally to other continuous-variable~\cite{Fluhmann2019,Wollman2015,Braunstein2005,Serafini2017,Albert2019,Gertler2023,Ma2021,Cai2021} and hybrid continuous-discrete-variable platforms~\cite{Hoshi2025,Liu2026,Bazavan2026}.
Two open questions concern the relation of these spectral witnesses to
information-theoretic measures~\cite{Breuer2009,Rivas2010,Liu2011}:
whether double-EP induce stronger entanglement backflow than
single-EP ones, and whether revival of channel-capacity~\cite{Ruskai2002,Bylicka2014} scales with the number of EPs.
Resolving these questions would elevate channel EPs from mathematical labels to physical resources for quantum communication and sensing.
Experimentally, Markovian dynamics has been digitally simulated~\cite{Han2021}.
Arbitrary channels can be efficiently simulated with a single ancilla qubit~\cite{Wang2013,Shen2017}, and these protocols have been demonstrated experimentally~\cite{Hu2018}.
Robust optimal control keeps channel implementations accurate on noisy hardware~\cite{Chen2025}.
The EP trajectories generated by parameter variation may thus serve as measurable signatures of interpolation-induced non-Markovianity without requiring quantum process tomography.

\textit{Acknowledgments}---This work was supported by the National Natural Science Foundation of China (Grant Nos. 12475015, 12274080, 12474356, 12505021, and 11875108), the Quantum Science and Technology-National Science and Technology Major Project (Grant No. 2021ZD0300200), and the Natural Science Foundation of Fujian Province (Grant No. 2025J01383).

\bibliography{Ref}

\begin{thebibliography}{75}%
\makeatletter
\providecommand \@ifxundefined [1]{%
 \@ifx{#1\undefined}
}%
\providecommand \@ifnum [1]{%
 \ifnum #1\expandafter \@firstoftwo
 \else \expandafter \@secondoftwo
 \fi
}%
\providecommand \@ifx [1]{%
 \ifx #1\expandafter \@firstoftwo
 \else \expandafter \@secondoftwo
 \fi
}%
\providecommand \natexlab [1]{#1}%
\providecommand \enquote  [1]{``#1''}%
\providecommand \bibnamefont  [1]{#1}%
\providecommand \bibfnamefont [1]{#1}%
\providecommand \citenamefont [1]{#1}%
\providecommand \href@noop [0]{\@secondoftwo}%
\providecommand \href [0]{\begingroup \@sanitize@url \@href}%
\providecommand \@href[1]{\@@startlink{#1}\@@href}%
\providecommand \@@href[1]{\endgroup#1\@@endlink}%
\providecommand \@sanitize@url [0]{\catcode `\\12\catcode `\$12\catcode `\&12\catcode `\#12\catcode `\^12\catcode `\_12\catcode `\%12\relax}%
\providecommand \@@startlink[1]{}%
\providecommand \@@endlink[0]{}%
\providecommand \url  [0]{\begingroup\@sanitize@url \@url }%
\providecommand \@url [1]{\endgroup\@href {#1}{\urlprefix }}%
\providecommand \urlprefix  [0]{URL }%
\providecommand \Eprint [0]{\href }%
\providecommand \doibase [0]{https://doi.org/}%
\providecommand \selectlanguage [0]{\@gobble}%
\providecommand \bibinfo  [0]{\@secondoftwo}%
\providecommand \bibfield  [0]{\@secondoftwo}%
\providecommand \translation [1]{[#1]}%
\providecommand \BibitemOpen [0]{}%
\providecommand \bibitemStop [0]{}%
\providecommand \bibitemNoStop [0]{.\EOS\space}%
\providecommand \EOS [0]{\spacefactor3000\relax}%
\providecommand \BibitemShut  [1]{\csname bibitem#1\endcsname}%
\let\auto@bib@innerbib\@empty
\bibitem [{\citenamefont {Gorini}\ \emph {et~al.}(1978)\citenamefont {Gorini}, \citenamefont {Frigerio}, \citenamefont {Verri}, \citenamefont {Kossakowski},\ and\ \citenamefont {Sudarshan}}]{Gorini1978}%
  \BibitemOpen
  \bibfield  {author} {\bibinfo {author} {\bibfnamefont {V.}~\bibnamefont {Gorini}}, \bibinfo {author} {\bibfnamefont {A.}~\bibnamefont {Frigerio}}, \bibinfo {author} {\bibfnamefont {M.}~\bibnamefont {Verri}}, \bibinfo {author} {\bibfnamefont {A.}~\bibnamefont {Kossakowski}},\ and\ \bibinfo {author} {\bibfnamefont {E.}~\bibnamefont {Sudarshan}},\ }\bibfield  {title} {\bibinfo {title} {Properties of quantum markovian master equations},\ }\href {https://doi.org/10.1016/0034-4877(78)90050-2} {\bibfield  {journal} {\bibinfo  {journal} {Rep. Math. Phys.}\ }\textbf {\bibinfo {volume} {13}},\ \bibinfo {pages} {149} (\bibinfo {year} {1978})}\BibitemShut {NoStop}%
\bibitem [{\citenamefont {Breuer}\ and\ \citenamefont {Petruccione}(2007)}]{Breuer2007}%
  \BibitemOpen
  \bibfield  {author} {\bibinfo {author} {\bibfnamefont {H.-P.}\ \bibnamefont {Breuer}}\ and\ \bibinfo {author} {\bibfnamefont {F.}~\bibnamefont {Petruccione}},\ }\href@noop {} {\emph {\bibinfo {title} {The Theory of Open Quantum Systems}}}\ (\bibinfo  {publisher} {Oxford University Press},\ \bibinfo {year} {2007})\BibitemShut {NoStop}%
\bibitem [{\citenamefont {{Beth Ruskai}}\ \emph {et~al.}(2002)\citenamefont {{Beth Ruskai}}, \citenamefont {Szarek},\ and\ \citenamefont {Werner}}]{Ruskai2002}%
  \BibitemOpen
  \bibfield  {author} {\bibinfo {author} {\bibfnamefont {M.}~\bibnamefont {{Beth Ruskai}}}, \bibinfo {author} {\bibfnamefont {S.}~\bibnamefont {Szarek}},\ and\ \bibinfo {author} {\bibfnamefont {E.}~\bibnamefont {Werner}},\ }\bibfield  {title} {\bibinfo {title} {An analysis of completely-positive trace-preserving maps on m2},\ }\href {https://doi.org/https://doi.org/10.1016/S0024-3795(01)00547-X} {\bibfield  {journal} {\bibinfo  {journal} {Lin. Alg. Appl.}\ }\textbf {\bibinfo {volume} {347}},\ \bibinfo {pages} {159} (\bibinfo {year} {2002})}\BibitemShut {NoStop}%
\bibitem [{\citenamefont {Gorini}\ \emph {et~al.}(1976)\citenamefont {Gorini}, \citenamefont {Kossakowski},\ and\ \citenamefont {Sudarshan}}]{Gorini1976}%
  \BibitemOpen
  \bibfield  {author} {\bibinfo {author} {\bibfnamefont {V.}~\bibnamefont {Gorini}}, \bibinfo {author} {\bibfnamefont {A.}~\bibnamefont {Kossakowski}},\ and\ \bibinfo {author} {\bibfnamefont {E.~C.~G.}\ \bibnamefont {Sudarshan}},\ }\bibfield  {title} {\bibinfo {title} {Completely positive dynamical semigroups of n‐level systems},\ }\href {https://doi.org/10.1063/1.522979} {\bibfield  {journal} {\bibinfo  {journal} {J. Math. Phys.}\ }\textbf {\bibinfo {volume} {17}},\ \bibinfo {pages} {821} (\bibinfo {year} {1976})}\BibitemShut {NoStop}%
\bibitem [{\citenamefont {Lindblad}(1976)}]{Lindblad1976}%
  \BibitemOpen
  \bibfield  {author} {\bibinfo {author} {\bibfnamefont {G.}~\bibnamefont {Lindblad}},\ }\bibfield  {title} {\bibinfo {title} {On the generators of quantum dynamical semigroups},\ }\href {https://doi.org/10.1007/BF01608499} {\bibfield  {journal} {\bibinfo  {journal} {Commun. Math. Phys.}\ }\textbf {\bibinfo {volume} {48}},\ \bibinfo {pages} {119} (\bibinfo {year} {1976})}\BibitemShut {NoStop}%
\bibitem [{\citenamefont {Chru\ifmmode \acute{s}\else \'{s}\fi{}ci\ifmmode~\acute{n}\else \'{n}\fi{}ski}\ and\ \citenamefont {Kossakowski}(2010)}]{Chruscinski2010}%
  \BibitemOpen
  \bibfield  {author} {\bibinfo {author} {\bibfnamefont {D.}~\bibnamefont {Chru\ifmmode \acute{s}\else \'{s}\fi{}ci\ifmmode~\acute{n}\else \'{n}\fi{}ski}}\ and\ \bibinfo {author} {\bibfnamefont {A.}~\bibnamefont {Kossakowski}},\ }\bibfield  {title} {\bibinfo {title} {Non-markovian quantum dynamics: Local versus nonlocal},\ }\href {https://doi.org/10.1103/PhysRevLett.104.070406} {\bibfield  {journal} {\bibinfo  {journal} {Phys. Rev. Lett.}\ }\textbf {\bibinfo {volume} {104}},\ \bibinfo {pages} {070406} (\bibinfo {year} {2010})}\BibitemShut {NoStop}%
\bibitem [{\citenamefont {Breuer}\ \emph {et~al.}(2016)\citenamefont {Breuer}, \citenamefont {Laine}, \citenamefont {Piilo},\ and\ \citenamefont {Vacchini}}]{Breuer2016}%
  \BibitemOpen
  \bibfield  {author} {\bibinfo {author} {\bibfnamefont {H.-P.}\ \bibnamefont {Breuer}}, \bibinfo {author} {\bibfnamefont {E.-M.}\ \bibnamefont {Laine}}, \bibinfo {author} {\bibfnamefont {J.}~\bibnamefont {Piilo}},\ and\ \bibinfo {author} {\bibfnamefont {B.}~\bibnamefont {Vacchini}},\ }\bibfield  {title} {\bibinfo {title} {Colloquium: Non-markovian dynamics in open quantum systems},\ }\href {https://doi.org/10.1103/RevModPhys.88.021002} {\bibfield  {journal} {\bibinfo  {journal} {Rev. Mod. Phys.}\ }\textbf {\bibinfo {volume} {88}},\ \bibinfo {pages} {021002} (\bibinfo {year} {2016})}\BibitemShut {NoStop}%
\bibitem [{\citenamefont {Breuer}\ \emph {et~al.}(2009)\citenamefont {Breuer}, \citenamefont {Laine},\ and\ \citenamefont {Piilo}}]{Breuer2009}%
  \BibitemOpen
  \bibfield  {author} {\bibinfo {author} {\bibfnamefont {H.-P.}\ \bibnamefont {Breuer}}, \bibinfo {author} {\bibfnamefont {E.-M.}\ \bibnamefont {Laine}},\ and\ \bibinfo {author} {\bibfnamefont {J.}~\bibnamefont {Piilo}},\ }\bibfield  {title} {\bibinfo {title} {Measure for the degree of non-markovian behavior of quantum processes in open systems},\ }\href {https://doi.org/10.1103/PhysRevLett.103.210401} {\bibfield  {journal} {\bibinfo  {journal} {Phys. Rev. Lett.}\ }\textbf {\bibinfo {volume} {103}},\ \bibinfo {pages} {210401} (\bibinfo {year} {2009})}\BibitemShut {NoStop}%
\bibitem [{\citenamefont {Hall}\ \emph {et~al.}(2014)\citenamefont {Hall}, \citenamefont {Cresser}, \citenamefont {Li},\ and\ \citenamefont {Andersson}}]{Hall2014}%
  \BibitemOpen
  \bibfield  {author} {\bibinfo {author} {\bibfnamefont {M.~J.~W.}\ \bibnamefont {Hall}}, \bibinfo {author} {\bibfnamefont {J.~D.}\ \bibnamefont {Cresser}}, \bibinfo {author} {\bibfnamefont {L.}~\bibnamefont {Li}},\ and\ \bibinfo {author} {\bibfnamefont {E.}~\bibnamefont {Andersson}},\ }\bibfield  {title} {\bibinfo {title} {Canonical form of master equations and characterization of non-markovianity},\ }\href {https://doi.org/10.1103/PhysRevA.89.042120} {\bibfield  {journal} {\bibinfo  {journal} {Phys. Rev. A}\ }\textbf {\bibinfo {volume} {89}},\ \bibinfo {pages} {042120} (\bibinfo {year} {2014})}\BibitemShut {NoStop}%
\bibitem [{\citenamefont {Lin}\ \emph {et~al.}(2025)\citenamefont {Lin}, \citenamefont {Kuo}, \citenamefont {Lambert}, \citenamefont {Miranowicz}, \citenamefont {Nori},\ and\ \citenamefont {Chen}}]{Lin2025}%
  \BibitemOpen
  \bibfield  {author} {\bibinfo {author} {\bibfnamefont {J.-D.}\ \bibnamefont {Lin}}, \bibinfo {author} {\bibfnamefont {P.-C.}\ \bibnamefont {Kuo}}, \bibinfo {author} {\bibfnamefont {N.}~\bibnamefont {Lambert}}, \bibinfo {author} {\bibfnamefont {A.}~\bibnamefont {Miranowicz}}, \bibinfo {author} {\bibfnamefont {F.}~\bibnamefont {Nori}},\ and\ \bibinfo {author} {\bibfnamefont {Y.-N.}\ \bibnamefont {Chen}},\ }\bibfield  {title} {\bibinfo {title} {Non-markovian quantum exceptional points},\ }\href {https://doi.org/10.1038/s41467-025-56242-w} {\bibfield  {journal} {\bibinfo  {journal} {Nat. Commun.}\ }\textbf {\bibinfo {volume} {16}},\ \bibinfo {pages} {1289} (\bibinfo {year} {2025})}\BibitemShut {NoStop}%
\bibitem [{\citenamefont {Laine}\ \emph {et~al.}(2010)\citenamefont {Laine}, \citenamefont {Piilo},\ and\ \citenamefont {Breuer}}]{Laine2010}%
  \BibitemOpen
  \bibfield  {author} {\bibinfo {author} {\bibfnamefont {E.-M.}\ \bibnamefont {Laine}}, \bibinfo {author} {\bibfnamefont {J.}~\bibnamefont {Piilo}},\ and\ \bibinfo {author} {\bibfnamefont {H.-P.}\ \bibnamefont {Breuer}},\ }\bibfield  {title} {\bibinfo {title} {Measure for the non-markovianity of quantum processes},\ }\href {https://doi.org/10.1103/PhysRevA.81.062115} {\bibfield  {journal} {\bibinfo  {journal} {Phys. Rev. A}\ }\textbf {\bibinfo {volume} {81}},\ \bibinfo {pages} {062115} (\bibinfo {year} {2010})}\BibitemShut {NoStop}%
\bibitem [{\citenamefont {Chru\ifmmode \acute{s}\else \'{s}\fi{}ci\ifmmode~\acute{n}\else \'{n}\fi{}ski}\ \emph {et~al.}(2011)\citenamefont {Chru\ifmmode \acute{s}\else \'{s}\fi{}ci\ifmmode~\acute{n}\else \'{n}\fi{}ski}, \citenamefont {Kossakowski},\ and\ \citenamefont {Rivas}}]{Chruscinski2011}%
  \BibitemOpen
  \bibfield  {author} {\bibinfo {author} {\bibfnamefont {D.}~\bibnamefont {Chru\ifmmode \acute{s}\else \'{s}\fi{}ci\ifmmode~\acute{n}\else \'{n}\fi{}ski}}, \bibinfo {author} {\bibfnamefont {A.}~\bibnamefont {Kossakowski}},\ and\ \bibinfo {author} {\bibfnamefont {A.}~\bibnamefont {Rivas}},\ }\bibfield  {title} {\bibinfo {title} {Measures of non-markovianity: Divisibility versus backflow of information},\ }\href {https://doi.org/10.1103/PhysRevA.83.052128} {\bibfield  {journal} {\bibinfo  {journal} {Phys. Rev. A}\ }\textbf {\bibinfo {volume} {83}},\ \bibinfo {pages} {052128} (\bibinfo {year} {2011})}\BibitemShut {NoStop}%
\bibitem [{\citenamefont {Wolf}\ and\ \citenamefont {Cirac}(2008)}]{Wolf2008CMP}%
  \BibitemOpen
  \bibfield  {author} {\bibinfo {author} {\bibfnamefont {M.~M.}\ \bibnamefont {Wolf}}\ and\ \bibinfo {author} {\bibfnamefont {J.~I.}\ \bibnamefont {Cirac}},\ }\bibfield  {title} {\bibinfo {title} {Dividing quantum channels},\ }\href {https://doi.org/10.1007/s00220-008-0411-y} {\bibfield  {journal} {\bibinfo  {journal} {Commun. Math. Phys.}\ }\textbf {\bibinfo {volume} {279}},\ \bibinfo {pages} {147} (\bibinfo {year} {2008})}\BibitemShut {NoStop}%
\bibitem [{\citenamefont {Rivas}\ \emph {et~al.}(2010)\citenamefont {Rivas}, \citenamefont {Huelga},\ and\ \citenamefont {Plenio}}]{Rivas2010}%
  \BibitemOpen
  \bibfield  {author} {\bibinfo {author} {\bibfnamefont {{\'A}.}~\bibnamefont {Rivas}}, \bibinfo {author} {\bibfnamefont {S.~F.}\ \bibnamefont {Huelga}},\ and\ \bibinfo {author} {\bibfnamefont {M.~B.}\ \bibnamefont {Plenio}},\ }\bibfield  {title} {\bibinfo {title} {Entanglement and non-markovianity of quantum evolutions},\ }\href {https://doi.org/10.1103/PhysRevLett.105.050403} {\bibfield  {journal} {\bibinfo  {journal} {Phys. Rev. Lett.}\ }\textbf {\bibinfo {volume} {105}},\ \bibinfo {pages} {050403} (\bibinfo {year} {2010})}\BibitemShut {NoStop}%
\bibitem [{\citenamefont {Liu}\ \emph {et~al.}(2011)\citenamefont {Liu}, \citenamefont {Li}, \citenamefont {Huang}, \citenamefont {Li}, \citenamefont {Guo}, \citenamefont {Laine}, \citenamefont {Breuer},\ and\ \citenamefont {Piilo}}]{Liu2011}%
  \BibitemOpen
  \bibfield  {author} {\bibinfo {author} {\bibfnamefont {B.-H.}\ \bibnamefont {Liu}}, \bibinfo {author} {\bibfnamefont {L.}~\bibnamefont {Li}}, \bibinfo {author} {\bibfnamefont {Y.-F.}\ \bibnamefont {Huang}}, \bibinfo {author} {\bibfnamefont {C.-F.}\ \bibnamefont {Li}}, \bibinfo {author} {\bibfnamefont {G.-C.}\ \bibnamefont {Guo}}, \bibinfo {author} {\bibfnamefont {E.-M.}\ \bibnamefont {Laine}}, \bibinfo {author} {\bibfnamefont {H.-P.}\ \bibnamefont {Breuer}},\ and\ \bibinfo {author} {\bibfnamefont {J.}~\bibnamefont {Piilo}},\ }\bibfield  {title} {\bibinfo {title} {Experimental control of the transition from markovian to non-markovian dynamics of open quantum systems},\ }\href {https://doi.org/10.1038/nphys2085} {\bibfield  {journal} {\bibinfo  {journal} {Nat. Phys.}\ }\textbf {\bibinfo {volume} {7}},\ \bibinfo {pages} {931} (\bibinfo {year} {2011})}\BibitemShut {NoStop}%
\bibitem [{\citenamefont {Han}\ \emph {et~al.}(2024{\natexlab{a}})\citenamefont {Han}, \citenamefont {Wu}, \citenamefont {Huang}, \citenamefont {Wu}, \citenamefont {Yi}, \citenamefont {Wen}, \citenamefont {Yang},\ and\ \citenamefont {Zheng}}]{Han2024APL}%
  \BibitemOpen
  \bibfield  {author} {\bibinfo {author} {\bibfnamefont {P.-R.}\ \bibnamefont {Han}}, \bibinfo {author} {\bibfnamefont {F.}~\bibnamefont {Wu}}, \bibinfo {author} {\bibfnamefont {X.-J.}\ \bibnamefont {Huang}}, \bibinfo {author} {\bibfnamefont {H.-Z.}\ \bibnamefont {Wu}}, \bibinfo {author} {\bibfnamefont {W.}~\bibnamefont {Yi}}, \bibinfo {author} {\bibfnamefont {J.}~\bibnamefont {Wen}}, \bibinfo {author} {\bibfnamefont {Z.-B.}\ \bibnamefont {Yang}},\ and\ \bibinfo {author} {\bibfnamefont {S.-B.}\ \bibnamefont {Zheng}},\ }\bibfield  {title} {\bibinfo {title} {Characterization of non-markovianity with maximal extractable qubit-reservoir entanglement},\ }\href {https://doi.org/10.1063/5.0224195} {\bibfield  {journal} {\bibinfo  {journal} {Appl. Phys. Lett.}\ }\textbf {\bibinfo {volume} {125}},\ \bibinfo {pages} {124003} (\bibinfo {year} {2024}{\natexlab{a}})}\BibitemShut {NoStop}%
\bibitem [{\citenamefont {El-Ganainy}\ \emph {et~al.}(2018)\citenamefont {El-Ganainy}, \citenamefont {Makris}, \citenamefont {Khajavikhan}, \citenamefont {Musslimani}, \citenamefont {Rotter},\ and\ \citenamefont {Christodoulides}}]{Ganainy2018}%
  \BibitemOpen
  \bibfield  {author} {\bibinfo {author} {\bibfnamefont {R.}~\bibnamefont {El-Ganainy}}, \bibinfo {author} {\bibfnamefont {K.~G.}\ \bibnamefont {Makris}}, \bibinfo {author} {\bibfnamefont {M.}~\bibnamefont {Khajavikhan}}, \bibinfo {author} {\bibfnamefont {Z.~H.}\ \bibnamefont {Musslimani}}, \bibinfo {author} {\bibfnamefont {S.}~\bibnamefont {Rotter}},\ and\ \bibinfo {author} {\bibfnamefont {D.~N.}\ \bibnamefont {Christodoulides}},\ }\bibfield  {title} {\bibinfo {title} {Non-hermitian physics and pt symmetry},\ }\href {https://doi.org/10.1038/nphys4323} {\bibfield  {journal} {\bibinfo  {journal} {Nat. Phys.}\ }\textbf {\bibinfo {volume} {14}},\ \bibinfo {pages} {11} (\bibinfo {year} {2018})}\BibitemShut {NoStop}%
\bibitem [{\citenamefont {Miri}\ and\ \citenamefont {Alù}(2019)}]{Miri2019}%
  \BibitemOpen
  \bibfield  {author} {\bibinfo {author} {\bibfnamefont {M.-A.}\ \bibnamefont {Miri}}\ and\ \bibinfo {author} {\bibfnamefont {A.}~\bibnamefont {Alù}},\ }\bibfield  {title} {\bibinfo {title} {Exceptional points in optics and photonics},\ }\href {https://doi.org/10.1126/science.aar7709} {\bibfield  {journal} {\bibinfo  {journal} {Science}\ }\textbf {\bibinfo {volume} {363}},\ \bibinfo {pages} {eaar7709} (\bibinfo {year} {2019})}\BibitemShut {NoStop}%
\bibitem [{\citenamefont {{\"O}zdemir}\ \emph {et~al.}(2019)\citenamefont {{\"O}zdemir}, \citenamefont {Rotter}, \citenamefont {Nori},\ and\ \citenamefont {Yang}}]{Ozdemir2019}%
  \BibitemOpen
  \bibfield  {author} {\bibinfo {author} {\bibfnamefont {{\c{S}}.~K.}\ \bibnamefont {{\"O}zdemir}}, \bibinfo {author} {\bibfnamefont {S.}~\bibnamefont {Rotter}}, \bibinfo {author} {\bibfnamefont {F.}~\bibnamefont {Nori}},\ and\ \bibinfo {author} {\bibfnamefont {L.}~\bibnamefont {Yang}},\ }\bibfield  {title} {\bibinfo {title} {Parity--time symmetry and exceptional points in photonics},\ }\href {https://doi.org/10.1038/s41563-019-0304-9} {\bibfield  {journal} {\bibinfo  {journal} {Nat. Mater.}\ }\textbf {\bibinfo {volume} {18}},\ \bibinfo {pages} {783} (\bibinfo {year} {2019})}\BibitemShut {NoStop}%
\bibitem [{\citenamefont {Xue}(2026)}]{Xue2026PRL}%
  \BibitemOpen
  \bibfield  {author} {\bibinfo {author} {\bibfnamefont {P.}~\bibnamefont {Xue}},\ }\bibfield  {title} {\bibinfo {title} {Essay: Topological phases and exceptional points in non-hermitian systems},\ }\href {https://doi.org/10.1103/ll76-j2l5} {\bibfield  {journal} {\bibinfo  {journal} {Phys. Rev. Lett.}\ }\textbf {\bibinfo {volume} {136}},\ \bibinfo {pages} {170001} (\bibinfo {year} {2026})}\BibitemShut {NoStop}%
\bibitem [{\citenamefont {Han}\ \emph {et~al.}(2024{\natexlab{b}})\citenamefont {Han}, \citenamefont {Ning}, \citenamefont {Huang}, \citenamefont {Zheng}, \citenamefont {Yang}, \citenamefont {Wu}, \citenamefont {Yang}, \citenamefont {Su}, \citenamefont {Yang},\ and\ \citenamefont {Zheng}}]{Han2024NC}%
  \BibitemOpen
  \bibfield  {author} {\bibinfo {author} {\bibfnamefont {P.-R.}\ \bibnamefont {Han}}, \bibinfo {author} {\bibfnamefont {W.}~\bibnamefont {Ning}}, \bibinfo {author} {\bibfnamefont {X.-J.}\ \bibnamefont {Huang}}, \bibinfo {author} {\bibfnamefont {R.-H.}\ \bibnamefont {Zheng}}, \bibinfo {author} {\bibfnamefont {S.-B.}\ \bibnamefont {Yang}}, \bibinfo {author} {\bibfnamefont {F.}~\bibnamefont {Wu}}, \bibinfo {author} {\bibfnamefont {Z.-B.}\ \bibnamefont {Yang}}, \bibinfo {author} {\bibfnamefont {Q.-P.}\ \bibnamefont {Su}}, \bibinfo {author} {\bibfnamefont {C.-P.}\ \bibnamefont {Yang}},\ and\ \bibinfo {author} {\bibfnamefont {S.-B.}\ \bibnamefont {Zheng}},\ }\bibfield  {title} {\bibinfo {title} {Measuring topological invariants for higher-order exceptional points in quantum three-mode systems},\ }\href {https://doi.org/10.1038/s41467-024-54662-8} {\bibfield  {journal} {\bibinfo  {journal} {Nat. Commun.}\ }\textbf {\bibinfo {volume} {15}},\ \bibinfo {pages} {10293} (\bibinfo {year} {2024}{\natexlab{b}})}\BibitemShut
  {NoStop}%
\bibitem [{\citenamefont {Zhang}\ \emph {et~al.}(2025{\natexlab{a}})\citenamefont {Zhang}, \citenamefont {Han}, \citenamefont {Yu}, \citenamefont {Yang}, \citenamefont {Lü}, \citenamefont {Ning}, \citenamefont {Wu}, \citenamefont {Su}, \citenamefont {Yang}, \citenamefont {Yang},\ and\ \citenamefont {Zheng}}]{Zhang2025}%
  \BibitemOpen
  \bibfield  {author} {\bibinfo {author} {\bibfnamefont {H.-L.}\ \bibnamefont {Zhang}}, \bibinfo {author} {\bibfnamefont {P.-R.}\ \bibnamefont {Han}}, \bibinfo {author} {\bibfnamefont {X.-J.}\ \bibnamefont {Yu}}, \bibinfo {author} {\bibfnamefont {S.-B.}\ \bibnamefont {Yang}}, \bibinfo {author} {\bibfnamefont {J.-H.}\ \bibnamefont {Lü}}, \bibinfo {author} {\bibfnamefont {W.}~\bibnamefont {Ning}}, \bibinfo {author} {\bibfnamefont {F.}~\bibnamefont {Wu}}, \bibinfo {author} {\bibfnamefont {Q.-P.}\ \bibnamefont {Su}}, \bibinfo {author} {\bibfnamefont {C.-P.}\ \bibnamefont {Yang}}, \bibinfo {author} {\bibfnamefont {Z.-B.}\ \bibnamefont {Yang}},\ and\ \bibinfo {author} {\bibfnamefont {S.-B.}\ \bibnamefont {Zheng}},\ }\bibfield  {title} {\bibinfo {title} {Implementation and topological characterization of weyl exceptional rings in quantum-mechanical systems},\ }\href {https://doi.org/10.1016/j.scib.2025.05.041} {\bibfield  {journal} {\bibinfo  {journal} {Sci. Bull.}\ }\textbf {\bibinfo {volume} {70}},\ \bibinfo
  {pages} {2446} (\bibinfo {year} {2025}{\natexlab{a}})}\BibitemShut {NoStop}%
\bibitem [{\citenamefont {Abbasi}\ \emph {et~al.}(2022)\citenamefont {Abbasi}, \citenamefont {Chen}, \citenamefont {Naghiloo}, \citenamefont {Joglekar},\ and\ \citenamefont {Murch}}]{Abbasi2022}%
  \BibitemOpen
  \bibfield  {author} {\bibinfo {author} {\bibfnamefont {M.}~\bibnamefont {Abbasi}}, \bibinfo {author} {\bibfnamefont {W.}~\bibnamefont {Chen}}, \bibinfo {author} {\bibfnamefont {M.}~\bibnamefont {Naghiloo}}, \bibinfo {author} {\bibfnamefont {Y.~N.}\ \bibnamefont {Joglekar}},\ and\ \bibinfo {author} {\bibfnamefont {K.~W.}\ \bibnamefont {Murch}},\ }\bibfield  {title} {\bibinfo {title} {Topological quantum state control through exceptional-point proximity},\ }\href {https://doi.org/10.1103/PhysRevLett.128.160401} {\bibfield  {journal} {\bibinfo  {journal} {Phys. Rev. Lett.}\ }\textbf {\bibinfo {volume} {128}},\ \bibinfo {pages} {160401} (\bibinfo {year} {2022})}\BibitemShut {NoStop}%
\bibitem [{\citenamefont {Han}\ \emph {et~al.}(2023)\citenamefont {Han}, \citenamefont {Wu}, \citenamefont {Huang}, \citenamefont {Wu}, \citenamefont {Zou}, \citenamefont {Yi}, \citenamefont {Zhang}, \citenamefont {Li}, \citenamefont {Xu}, \citenamefont {Zheng}, \citenamefont {Fan}, \citenamefont {Wen}, \citenamefont {Yang},\ and\ \citenamefont {Zheng}}]{Han2023PRL}%
  \BibitemOpen
  \bibfield  {author} {\bibinfo {author} {\bibfnamefont {P.-R.}\ \bibnamefont {Han}}, \bibinfo {author} {\bibfnamefont {F.}~\bibnamefont {Wu}}, \bibinfo {author} {\bibfnamefont {X.-J.}\ \bibnamefont {Huang}}, \bibinfo {author} {\bibfnamefont {H.-Z.}\ \bibnamefont {Wu}}, \bibinfo {author} {\bibfnamefont {C.-L.}\ \bibnamefont {Zou}}, \bibinfo {author} {\bibfnamefont {W.}~\bibnamefont {Yi}}, \bibinfo {author} {\bibfnamefont {M.}~\bibnamefont {Zhang}}, \bibinfo {author} {\bibfnamefont {H.}~\bibnamefont {Li}}, \bibinfo {author} {\bibfnamefont {K.}~\bibnamefont {Xu}}, \bibinfo {author} {\bibfnamefont {D.}~\bibnamefont {Zheng}}, \bibinfo {author} {\bibfnamefont {H.}~\bibnamefont {Fan}}, \bibinfo {author} {\bibfnamefont {J.}~\bibnamefont {Wen}}, \bibinfo {author} {\bibfnamefont {Z.-B.}\ \bibnamefont {Yang}},\ and\ \bibinfo {author} {\bibfnamefont {S.-B.}\ \bibnamefont {Zheng}},\ }\bibfield  {title} {\bibinfo {title} {Exceptional entanglement phenomena: Non-hermiticity meeting nonclassicality},\ }\href
  {https://doi.org/10.1103/PhysRevLett.131.260201} {\bibfield  {journal} {\bibinfo  {journal} {Phys. Rev. Lett.}\ }\textbf {\bibinfo {volume} {131}},\ \bibinfo {pages} {260201} (\bibinfo {year} {2023})}\BibitemShut {NoStop}%
\bibitem [{\citenamefont {Wu}\ \emph {et~al.}(2024)\citenamefont {Wu}, \citenamefont {Wang}, \citenamefont {Ye}, \citenamefont {Liu}, \citenamefont {Niu}, \citenamefont {Duan}, \citenamefont {Wang}, \citenamefont {Rong},\ and\ \citenamefont {Du}}]{Wu2024}%
  \BibitemOpen
  \bibfield  {author} {\bibinfo {author} {\bibfnamefont {Y.}~\bibnamefont {Wu}}, \bibinfo {author} {\bibfnamefont {Y.}~\bibnamefont {Wang}}, \bibinfo {author} {\bibfnamefont {X.}~\bibnamefont {Ye}}, \bibinfo {author} {\bibfnamefont {W.}~\bibnamefont {Liu}}, \bibinfo {author} {\bibfnamefont {Z.}~\bibnamefont {Niu}}, \bibinfo {author} {\bibfnamefont {C.-K.}\ \bibnamefont {Duan}}, \bibinfo {author} {\bibfnamefont {Y.}~\bibnamefont {Wang}}, \bibinfo {author} {\bibfnamefont {X.}~\bibnamefont {Rong}},\ and\ \bibinfo {author} {\bibfnamefont {J.}~\bibnamefont {Du}},\ }\bibfield  {title} {\bibinfo {title} {Third-order exceptional line in a nitrogen-vacancy spin system},\ }\href {https://doi.org/10.1038/s41565-023-01583-0} {\bibfield  {journal} {\bibinfo  {journal} {Nat. Nanotechnol.}\ }\textbf {\bibinfo {volume} {19}},\ \bibinfo {pages} {160} (\bibinfo {year} {2024})}\BibitemShut {NoStop}%
\bibitem [{\citenamefont {Yang}\ \emph {et~al.}(2026{\natexlab{a}})\citenamefont {Yang}, \citenamefont {Han}, \citenamefont {Ning}, \citenamefont {Wu}, \citenamefont {Yang},\ and\ \citenamefont {Zheng}}]{Yang2026SCPMA}%
  \BibitemOpen
  \bibfield  {author} {\bibinfo {author} {\bibfnamefont {S.-B.}\ \bibnamefont {Yang}}, \bibinfo {author} {\bibfnamefont {P.-R.}\ \bibnamefont {Han}}, \bibinfo {author} {\bibfnamefont {W.}~\bibnamefont {Ning}}, \bibinfo {author} {\bibfnamefont {F.}~\bibnamefont {Wu}}, \bibinfo {author} {\bibfnamefont {Z.-B.}\ \bibnamefont {Yang}},\ and\ \bibinfo {author} {\bibfnamefont {S.-B.}\ \bibnamefont {Zheng}},\ }\bibfield  {title} {\bibinfo {title} {An exceptional surface and its topology},\ }\href {https://doi.org/10.1007/s11433-025-2851-8} {\bibfield  {journal} {\bibinfo  {journal} {Sci. China Phys. Mech. Astron.}\ }\textbf {\bibinfo {volume} {69}},\ \bibinfo {pages} {230313} (\bibinfo {year} {2026}{\natexlab{a}})}\BibitemShut {NoStop}%
\bibitem [{\citenamefont {Yang}\ \emph {et~al.}(2026{\natexlab{b}})\citenamefont {Yang}, \citenamefont {Han}, \citenamefont {Ning}, \citenamefont {Wu}, \citenamefont {Yang},\ and\ \citenamefont {Zheng}}]{Yang2026SCPMA-Hyper}%
  \BibitemOpen
  \bibfield  {author} {\bibinfo {author} {\bibfnamefont {S.-B.}\ \bibnamefont {Yang}}, \bibinfo {author} {\bibfnamefont {P.-R.}\ \bibnamefont {Han}}, \bibinfo {author} {\bibfnamefont {W.}~\bibnamefont {Ning}}, \bibinfo {author} {\bibfnamefont {F.}~\bibnamefont {Wu}}, \bibinfo {author} {\bibfnamefont {Z.-B.}\ \bibnamefont {Yang}},\ and\ \bibinfo {author} {\bibfnamefont {S.-B.}\ \bibnamefont {Zheng}},\ }\bibfield  {title} {\bibinfo {title} {Hypersphere-like non-abelian {Y}ang monopole and its topological characterization},\ }\href {https://doi.org/10.1007/s11433-025-2936-1} {\bibfield  {journal} {\bibinfo  {journal} {Sci. China Phys. Mech. Astron.}\ }\textbf {\bibinfo {volume} {69}},\ \bibinfo {pages} {260314} (\bibinfo {year} {2026}{\natexlab{b}})}\BibitemShut {NoStop}%
\bibitem [{\citenamefont {Yan}\ \emph {et~al.}(2026)\citenamefont {Yan}, \citenamefont {Shi}, \citenamefont {Chen},\ and\ \citenamefont {Xia}}]{Yan2026}%
  \BibitemOpen
  \bibfield  {author} {\bibinfo {author} {\bibfnamefont {K.-X.}\ \bibnamefont {Yan}}, \bibinfo {author} {\bibfnamefont {Z.-C.}\ \bibnamefont {Shi}}, \bibinfo {author} {\bibfnamefont {Y.-H.}\ \bibnamefont {Chen}},\ and\ \bibinfo {author} {\bibfnamefont {Y.}~\bibnamefont {Xia}},\ }\bibfield  {title} {\bibinfo {title} {Controllable non-hermiticity in continuous-variable qubits},\ }\href {https://doi.org/10.1103/jzys-kr51} {\bibfield  {journal} {\bibinfo  {journal} {Phys. Rev. A}\ }\textbf {\bibinfo {volume} {113}},\ \bibinfo {pages} {052406} (\bibinfo {year} {2026})}\BibitemShut {NoStop}%
\bibitem [{\citenamefont {Minganti}\ \emph {et~al.}(2020)\citenamefont {Minganti}, \citenamefont {Miranowicz}, \citenamefont {Chhajlany}, \citenamefont {Arkhipov},\ and\ \citenamefont {Nori}}]{Minganti2020}%
  \BibitemOpen
  \bibfield  {author} {\bibinfo {author} {\bibfnamefont {F.}~\bibnamefont {Minganti}}, \bibinfo {author} {\bibfnamefont {A.}~\bibnamefont {Miranowicz}}, \bibinfo {author} {\bibfnamefont {R.~W.}\ \bibnamefont {Chhajlany}}, \bibinfo {author} {\bibfnamefont {I.~I.}\ \bibnamefont {Arkhipov}},\ and\ \bibinfo {author} {\bibfnamefont {F.}~\bibnamefont {Nori}},\ }\bibfield  {title} {\bibinfo {title} {Hybrid-liouvillian formalism connecting exceptional points of non-hermitian hamiltonians and liouvillians via postselection of quantum trajectories},\ }\href {https://doi.org/10.1103/PhysRevA.101.062112} {\bibfield  {journal} {\bibinfo  {journal} {Phys. Rev. A}\ }\textbf {\bibinfo {volume} {101}},\ \bibinfo {pages} {062112} (\bibinfo {year} {2020})}\BibitemShut {NoStop}%
\bibitem [{\citenamefont {Chen}\ \emph {et~al.}(2022)\citenamefont {Chen}, \citenamefont {Abbasi}, \citenamefont {Ha}, \citenamefont {Erdamar}, \citenamefont {Joglekar},\ and\ \citenamefont {Murch}}]{Chen2022}%
  \BibitemOpen
  \bibfield  {author} {\bibinfo {author} {\bibfnamefont {W.}~\bibnamefont {Chen}}, \bibinfo {author} {\bibfnamefont {M.}~\bibnamefont {Abbasi}}, \bibinfo {author} {\bibfnamefont {B.}~\bibnamefont {Ha}}, \bibinfo {author} {\bibfnamefont {S.}~\bibnamefont {Erdamar}}, \bibinfo {author} {\bibfnamefont {Y.~N.}\ \bibnamefont {Joglekar}},\ and\ \bibinfo {author} {\bibfnamefont {K.~W.}\ \bibnamefont {Murch}},\ }\bibfield  {title} {\bibinfo {title} {Decoherence-induced exceptional points in a dissipative superconducting qubit},\ }\href {https://doi.org/10.1103/PhysRevLett.128.110402} {\bibfield  {journal} {\bibinfo  {journal} {Phys. Rev. Lett.}\ }\textbf {\bibinfo {volume} {128}},\ \bibinfo {pages} {110402} (\bibinfo {year} {2022})}\BibitemShut {NoStop}%
\bibitem [{\citenamefont {Zhang}\ \emph {et~al.}(2022)\citenamefont {Zhang}, \citenamefont {Zhang}, \citenamefont {Ding}, \citenamefont {Li}, \citenamefont {Bu}, \citenamefont {Wang}, \citenamefont {Yan}, \citenamefont {Su}, \citenamefont {Chen}, \citenamefont {Nori}, \citenamefont {{\"O}zdemir}, \citenamefont {Zhou}, \citenamefont {Jing},\ and\ \citenamefont {Feng}}]{Zhang2022}%
  \BibitemOpen
  \bibfield  {author} {\bibinfo {author} {\bibfnamefont {J.-W.}\ \bibnamefont {Zhang}}, \bibinfo {author} {\bibfnamefont {J.-Q.}\ \bibnamefont {Zhang}}, \bibinfo {author} {\bibfnamefont {G.-Y.}\ \bibnamefont {Ding}}, \bibinfo {author} {\bibfnamefont {J.-C.}\ \bibnamefont {Li}}, \bibinfo {author} {\bibfnamefont {J.-T.}\ \bibnamefont {Bu}}, \bibinfo {author} {\bibfnamefont {B.}~\bibnamefont {Wang}}, \bibinfo {author} {\bibfnamefont {L.-L.}\ \bibnamefont {Yan}}, \bibinfo {author} {\bibfnamefont {S.-L.}\ \bibnamefont {Su}}, \bibinfo {author} {\bibfnamefont {L.}~\bibnamefont {Chen}}, \bibinfo {author} {\bibfnamefont {F.}~\bibnamefont {Nori}}, \bibinfo {author} {\bibfnamefont {{\c{S}}.~K.}\ \bibnamefont {{\"O}zdemir}}, \bibinfo {author} {\bibfnamefont {F.}~\bibnamefont {Zhou}}, \bibinfo {author} {\bibfnamefont {H.}~\bibnamefont {Jing}},\ and\ \bibinfo {author} {\bibfnamefont {M.}~\bibnamefont {Feng}},\ }\bibfield  {title} {\bibinfo {title} {Dynamical control of quantum heat engines using exceptional points},\
  }\href {https://doi.org/10.1038/s41467-022-33667-1} {\bibfield  {journal} {\bibinfo  {journal} {Nat. Commun.}\ }\textbf {\bibinfo {volume} {13}},\ \bibinfo {pages} {6225} (\bibinfo {year} {2022})}\BibitemShut {NoStop}%
\bibitem [{\citenamefont {Sun}\ and\ \citenamefont {Yi}(2024)}]{Sun2024}%
  \BibitemOpen
  \bibfield  {author} {\bibinfo {author} {\bibfnamefont {K.}~\bibnamefont {Sun}}\ and\ \bibinfo {author} {\bibfnamefont {W.}~\bibnamefont {Yi}},\ }\bibfield  {title} {\bibinfo {title} {Encircling the liouvillian exceptional points: a brief review},\ }\href {https://doi.org/10.1007/s43673-024-00129-3} {\bibfield  {journal} {\bibinfo  {journal} {AAPPS Bull.}\ }\textbf {\bibinfo {volume} {34}},\ \bibinfo {pages} {22} (\bibinfo {year} {2024})}\BibitemShut {NoStop}%
\bibitem [{\citenamefont {Khandelwal}\ \emph {et~al.}(2021)\citenamefont {Khandelwal}, \citenamefont {Brunner},\ and\ \citenamefont {Haack}}]{Khandelwal2021}%
  \BibitemOpen
  \bibfield  {author} {\bibinfo {author} {\bibfnamefont {S.}~\bibnamefont {Khandelwal}}, \bibinfo {author} {\bibfnamefont {N.}~\bibnamefont {Brunner}},\ and\ \bibinfo {author} {\bibfnamefont {G.}~\bibnamefont {Haack}},\ }\bibfield  {title} {\bibinfo {title} {Signatures of liouvillian exceptional points in a quantum thermal machine},\ }\href {https://doi.org/10.1103/PRXQuantum.2.040346} {\bibfield  {journal} {\bibinfo  {journal} {PRX Quantum}\ }\textbf {\bibinfo {volume} {2}},\ \bibinfo {pages} {040346} (\bibinfo {year} {2021})}\BibitemShut {NoStop}%
\bibitem [{\citenamefont {Bu}\ \emph {et~al.}(2023)\citenamefont {Bu}, \citenamefont {Zhang}, \citenamefont {Ding}, \citenamefont {Li}, \citenamefont {Zhang}, \citenamefont {Wang}, \citenamefont {Ding}, \citenamefont {Yuan}, \citenamefont {Chen}, \citenamefont {\"Ozdemir}, \citenamefont {Zhou}, \citenamefont {Jing},\ and\ \citenamefont {Feng}}]{Bu2023}%
  \BibitemOpen
  \bibfield  {author} {\bibinfo {author} {\bibfnamefont {J.-T.}\ \bibnamefont {Bu}}, \bibinfo {author} {\bibfnamefont {J.-Q.}\ \bibnamefont {Zhang}}, \bibinfo {author} {\bibfnamefont {G.-Y.}\ \bibnamefont {Ding}}, \bibinfo {author} {\bibfnamefont {J.-C.}\ \bibnamefont {Li}}, \bibinfo {author} {\bibfnamefont {J.-W.}\ \bibnamefont {Zhang}}, \bibinfo {author} {\bibfnamefont {B.}~\bibnamefont {Wang}}, \bibinfo {author} {\bibfnamefont {W.-Q.}\ \bibnamefont {Ding}}, \bibinfo {author} {\bibfnamefont {W.-F.}\ \bibnamefont {Yuan}}, \bibinfo {author} {\bibfnamefont {L.}~\bibnamefont {Chen}}, \bibinfo {author} {\bibfnamefont {i.~m. c.~K.}\ \bibnamefont {\"Ozdemir}}, \bibinfo {author} {\bibfnamefont {F.}~\bibnamefont {Zhou}}, \bibinfo {author} {\bibfnamefont {H.}~\bibnamefont {Jing}},\ and\ \bibinfo {author} {\bibfnamefont {M.}~\bibnamefont {Feng}},\ }\bibfield  {title} {\bibinfo {title} {Enhancement of quantum heat engine by encircling a liouvillian exceptional point},\ }\href
  {https://doi.org/10.1103/PhysRevLett.130.110402} {\bibfield  {journal} {\bibinfo  {journal} {Phys. Rev. Lett.}\ }\textbf {\bibinfo {volume} {130}},\ \bibinfo {pages} {110402} (\bibinfo {year} {2023})}\BibitemShut {NoStop}%
\bibitem [{\citenamefont {Han}\ \emph {et~al.}(2026{\natexlab{a}})\citenamefont {Han}, \citenamefont {Qiu}, \citenamefont {Zhang}, \citenamefont {Ning}, \citenamefont {Yang},\ and\ \citenamefont {Zheng}}]{Han2026SCPMA}%
  \BibitemOpen
  \bibfield  {author} {\bibinfo {author} {\bibfnamefont {P.-R.}\ \bibnamefont {Han}}, \bibinfo {author} {\bibfnamefont {H.}~\bibnamefont {Qiu}}, \bibinfo {author} {\bibfnamefont {H.-L.}\ \bibnamefont {Zhang}}, \bibinfo {author} {\bibfnamefont {W.}~\bibnamefont {Ning}}, \bibinfo {author} {\bibfnamefont {Z.-B.}\ \bibnamefont {Yang}},\ and\ \bibinfo {author} {\bibfnamefont {S.-B.}\ \bibnamefont {Zheng}},\ }\bibfield  {title} {\bibinfo {title} {Non-hermitian topology in a single driven-dissipative kerr-cat qubit},\ }\href {https://doi.org/10.1007/s11433-025-2933-x} {\bibfield  {journal} {\bibinfo  {journal} {Sci. China Phys. Mech. Astron.}\ }\textbf {\bibinfo {volume} {69}},\ \bibinfo {pages} {260313} (\bibinfo {year} {2026}{\natexlab{a}})}\BibitemShut {NoStop}%
\bibitem [{\citenamefont {Yang}\ \emph {et~al.}(2026{\natexlab{c}})\citenamefont {Yang}, \citenamefont {Han}, \citenamefont {Yang},\ and\ \citenamefont {Zheng}}]{Yang2026}%
  \BibitemOpen
  \bibfield  {author} {\bibinfo {author} {\bibfnamefont {T.-L.}\ \bibnamefont {Yang}}, \bibinfo {author} {\bibfnamefont {P.-R.}\ \bibnamefont {Han}}, \bibinfo {author} {\bibfnamefont {Z.-B.}\ \bibnamefont {Yang}},\ and\ \bibinfo {author} {\bibfnamefont {S.-B.}\ \bibnamefont {Zheng}},\ }\bibfield  {title} {\bibinfo {title} {Controllable non-hermitian topology in a dynamically protected cat qubit},\ }\href {https://doi.org/10.48550/arXiv.2604.20680} {\bibfield  {journal} {\bibinfo  {journal} {arXiv}\ }\textbf {\bibinfo {volume} {2604.20680}} (\bibinfo {year} {2026}{\natexlab{c}})}\BibitemShut {NoStop}%
\bibitem [{\citenamefont {Nakajima}(1958)}]{Nakajima1958}%
  \BibitemOpen
  \bibfield  {author} {\bibinfo {author} {\bibfnamefont {S.}~\bibnamefont {Nakajima}},\ }\bibfield  {title} {\bibinfo {title} {On quantum theory of transport phenomena: Steady diffusion},\ }\href {https://doi.org/10.1143/PTP.20.948} {\bibfield  {journal} {\bibinfo  {journal} {Prog. Theor. Phys.}\ }\textbf {\bibinfo {volume} {20}},\ \bibinfo {pages} {948} (\bibinfo {year} {1958})}\BibitemShut {NoStop}%
\bibitem [{\citenamefont {Zwanzig}(1960)}]{Zwanzig1960}%
  \BibitemOpen
  \bibfield  {author} {\bibinfo {author} {\bibfnamefont {R.}~\bibnamefont {Zwanzig}},\ }\bibfield  {title} {\bibinfo {title} {Ensemble method in the theory of irreversibility},\ }\href {https://doi.org/10.1063/1.1731409} {\bibfield  {journal} {\bibinfo  {journal} {J. Chem. Phys.}\ }\textbf {\bibinfo {volume} {33}},\ \bibinfo {pages} {1338} (\bibinfo {year} {1960})}\BibitemShut {NoStop}%
\bibitem [{\citenamefont {Zhang}\ \emph {et~al.}(2025{\natexlab{b}})\citenamefont {Zhang}, \citenamefont {Han}, \citenamefont {Wu}, \citenamefont {Ning}, \citenamefont {Yang},\ and\ \citenamefont {Zheng}}]{Zhang2025PRL}%
  \BibitemOpen
  \bibfield  {author} {\bibinfo {author} {\bibfnamefont {H.-L.}\ \bibnamefont {Zhang}}, \bibinfo {author} {\bibfnamefont {P.-R.}\ \bibnamefont {Han}}, \bibinfo {author} {\bibfnamefont {F.}~\bibnamefont {Wu}}, \bibinfo {author} {\bibfnamefont {W.}~\bibnamefont {Ning}}, \bibinfo {author} {\bibfnamefont {Z.-B.}\ \bibnamefont {Yang}},\ and\ \bibinfo {author} {\bibfnamefont {S.-B.}\ \bibnamefont {Zheng}},\ }\bibfield  {title} {\bibinfo {title} {Experimental observation of non-markovian quantum exceptional points},\ }\href {https://doi.org/10.1103/jk6y-55xp} {\bibfield  {journal} {\bibinfo  {journal} {Phys. Rev. Lett.}\ }\textbf {\bibinfo {volume} {135}},\ \bibinfo {pages} {230203} (\bibinfo {year} {2025}{\natexlab{b}})}\BibitemShut {NoStop}%
\bibitem [{\citenamefont {Holst}\ \emph {et~al.}(2026)\citenamefont {Holst}, \citenamefont {Blasi},\ and\ \citenamefont {Khandelwal}}]{Holst2026}%
  \BibitemOpen
  \bibfield  {author} {\bibinfo {author} {\bibfnamefont {D.}~\bibnamefont {Holst}}, \bibinfo {author} {\bibfnamefont {G.}~\bibnamefont {Blasi}},\ and\ \bibinfo {author} {\bibfnamefont {S.}~\bibnamefont {Khandelwal}},\ }\bibfield  {title} {\bibinfo {title} {Non-markovian exceptional points and information backflow: a heisenberg reaction-coordinate approach},\ }\href {https://arxiv.org/abs/2610.26620} {\bibfield  {journal} {\bibinfo  {journal} {arXiv}\ }\textbf {\bibinfo {volume} {2610.26620}} (\bibinfo {year} {2026})}\BibitemShut {NoStop}%
\bibitem [{\citenamefont {Chen}\ \emph {et~al.}(2026)\citenamefont {Chen}, \citenamefont {Liu}, \citenamefont {Xiong}, \citenamefont {Ding},\ and\ \citenamefont {Guo}}]{Chen2026NMQSD}%
  \BibitemOpen
  \bibfield  {author} {\bibinfo {author} {\bibfnamefont {Y.}~\bibnamefont {Chen}}, \bibinfo {author} {\bibfnamefont {H.}~\bibnamefont {Liu}}, \bibinfo {author} {\bibfnamefont {K.}~\bibnamefont {Xiong}}, \bibinfo {author} {\bibfnamefont {Q.}~\bibnamefont {Ding}},\ and\ \bibinfo {author} {\bibfnamefont {E.}~\bibnamefont {Guo}},\ }\bibfield  {title} {\bibinfo {title} {Exceptional points in non-markovian quantum-state diffusion: From coefficient dynamics to state signatures},\ }\href {https://arxiv.org/abs/2611.13221} {\bibfield  {journal} {\bibinfo  {journal} {arXiv}\ } (\bibinfo {year} {2026})},\ \Eprint {https://arxiv.org/abs/2611.13221} {arXiv:2611.13221} \BibitemShut {NoStop}%
\bibitem [{\citenamefont {Wong}\ \emph {et~al.}(2026)\citenamefont {Wong}, \citenamefont {Zeng},\ and\ \citenamefont {Li}}]{Wong2026}%
  \BibitemOpen
  \bibfield  {author} {\bibinfo {author} {\bibfnamefont {W.~C.}\ \bibnamefont {Wong}}, \bibinfo {author} {\bibfnamefont {B.}~\bibnamefont {Zeng}},\ and\ \bibinfo {author} {\bibfnamefont {J.}~\bibnamefont {Li}},\ }\bibfield  {title} {\bibinfo {title} {Non-markovian exceptional points by interpolating quantum channels},\ }\href {https://doi.org/10.1038/s41534-026-01205-2} {\bibfield  {journal} {\bibinfo  {journal} {npj Quantum Inf.}\ }\textbf {\bibinfo {volume} {12}},\ \bibinfo {pages} {63} (\bibinfo {year} {2026})}\BibitemShut {NoStop}%
\bibitem [{\citenamefont {Wielinga}\ and\ \citenamefont {Milburn}(1993)}]{Wielinga1993}%
  \BibitemOpen
  \bibfield  {author} {\bibinfo {author} {\bibfnamefont {B.}~\bibnamefont {Wielinga}}\ and\ \bibinfo {author} {\bibfnamefont {G.~J.}\ \bibnamefont {Milburn}},\ }\bibfield  {title} {\bibinfo {title} {Quantum tunneling in a kerr medium with parametric pumping},\ }\href {https://doi.org/10.1103/PhysRevA.48.2494} {\bibfield  {journal} {\bibinfo  {journal} {Phys. Rev. A}\ }\textbf {\bibinfo {volume} {48}},\ \bibinfo {pages} {2494} (\bibinfo {year} {1993})}\BibitemShut {NoStop}%
\bibitem [{\citenamefont {Goto}(2016)}]{Goto2016}%
  \BibitemOpen
  \bibfield  {author} {\bibinfo {author} {\bibfnamefont {H.}~\bibnamefont {Goto}},\ }\bibfield  {title} {\bibinfo {title} {Universal quantum computation with a nonlinear oscillator network},\ }\href {https://doi.org/10.1103/PhysRevA.93.050301} {\bibfield  {journal} {\bibinfo  {journal} {Phys. Rev. A}\ }\textbf {\bibinfo {volume} {93}},\ \bibinfo {pages} {050301(R)} (\bibinfo {year} {2016})}\BibitemShut {NoStop}%
\bibitem [{\citenamefont {Minganti}\ \emph {et~al.}(2016)\citenamefont {Minganti}, \citenamefont {Bartolo}, \citenamefont {Lolli}, \citenamefont {Casteels},\ and\ \citenamefont {Ciuti}}]{Minganti2016}%
  \BibitemOpen
  \bibfield  {author} {\bibinfo {author} {\bibfnamefont {F.}~\bibnamefont {Minganti}}, \bibinfo {author} {\bibfnamefont {N.}~\bibnamefont {Bartolo}}, \bibinfo {author} {\bibfnamefont {J.}~\bibnamefont {Lolli}}, \bibinfo {author} {\bibfnamefont {W.}~\bibnamefont {Casteels}},\ and\ \bibinfo {author} {\bibfnamefont {C.}~\bibnamefont {Ciuti}},\ }\bibfield  {title} {\bibinfo {title} {Exact results for schr{\"o}dinger cats in driven-dissipative systems and their feedback control},\ }\href {https://doi.org/10.1038/srep26987} {\bibfield  {journal} {\bibinfo  {journal} {Sci. Rep.}\ }\textbf {\bibinfo {volume} {6}},\ \bibinfo {pages} {26987} (\bibinfo {year} {2016})}\BibitemShut {NoStop}%
\bibitem [{\citenamefont {Puri}\ \emph {et~al.}(2017)\citenamefont {Puri}, \citenamefont {Boutin},\ and\ \citenamefont {Blais}}]{Puri2017}%
  \BibitemOpen
  \bibfield  {author} {\bibinfo {author} {\bibfnamefont {S.}~\bibnamefont {Puri}}, \bibinfo {author} {\bibfnamefont {S.}~\bibnamefont {Boutin}},\ and\ \bibinfo {author} {\bibfnamefont {A.}~\bibnamefont {Blais}},\ }\bibfield  {title} {\bibinfo {title} {Engineering the quantum states of light in a kerr-nonlinear resonator by two-photon driving},\ }\href {https://doi.org/10.1038/s41534-017-0019-1} {\bibfield  {journal} {\bibinfo  {journal} {npj Quantum Inf.}\ }\textbf {\bibinfo {volume} {3}},\ \bibinfo {pages} {18} (\bibinfo {year} {2017})}\BibitemShut {NoStop}%
\bibitem [{\citenamefont {Puri}\ \emph {et~al.}(2019)\citenamefont {Puri}, \citenamefont {Grimm}, \citenamefont {Campagne-Ibarcq}, \citenamefont {Eickbusch}, \citenamefont {Noh}, \citenamefont {Roberts}, \citenamefont {Jiang}, \citenamefont {Mirrahimi}, \citenamefont {Devoret},\ and\ \citenamefont {Girvin}}]{Puri2019}%
  \BibitemOpen
  \bibfield  {author} {\bibinfo {author} {\bibfnamefont {S.}~\bibnamefont {Puri}}, \bibinfo {author} {\bibfnamefont {A.}~\bibnamefont {Grimm}}, \bibinfo {author} {\bibfnamefont {P.}~\bibnamefont {Campagne-Ibarcq}}, \bibinfo {author} {\bibfnamefont {A.}~\bibnamefont {Eickbusch}}, \bibinfo {author} {\bibfnamefont {K.}~\bibnamefont {Noh}}, \bibinfo {author} {\bibfnamefont {G.}~\bibnamefont {Roberts}}, \bibinfo {author} {\bibfnamefont {L.}~\bibnamefont {Jiang}}, \bibinfo {author} {\bibfnamefont {M.}~\bibnamefont {Mirrahimi}}, \bibinfo {author} {\bibfnamefont {M.~H.}\ \bibnamefont {Devoret}},\ and\ \bibinfo {author} {\bibfnamefont {S.~M.}\ \bibnamefont {Girvin}},\ }\bibfield  {title} {\bibinfo {title} {Stabilized cat in a driven nonlinear cavity: A fault-tolerant error syndrome detector},\ }\href {https://doi.org/10.1103/PhysRevX.9.041009} {\bibfield  {journal} {\bibinfo  {journal} {Phys. Rev. X}\ }\textbf {\bibinfo {volume} {9}},\ \bibinfo {pages} {041009} (\bibinfo {year} {2019})}\BibitemShut {NoStop}%
\bibitem [{\citenamefont {Grimm}\ \emph {et~al.}(2020)\citenamefont {Grimm}, \citenamefont {Frattini}, \citenamefont {Puri}, \citenamefont {Mundhada}, \citenamefont {Touzard}, \citenamefont {Mirrahimi}, \citenamefont {Girvin}, \citenamefont {Shankar},\ and\ \citenamefont {Devoret}}]{Grimm2020}%
  \BibitemOpen
  \bibfield  {author} {\bibinfo {author} {\bibfnamefont {A.}~\bibnamefont {Grimm}}, \bibinfo {author} {\bibfnamefont {N.~E.}\ \bibnamefont {Frattini}}, \bibinfo {author} {\bibfnamefont {S.}~\bibnamefont {Puri}}, \bibinfo {author} {\bibfnamefont {S.~O.}\ \bibnamefont {Mundhada}}, \bibinfo {author} {\bibfnamefont {S.}~\bibnamefont {Touzard}}, \bibinfo {author} {\bibfnamefont {M.}~\bibnamefont {Mirrahimi}}, \bibinfo {author} {\bibfnamefont {S.~M.}\ \bibnamefont {Girvin}}, \bibinfo {author} {\bibfnamefont {S.}~\bibnamefont {Shankar}},\ and\ \bibinfo {author} {\bibfnamefont {M.~H.}\ \bibnamefont {Devoret}},\ }\bibfield  {title} {\bibinfo {title} {Stabilization and operation of a kerr-cat qubit},\ }\href {https://doi.org/10.1038/s41586-020-2587-z} {\bibfield  {journal} {\bibinfo  {journal} {Nature (London)}\ }\textbf {\bibinfo {volume} {584}},\ \bibinfo {pages} {205} (\bibinfo {year} {2020})}\BibitemShut {NoStop}%
\bibitem [{\citenamefont {Lin}\ \emph {et~al.}(2024)\citenamefont {Lin}, \citenamefont {Yang}, \citenamefont {Wu},\ and\ \citenamefont {Yang}}]{Lin2024}%
  \BibitemOpen
  \bibfield  {author} {\bibinfo {author} {\bibfnamefont {J.}~\bibnamefont {Lin}}, \bibinfo {author} {\bibfnamefont {S.-B.}\ \bibnamefont {Yang}}, \bibinfo {author} {\bibfnamefont {F.}~\bibnamefont {Wu}},\ and\ \bibinfo {author} {\bibfnamefont {Z.-B.}\ \bibnamefont {Yang}},\ }\bibfield  {title} {\bibinfo {title} {Topological transitions in a kerr nonlinear oscillator},\ }\href {https://doi.org/10.34133/icomputing.0099} {\bibfield  {journal} {\bibinfo  {journal} {Intell. Comput.}\ }\textbf {\bibinfo {volume} {3}},\ \bibinfo {pages} {0099} (\bibinfo {year} {2024})}\BibitemShut {NoStop}%
\bibitem [{\citenamefont {Wang}\ \emph {et~al.}(2019)\citenamefont {Wang}, \citenamefont {Pechal}, \citenamefont {Wollack}, \citenamefont {Arrangoiz-Arriola}, \citenamefont {Gao}, \citenamefont {Lee},\ and\ \citenamefont {Safavi-Naeini}}]{Wang2019}%
  \BibitemOpen
  \bibfield  {author} {\bibinfo {author} {\bibfnamefont {Z.}~\bibnamefont {Wang}}, \bibinfo {author} {\bibfnamefont {M.}~\bibnamefont {Pechal}}, \bibinfo {author} {\bibfnamefont {E.~A.}\ \bibnamefont {Wollack}}, \bibinfo {author} {\bibfnamefont {P.}~\bibnamefont {Arrangoiz-Arriola}}, \bibinfo {author} {\bibfnamefont {M.}~\bibnamefont {Gao}}, \bibinfo {author} {\bibfnamefont {N.~R.}\ \bibnamefont {Lee}},\ and\ \bibinfo {author} {\bibfnamefont {A.~H.}\ \bibnamefont {Safavi-Naeini}},\ }\bibfield  {title} {\bibinfo {title} {Quantum dynamics of a few-photon parametric oscillator},\ }\href {https://doi.org/10.1103/PhysRevX.9.021049} {\bibfield  {journal} {\bibinfo  {journal} {Phys. Rev. X}\ }\textbf {\bibinfo {volume} {9}},\ \bibinfo {pages} {021049} (\bibinfo {year} {2019})}\BibitemShut {NoStop}%
\bibitem [{sup()}]{supplement}%
  \BibitemOpen
  \href@noop {} {}\bibinfo {note} {See Supplemental Material at xxx for the calculation of the quantum-channel exceptional points, proofs of the three certification criteria (non-HP, non-cCP, and cCP), a discussion of the criteria's scope, and the time evolution of the ICEP landscapes.}\BibitemShut {Stop}%
\bibitem [{\citenamefont {Han}\ \emph {et~al.}(2026{\natexlab{b}})\citenamefont {Han}, \citenamefont {Yang}, \citenamefont {Ning}, \citenamefont {Zhang}, \citenamefont {Kang}, \citenamefont {Qiu},\ and\ \citenamefont {Yang}}]{Han2026}%
  \BibitemOpen
  \bibfield  {author} {\bibinfo {author} {\bibfnamefont {P.-R.}\ \bibnamefont {Han}}, \bibinfo {author} {\bibfnamefont {T.-L.}\ \bibnamefont {Yang}}, \bibinfo {author} {\bibfnamefont {W.}~\bibnamefont {Ning}}, \bibinfo {author} {\bibfnamefont {H.-L.}\ \bibnamefont {Zhang}}, \bibinfo {author} {\bibfnamefont {H.}~\bibnamefont {Kang}}, \bibinfo {author} {\bibfnamefont {H.}~\bibnamefont {Qiu}},\ and\ \bibinfo {author} {\bibfnamefont {Z.-B.}\ \bibnamefont {Yang}},\ }\bibfield  {title} {\bibinfo {title} {Liouvillian exceptional-point-induced phase transition in a single kerr-cat qubit},\ }\bibfield  {journal} {\bibinfo  {journal} {Commun. Phys.}\ }\href {https://doi.org/10.1038/s42005-026-02919-7} {10.1038/s42005-026-02919-7} (\bibinfo {year} {2026}{\natexlab{b}})\BibitemShut {NoStop}%
\bibitem [{\citenamefont {Wolf}\ \emph {et~al.}(2008)\citenamefont {Wolf}, \citenamefont {Eisert}, \citenamefont {Cubitt},\ and\ \citenamefont {Cirac}}]{Wolf2008}%
  \BibitemOpen
  \bibfield  {author} {\bibinfo {author} {\bibfnamefont {M.~M.}\ \bibnamefont {Wolf}}, \bibinfo {author} {\bibfnamefont {J.}~\bibnamefont {Eisert}}, \bibinfo {author} {\bibfnamefont {T.~S.}\ \bibnamefont {Cubitt}},\ and\ \bibinfo {author} {\bibfnamefont {J.~I.}\ \bibnamefont {Cirac}},\ }\bibfield  {title} {\bibinfo {title} {Assessing non-markovian quantum dynamics},\ }\href {https://doi.org/10.1103/PhysRevLett.101.150402} {\bibfield  {journal} {\bibinfo  {journal} {Phys. Rev. Lett.}\ }\textbf {\bibinfo {volume} {101}},\ \bibinfo {pages} {150402} (\bibinfo {year} {2008})}\BibitemShut {NoStop}%
\bibitem [{\citenamefont {Minganti}\ \emph {et~al.}(2018)\citenamefont {Minganti}, \citenamefont {Biella}, \citenamefont {Bartolo},\ and\ \citenamefont {Ciuti}}]{Minganti2018}%
  \BibitemOpen
  \bibfield  {author} {\bibinfo {author} {\bibfnamefont {F.}~\bibnamefont {Minganti}}, \bibinfo {author} {\bibfnamefont {A.}~\bibnamefont {Biella}}, \bibinfo {author} {\bibfnamefont {N.}~\bibnamefont {Bartolo}},\ and\ \bibinfo {author} {\bibfnamefont {C.}~\bibnamefont {Ciuti}},\ }\bibfield  {title} {\bibinfo {title} {Spectral theory of liouvillians for dissipative phase transitions},\ }\href {https://doi.org/10.1103/PhysRevA.98.042118} {\bibfield  {journal} {\bibinfo  {journal} {Phys. Rev. A}\ }\textbf {\bibinfo {volume} {98}},\ \bibinfo {pages} {042118} (\bibinfo {year} {2018})}\BibitemShut {NoStop}%
\bibitem [{\citenamefont {Brandão}(2026)}]{Brandao2026}%
  \BibitemOpen
  \bibfield  {author} {\bibinfo {author} {\bibfnamefont {P.~A.}\ \bibnamefont {Brandão}},\ }\bibfield  {title} {\bibinfo {title} {Operational signatures of exceptional quantum channels},\ }\href {https://doi.org/10.48550/arXiv.2610.13622} {\bibfield  {journal} {\bibinfo  {journal} {arXiv}\ }\textbf {\bibinfo {volume} {2610.13622}} (\bibinfo {year} {2026})}\BibitemShut {NoStop}%
\bibitem [{\citenamefont {Leghtas}\ \emph {et~al.}(2015)\citenamefont {Leghtas}, \citenamefont {Touzard}, \citenamefont {Pop}, \citenamefont {Kou}, \citenamefont {Vlastakis}, \citenamefont {Petrenko}, \citenamefont {Sliwa}, \citenamefont {Narla}, \citenamefont {Shankar}, \citenamefont {Hatridge}, \citenamefont {Reagor}, \citenamefont {Frunzio}, \citenamefont {Schoelkopf}, \citenamefont {Mirrahimi},\ and\ \citenamefont {Devoret}}]{Leghtas2015}%
  \BibitemOpen
  \bibfield  {author} {\bibinfo {author} {\bibfnamefont {Z.}~\bibnamefont {Leghtas}}, \bibinfo {author} {\bibfnamefont {S.}~\bibnamefont {Touzard}}, \bibinfo {author} {\bibfnamefont {I.~M.}\ \bibnamefont {Pop}}, \bibinfo {author} {\bibfnamefont {A.}~\bibnamefont {Kou}}, \bibinfo {author} {\bibfnamefont {B.}~\bibnamefont {Vlastakis}}, \bibinfo {author} {\bibfnamefont {A.}~\bibnamefont {Petrenko}}, \bibinfo {author} {\bibfnamefont {K.~M.}\ \bibnamefont {Sliwa}}, \bibinfo {author} {\bibfnamefont {A.}~\bibnamefont {Narla}}, \bibinfo {author} {\bibfnamefont {S.}~\bibnamefont {Shankar}}, \bibinfo {author} {\bibfnamefont {M.~J.}\ \bibnamefont {Hatridge}}, \bibinfo {author} {\bibfnamefont {M.}~\bibnamefont {Reagor}}, \bibinfo {author} {\bibfnamefont {L.}~\bibnamefont {Frunzio}}, \bibinfo {author} {\bibfnamefont {R.~J.}\ \bibnamefont {Schoelkopf}}, \bibinfo {author} {\bibfnamefont {M.}~\bibnamefont {Mirrahimi}},\ and\ \bibinfo {author} {\bibfnamefont {M.~H.}\ \bibnamefont {Devoret}},\ }\bibfield  {title} {\bibinfo
  {title} {Confining the state of light to a quantum manifold by engineered two-photon loss},\ }\href {https://doi.org/10.1126/science.aaa2085} {\bibfield  {journal} {\bibinfo  {journal} {Science}\ }\textbf {\bibinfo {volume} {347}},\ \bibinfo {pages} {853} (\bibinfo {year} {2015})}\BibitemShut {NoStop}%
\bibitem [{\citenamefont {He}\ \emph {et~al.}(2023)\citenamefont {He}, \citenamefont {Lu}, \citenamefont {Bao}, \citenamefont {Xue}, \citenamefont {Jiang}, \citenamefont {Wang}, \citenamefont {Roudsari}, \citenamefont {Delsing}, \citenamefont {Tsai},\ and\ \citenamefont {Lin}}]{He2023}%
  \BibitemOpen
  \bibfield  {author} {\bibinfo {author} {\bibfnamefont {X.~L.}\ \bibnamefont {He}}, \bibinfo {author} {\bibfnamefont {Y.}~\bibnamefont {Lu}}, \bibinfo {author} {\bibfnamefont {D.~Q.}\ \bibnamefont {Bao}}, \bibinfo {author} {\bibfnamefont {H.}~\bibnamefont {Xue}}, \bibinfo {author} {\bibfnamefont {W.~B.}\ \bibnamefont {Jiang}}, \bibinfo {author} {\bibfnamefont {Z.}~\bibnamefont {Wang}}, \bibinfo {author} {\bibfnamefont {A.~F.}\ \bibnamefont {Roudsari}}, \bibinfo {author} {\bibfnamefont {P.}~\bibnamefont {Delsing}}, \bibinfo {author} {\bibfnamefont {J.~S.}\ \bibnamefont {Tsai}},\ and\ \bibinfo {author} {\bibfnamefont {Z.~R.}\ \bibnamefont {Lin}},\ }\bibfield  {title} {\bibinfo {title} {Fast generation of schr{\"o}dinger cat states using a kerr-tunable superconducting resonator},\ }\href {https://doi.org/10.1038/s41467-023-42057-0} {\bibfield  {journal} {\bibinfo  {journal} {Nat. Commun.}\ }\textbf {\bibinfo {volume} {14}},\ \bibinfo {pages} {6358} (\bibinfo {year} {2023})}\BibitemShut {NoStop}%
\bibitem [{\citenamefont {Xu}\ \emph {et~al.}(2025)\citenamefont {Xu}, \citenamefont {Hua}, \citenamefont {Wang}, \citenamefont {Ma}, \citenamefont {Li}, \citenamefont {Chen}, \citenamefont {Zhou}, \citenamefont {Pan}, \citenamefont {Xiao}, \citenamefont {Huang}, \citenamefont {Cai}, \citenamefont {Ai}, \citenamefont {Liu}, \citenamefont {Zou},\ and\ \citenamefont {Sun}}]{Xu2025}%
  \BibitemOpen
  \bibfield  {author} {\bibinfo {author} {\bibfnamefont {Y.}~\bibnamefont {Xu}}, \bibinfo {author} {\bibfnamefont {Z.}~\bibnamefont {Hua}}, \bibinfo {author} {\bibfnamefont {W.}~\bibnamefont {Wang}}, \bibinfo {author} {\bibfnamefont {Y.}~\bibnamefont {Ma}}, \bibinfo {author} {\bibfnamefont {M.}~\bibnamefont {Li}}, \bibinfo {author} {\bibfnamefont {J.}~\bibnamefont {Chen}}, \bibinfo {author} {\bibfnamefont {J.}~\bibnamefont {Zhou}}, \bibinfo {author} {\bibfnamefont {X.}~\bibnamefont {Pan}}, \bibinfo {author} {\bibfnamefont {L.}~\bibnamefont {Xiao}}, \bibinfo {author} {\bibfnamefont {H.}~\bibnamefont {Huang}}, \bibinfo {author} {\bibfnamefont {W.}~\bibnamefont {Cai}}, \bibinfo {author} {\bibfnamefont {H.}~\bibnamefont {Ai}}, \bibinfo {author} {\bibfnamefont {Y.-x.}\ \bibnamefont {Liu}}, \bibinfo {author} {\bibfnamefont {C.-L.}\ \bibnamefont {Zou}},\ and\ \bibinfo {author} {\bibfnamefont {L.}~\bibnamefont {Sun}},\ }\bibfield  {title} {\bibinfo {title} {Dynamic compensation for pump-induced frequency shift in
  kerr-cat qubit initialization},\ }\href {https://doi.org/10.1103/PhysRevApplied.23.034060} {\bibfield  {journal} {\bibinfo  {journal} {Phys. Rev. Appl.}\ }\textbf {\bibinfo {volume} {23}},\ \bibinfo {pages} {034060} (\bibinfo {year} {2025})}\BibitemShut {NoStop}%
\bibitem [{\citenamefont {Fl{\"u}hmann}\ \emph {et~al.}(2019)\citenamefont {Fl{\"u}hmann}, \citenamefont {Nguyen}, \citenamefont {Marinelli}, \citenamefont {Negnevitsky}, \citenamefont {Mehta},\ and\ \citenamefont {Home}}]{Fluhmann2019}%
  \BibitemOpen
  \bibfield  {author} {\bibinfo {author} {\bibfnamefont {C.}~\bibnamefont {Fl{\"u}hmann}}, \bibinfo {author} {\bibfnamefont {T.~L.}\ \bibnamefont {Nguyen}}, \bibinfo {author} {\bibfnamefont {M.}~\bibnamefont {Marinelli}}, \bibinfo {author} {\bibfnamefont {V.}~\bibnamefont {Negnevitsky}}, \bibinfo {author} {\bibfnamefont {K.}~\bibnamefont {Mehta}},\ and\ \bibinfo {author} {\bibfnamefont {J.~P.}\ \bibnamefont {Home}},\ }\bibfield  {title} {\bibinfo {title} {Encoding a qubit in a trapped-ion mechanical oscillator},\ }\href {https://doi.org/10.1038/s41586-019-0960-6} {\bibfield  {journal} {\bibinfo  {journal} {Nature (London)}\ }\textbf {\bibinfo {volume} {566}},\ \bibinfo {pages} {513} (\bibinfo {year} {2019})}\BibitemShut {NoStop}%
\bibitem [{\citenamefont {Wollman}\ \emph {et~al.}(2015)\citenamefont {Wollman}, \citenamefont {Lei}, \citenamefont {Weinstein}, \citenamefont {Suh}, \citenamefont {Kronwald}, \citenamefont {Marquardt}, \citenamefont {Clerk},\ and\ \citenamefont {Schwab}}]{Wollman2015}%
  \BibitemOpen
  \bibfield  {author} {\bibinfo {author} {\bibfnamefont {E.~E.}\ \bibnamefont {Wollman}}, \bibinfo {author} {\bibfnamefont {C.~U.}\ \bibnamefont {Lei}}, \bibinfo {author} {\bibfnamefont {A.~J.}\ \bibnamefont {Weinstein}}, \bibinfo {author} {\bibfnamefont {J.}~\bibnamefont {Suh}}, \bibinfo {author} {\bibfnamefont {A.}~\bibnamefont {Kronwald}}, \bibinfo {author} {\bibfnamefont {F.}~\bibnamefont {Marquardt}}, \bibinfo {author} {\bibfnamefont {A.~A.}\ \bibnamefont {Clerk}},\ and\ \bibinfo {author} {\bibfnamefont {K.~C.}\ \bibnamefont {Schwab}},\ }\bibfield  {title} {\bibinfo {title} {Quantum squeezing of motion in a mechanical resonator},\ }\href {https://doi.org/10.1126/science.aac5138} {\bibfield  {journal} {\bibinfo  {journal} {Science}\ }\textbf {\bibinfo {volume} {349}},\ \bibinfo {pages} {952} (\bibinfo {year} {2015})}\BibitemShut {NoStop}%
\bibitem [{\citenamefont {Braunstein}\ and\ \citenamefont {van Loock}(2005)}]{Braunstein2005}%
  \BibitemOpen
  \bibfield  {author} {\bibinfo {author} {\bibfnamefont {S.~L.}\ \bibnamefont {Braunstein}}\ and\ \bibinfo {author} {\bibfnamefont {P.}~\bibnamefont {van Loock}},\ }\bibfield  {title} {\bibinfo {title} {Quantum information with continuous variables},\ }\href {https://doi.org/10.1103/RevModPhys.77.513} {\bibfield  {journal} {\bibinfo  {journal} {Rev. Mod. Phys.}\ }\textbf {\bibinfo {volume} {77}},\ \bibinfo {pages} {513} (\bibinfo {year} {2005})}\BibitemShut {NoStop}%
\bibitem [{\citenamefont {Serafini}(2017)}]{Serafini2017}%
  \BibitemOpen
  \bibfield  {author} {\bibinfo {author} {\bibfnamefont {A.}~\bibnamefont {Serafini}},\ }\href {https://doi.org/10.1201/9781315118727} {\emph {\bibinfo {title} {Quantum Continuous Variables: A Primer of Theoretical Methods}}},\ \bibinfo {edition} {1st}\ ed.\ (\bibinfo  {publisher} {CRC Press},\ \bibinfo {year} {2017})\BibitemShut {NoStop}%
\bibitem [{\citenamefont {Albert}\ \emph {et~al.}(2019)\citenamefont {Albert}, \citenamefont {Mundhada}, \citenamefont {Grimm}, \citenamefont {Touzard}, \citenamefont {Devoret},\ and\ \citenamefont {Jiang}}]{Albert2019}%
  \BibitemOpen
  \bibfield  {author} {\bibinfo {author} {\bibfnamefont {V.~V.}\ \bibnamefont {Albert}}, \bibinfo {author} {\bibfnamefont {S.~O.}\ \bibnamefont {Mundhada}}, \bibinfo {author} {\bibfnamefont {A.}~\bibnamefont {Grimm}}, \bibinfo {author} {\bibfnamefont {S.}~\bibnamefont {Touzard}}, \bibinfo {author} {\bibfnamefont {M.~H.}\ \bibnamefont {Devoret}},\ and\ \bibinfo {author} {\bibfnamefont {L.}~\bibnamefont {Jiang}},\ }\bibfield  {title} {\bibinfo {title} {Pair-cat codes: autonomous error-correction with low-order nonlinearity},\ }\href {https://doi.org/10.1088/2058-9565/ab1e69} {\bibfield  {journal} {\bibinfo  {journal} {Quantum Sci. Technol.}\ }\textbf {\bibinfo {volume} {4}},\ \bibinfo {pages} {035007} (\bibinfo {year} {2019})}\BibitemShut {NoStop}%
\bibitem [{\citenamefont {Gertler}\ \emph {et~al.}(2023)\citenamefont {Gertler}, \citenamefont {van Geldern}, \citenamefont {Shirol}, \citenamefont {Jiang},\ and\ \citenamefont {Wang}}]{Gertler2023}%
  \BibitemOpen
  \bibfield  {author} {\bibinfo {author} {\bibfnamefont {J.~M.}\ \bibnamefont {Gertler}}, \bibinfo {author} {\bibfnamefont {S.}~\bibnamefont {van Geldern}}, \bibinfo {author} {\bibfnamefont {S.}~\bibnamefont {Shirol}}, \bibinfo {author} {\bibfnamefont {L.}~\bibnamefont {Jiang}},\ and\ \bibinfo {author} {\bibfnamefont {C.}~\bibnamefont {Wang}},\ }\bibfield  {title} {\bibinfo {title} {Experimental realization and characterization of stabilized pair-coherent states},\ }\href {https://doi.org/10.1103/PRXQuantum.4.020319} {\bibfield  {journal} {\bibinfo  {journal} {PRX Quantum}\ }\textbf {\bibinfo {volume} {4}},\ \bibinfo {pages} {020319} (\bibinfo {year} {2023})}\BibitemShut {NoStop}%
\bibitem [{\citenamefont {Ma}\ \emph {et~al.}(2021)\citenamefont {Ma}, \citenamefont {Puri}, \citenamefont {Schoelkopf}, \citenamefont {Devoret}, \citenamefont {Girvin},\ and\ \citenamefont {Jiang}}]{Ma2021}%
  \BibitemOpen
  \bibfield  {author} {\bibinfo {author} {\bibfnamefont {W.-L.}\ \bibnamefont {Ma}}, \bibinfo {author} {\bibfnamefont {S.}~\bibnamefont {Puri}}, \bibinfo {author} {\bibfnamefont {R.~J.}\ \bibnamefont {Schoelkopf}}, \bibinfo {author} {\bibfnamefont {M.~H.}\ \bibnamefont {Devoret}}, \bibinfo {author} {\bibfnamefont {S.}~\bibnamefont {Girvin}},\ and\ \bibinfo {author} {\bibfnamefont {L.}~\bibnamefont {Jiang}},\ }\bibfield  {title} {\bibinfo {title} {Quantum control of bosonic modes with superconducting circuits},\ }\href {https://doi.org/https://doi.org/10.1016/j.scib.2021.05.024} {\bibfield  {journal} {\bibinfo  {journal} {Sci. Bull.}\ }\textbf {\bibinfo {volume} {66}},\ \bibinfo {pages} {1789} (\bibinfo {year} {2021})}\BibitemShut {NoStop}%
\bibitem [{\citenamefont {Cai}\ \emph {et~al.}(2021)\citenamefont {Cai}, \citenamefont {Ma}, \citenamefont {Wang}, \citenamefont {Zou},\ and\ \citenamefont {Sun}}]{Cai2021}%
  \BibitemOpen
  \bibfield  {author} {\bibinfo {author} {\bibfnamefont {W.}~\bibnamefont {Cai}}, \bibinfo {author} {\bibfnamefont {Y.}~\bibnamefont {Ma}}, \bibinfo {author} {\bibfnamefont {W.}~\bibnamefont {Wang}}, \bibinfo {author} {\bibfnamefont {C.-L.}\ \bibnamefont {Zou}},\ and\ \bibinfo {author} {\bibfnamefont {L.}~\bibnamefont {Sun}},\ }\bibfield  {title} {\bibinfo {title} {Bosonic quantum error correction codes in superconducting quantum circuits},\ }\href {https://doi.org/https://doi.org/10.1016/j.fmre.2020.12.006} {\bibfield  {journal} {\bibinfo  {journal} {Funda. Res.}\ }\textbf {\bibinfo {volume} {1}},\ \bibinfo {pages} {50} (\bibinfo {year} {2021})}\BibitemShut {NoStop}%
\bibitem [{\citenamefont {Hoshi}\ \emph {et~al.}(2025)\citenamefont {Hoshi}, \citenamefont {Nagase}, \citenamefont {Kwon}, \citenamefont {Iyama}, \citenamefont {Kamiya}, \citenamefont {Fujii}, \citenamefont {Mukai}, \citenamefont {Ahmed}, \citenamefont {Kockum}, \citenamefont {Watabe}, \citenamefont {Yoshihara},\ and\ \citenamefont {Tsai}}]{Hoshi2025}%
  \BibitemOpen
  \bibfield  {author} {\bibinfo {author} {\bibfnamefont {D.}~\bibnamefont {Hoshi}}, \bibinfo {author} {\bibfnamefont {T.}~\bibnamefont {Nagase}}, \bibinfo {author} {\bibfnamefont {S.}~\bibnamefont {Kwon}}, \bibinfo {author} {\bibfnamefont {D.}~\bibnamefont {Iyama}}, \bibinfo {author} {\bibfnamefont {T.}~\bibnamefont {Kamiya}}, \bibinfo {author} {\bibfnamefont {S.}~\bibnamefont {Fujii}}, \bibinfo {author} {\bibfnamefont {H.}~\bibnamefont {Mukai}}, \bibinfo {author} {\bibfnamefont {S.}~\bibnamefont {Ahmed}}, \bibinfo {author} {\bibfnamefont {A.~F.}\ \bibnamefont {Kockum}}, \bibinfo {author} {\bibfnamefont {S.}~\bibnamefont {Watabe}}, \bibinfo {author} {\bibfnamefont {F.}~\bibnamefont {Yoshihara}},\ and\ \bibinfo {author} {\bibfnamefont {J.-S.}\ \bibnamefont {Tsai}},\ }\bibfield  {title} {\bibinfo {title} {Entangling schr{\"o}dinger's cat states by bridging discrete- and continuous-variable encoding},\ }\href {https://doi.org/10.1038/s41467-025-56503-8} {\bibfield  {journal} {\bibinfo  {journal} {Nat. Commun.}\
  }\textbf {\bibinfo {volume} {16}},\ \bibinfo {pages} {1309} (\bibinfo {year} {2025})}\BibitemShut {NoStop}%
\bibitem [{\citenamefont {Liu}\ \emph {et~al.}(2026)\citenamefont {Liu}, \citenamefont {Singh}, \citenamefont {Smith}, \citenamefont {Crane}, \citenamefont {Martyn}, \citenamefont {Eickbusch}, \citenamefont {Schuckert}, \citenamefont {Li}, \citenamefont {Sinanan-Singh}, \citenamefont {Soley}, \citenamefont {Tsunoda}, \citenamefont {Chuang}, \citenamefont {Wiebe},\ and\ \citenamefont {Girvin}}]{Liu2026}%
  \BibitemOpen
  \bibfield  {author} {\bibinfo {author} {\bibfnamefont {Y.}~\bibnamefont {Liu}}, \bibinfo {author} {\bibfnamefont {S.}~\bibnamefont {Singh}}, \bibinfo {author} {\bibfnamefont {K.~C.}\ \bibnamefont {Smith}}, \bibinfo {author} {\bibfnamefont {E.}~\bibnamefont {Crane}}, \bibinfo {author} {\bibfnamefont {J.~M.}\ \bibnamefont {Martyn}}, \bibinfo {author} {\bibfnamefont {A.}~\bibnamefont {Eickbusch}}, \bibinfo {author} {\bibfnamefont {A.}~\bibnamefont {Schuckert}}, \bibinfo {author} {\bibfnamefont {R.~D.}\ \bibnamefont {Li}}, \bibinfo {author} {\bibfnamefont {J.}~\bibnamefont {Sinanan-Singh}}, \bibinfo {author} {\bibfnamefont {M.~B.}\ \bibnamefont {Soley}}, \bibinfo {author} {\bibfnamefont {T.}~\bibnamefont {Tsunoda}}, \bibinfo {author} {\bibfnamefont {I.~L.}\ \bibnamefont {Chuang}}, \bibinfo {author} {\bibfnamefont {N.}~\bibnamefont {Wiebe}},\ and\ \bibinfo {author} {\bibfnamefont {S.~M.}\ \bibnamefont {Girvin}},\ }\bibfield  {title} {\bibinfo {title} {Hybrid oscillator-qubit quantum processors: Instruction set
  architectures, abstract machine models, and applications},\ }\href {https://doi.org/10.1103/4rf7-9tfx} {\bibfield  {journal} {\bibinfo  {journal} {PRX Quantum}\ }\textbf {\bibinfo {volume} {7}},\ \bibinfo {pages} {010201} (\bibinfo {year} {2026})}\BibitemShut {NoStop}%
\bibitem [{\citenamefont {B{\u{a}}z{\u{a}}van}\ \emph {et~al.}(2026)\citenamefont {B{\u{a}}z{\u{a}}van}, \citenamefont {Saner}, \citenamefont {Webb}, \citenamefont {Ainley}, \citenamefont {Drmota}, \citenamefont {Nadlinger}, \citenamefont {Araneda}, \citenamefont {Lucas}, \citenamefont {Ballance},\ and\ \citenamefont {Srinivas}}]{Bazavan2026}%
  \BibitemOpen
  \bibfield  {author} {\bibinfo {author} {\bibfnamefont {O.}~\bibnamefont {B{\u{a}}z{\u{a}}van}}, \bibinfo {author} {\bibfnamefont {S.}~\bibnamefont {Saner}}, \bibinfo {author} {\bibfnamefont {D.~J.}\ \bibnamefont {Webb}}, \bibinfo {author} {\bibfnamefont {E.~M.}\ \bibnamefont {Ainley}}, \bibinfo {author} {\bibfnamefont {P.}~\bibnamefont {Drmota}}, \bibinfo {author} {\bibfnamefont {D.~P.}\ \bibnamefont {Nadlinger}}, \bibinfo {author} {\bibfnamefont {G.}~\bibnamefont {Araneda}}, \bibinfo {author} {\bibfnamefont {D.~M.}\ \bibnamefont {Lucas}}, \bibinfo {author} {\bibfnamefont {C.~J.}\ \bibnamefont {Ballance}},\ and\ \bibinfo {author} {\bibfnamefont {R.}~\bibnamefont {Srinivas}},\ }\bibfield  {title} {\bibinfo {title} {Squeezing, trisqueezing and quadsqueezing in a hybrid oscillator--spin system},\ }\href {https://doi.org/10.1038/s41567-026-03222-6} {\bibfield  {journal} {\bibinfo  {journal} {Nat. Phys.}\ }\textbf {\bibinfo {volume} {22}},\ \bibinfo {pages} {757} (\bibinfo {year} {2026})}\BibitemShut {NoStop}%
\bibitem [{\citenamefont {Bylicka}\ \emph {et~al.}(2014)\citenamefont {Bylicka}, \citenamefont {Chru{\'{s}}ci{\'{n}}ski},\ and\ \citenamefont {Maniscalco}}]{Bylicka2014}%
  \BibitemOpen
  \bibfield  {author} {\bibinfo {author} {\bibfnamefont {B.}~\bibnamefont {Bylicka}}, \bibinfo {author} {\bibfnamefont {D.}~\bibnamefont {Chru{\'{s}}ci{\'{n}}ski}},\ and\ \bibinfo {author} {\bibfnamefont {S.}~\bibnamefont {Maniscalco}},\ }\bibfield  {title} {\bibinfo {title} {Non-markovianity and reservoir memory of quantum channels: a quantum information theory perspective},\ }\href {https://doi.org/10.1038/srep05720} {\bibfield  {journal} {\bibinfo  {journal} {Sci. Rep.}\ }\textbf {\bibinfo {volume} {4}},\ \bibinfo {pages} {5720} (\bibinfo {year} {2014})}\BibitemShut {NoStop}%
\bibitem [{\citenamefont {Han}\ \emph {et~al.}(2021)\citenamefont {Han}, \citenamefont {Cai}, \citenamefont {Hu}, \citenamefont {Mu}, \citenamefont {Ma}, \citenamefont {Xu}, \citenamefont {Wang}, \citenamefont {Wang}, \citenamefont {Song}, \citenamefont {Zou},\ and\ \citenamefont {Sun}}]{Han2021}%
  \BibitemOpen
  \bibfield  {author} {\bibinfo {author} {\bibfnamefont {J.}~\bibnamefont {Han}}, \bibinfo {author} {\bibfnamefont {W.}~\bibnamefont {Cai}}, \bibinfo {author} {\bibfnamefont {L.}~\bibnamefont {Hu}}, \bibinfo {author} {\bibfnamefont {X.}~\bibnamefont {Mu}}, \bibinfo {author} {\bibfnamefont {Y.}~\bibnamefont {Ma}}, \bibinfo {author} {\bibfnamefont {Y.}~\bibnamefont {Xu}}, \bibinfo {author} {\bibfnamefont {W.}~\bibnamefont {Wang}}, \bibinfo {author} {\bibfnamefont {H.}~\bibnamefont {Wang}}, \bibinfo {author} {\bibfnamefont {Y.~P.}\ \bibnamefont {Song}}, \bibinfo {author} {\bibfnamefont {C.-L.}\ \bibnamefont {Zou}},\ and\ \bibinfo {author} {\bibfnamefont {L.}~\bibnamefont {Sun}},\ }\bibfield  {title} {\bibinfo {title} {Experimental simulation of open quantum system dynamics via trotterization},\ }\href {https://doi.org/10.1103/PhysRevLett.127.020504} {\bibfield  {journal} {\bibinfo  {journal} {Phys. Rev. Lett.}\ }\textbf {\bibinfo {volume} {127}},\ \bibinfo {pages} {020504} (\bibinfo {year} {2021})}\BibitemShut
  {NoStop}%
\bibitem [{\citenamefont {Wang}\ \emph {et~al.}(2013)\citenamefont {Wang}, \citenamefont {Berry}, \citenamefont {de~Oliveira},\ and\ \citenamefont {Sanders}}]{Wang2013}%
  \BibitemOpen
  \bibfield  {author} {\bibinfo {author} {\bibfnamefont {D.-S.}\ \bibnamefont {Wang}}, \bibinfo {author} {\bibfnamefont {D.~W.}\ \bibnamefont {Berry}}, \bibinfo {author} {\bibfnamefont {M.~C.}\ \bibnamefont {de~Oliveira}},\ and\ \bibinfo {author} {\bibfnamefont {B.~C.}\ \bibnamefont {Sanders}},\ }\bibfield  {title} {\bibinfo {title} {Solovay-kitaev decomposition strategy for single-qubit channels},\ }\href {https://doi.org/10.1103/PhysRevLett.111.130504} {\bibfield  {journal} {\bibinfo  {journal} {Phys. Rev. Lett.}\ }\textbf {\bibinfo {volume} {111}},\ \bibinfo {pages} {130504} (\bibinfo {year} {2013})}\BibitemShut {NoStop}%
\bibitem [{\citenamefont {Shen}\ \emph {et~al.}(2017)\citenamefont {Shen}, \citenamefont {Noh}, \citenamefont {Albert}, \citenamefont {Krastanov}, \citenamefont {Devoret}, \citenamefont {Schoelkopf}, \citenamefont {Girvin},\ and\ \citenamefont {Jiang}}]{Shen2017}%
  \BibitemOpen
  \bibfield  {author} {\bibinfo {author} {\bibfnamefont {C.}~\bibnamefont {Shen}}, \bibinfo {author} {\bibfnamefont {K.}~\bibnamefont {Noh}}, \bibinfo {author} {\bibfnamefont {V.~V.}\ \bibnamefont {Albert}}, \bibinfo {author} {\bibfnamefont {S.}~\bibnamefont {Krastanov}}, \bibinfo {author} {\bibfnamefont {M.~H.}\ \bibnamefont {Devoret}}, \bibinfo {author} {\bibfnamefont {R.~J.}\ \bibnamefont {Schoelkopf}}, \bibinfo {author} {\bibfnamefont {S.~M.}\ \bibnamefont {Girvin}},\ and\ \bibinfo {author} {\bibfnamefont {L.}~\bibnamefont {Jiang}},\ }\bibfield  {title} {\bibinfo {title} {Quantum channel construction with circuit quantum electrodynamics},\ }\href {https://doi.org/10.1103/PhysRevB.95.134501} {\bibfield  {journal} {\bibinfo  {journal} {Phys. Rev. B}\ }\textbf {\bibinfo {volume} {95}},\ \bibinfo {pages} {134501} (\bibinfo {year} {2017})}\BibitemShut {NoStop}%
\bibitem [{\citenamefont {Hu}\ \emph {et~al.}(2018)\citenamefont {Hu}, \citenamefont {Mu}, \citenamefont {Cai}, \citenamefont {Ma}, \citenamefont {Xu}, \citenamefont {Wang}, \citenamefont {Song}, \citenamefont {Zou},\ and\ \citenamefont {Sun}}]{Hu2018}%
  \BibitemOpen
  \bibfield  {author} {\bibinfo {author} {\bibfnamefont {L.}~\bibnamefont {Hu}}, \bibinfo {author} {\bibfnamefont {X.}~\bibnamefont {Mu}}, \bibinfo {author} {\bibfnamefont {W.}~\bibnamefont {Cai}}, \bibinfo {author} {\bibfnamefont {Y.}~\bibnamefont {Ma}}, \bibinfo {author} {\bibfnamefont {Y.}~\bibnamefont {Xu}}, \bibinfo {author} {\bibfnamefont {H.}~\bibnamefont {Wang}}, \bibinfo {author} {\bibfnamefont {Y.}~\bibnamefont {Song}}, \bibinfo {author} {\bibfnamefont {C.-L.}\ \bibnamefont {Zou}},\ and\ \bibinfo {author} {\bibfnamefont {L.}~\bibnamefont {Sun}},\ }\bibfield  {title} {\bibinfo {title} {Experimental repetitive quantum channel simulation},\ }\href {https://doi.org/10.1016/j.scib.2018.11.010} {\bibfield  {journal} {\bibinfo  {journal} {Sci. Bull.}\ }\textbf {\bibinfo {volume} {63}},\ \bibinfo {pages} {1551} (\bibinfo {year} {2018})}\BibitemShut {NoStop}%
\bibitem [{\citenamefont {Chen}\ \emph {et~al.}(2025)\citenamefont {Chen}, \citenamefont {Huang}, \citenamefont {Sun}, \citenamefont {Jie}, \citenamefont {Zhou}, \citenamefont {Hua}, \citenamefont {Xu}, \citenamefont {Wang}, \citenamefont {Guo}, \citenamefont {Zou}, \citenamefont {Sun},\ and\ \citenamefont {Zou}}]{Chen2025}%
  \BibitemOpen
  \bibfield  {author} {\bibinfo {author} {\bibfnamefont {Z.-J.}\ \bibnamefont {Chen}}, \bibinfo {author} {\bibfnamefont {H.}~\bibnamefont {Huang}}, \bibinfo {author} {\bibfnamefont {L.}~\bibnamefont {Sun}}, \bibinfo {author} {\bibfnamefont {Q.-X.}\ \bibnamefont {Jie}}, \bibinfo {author} {\bibfnamefont {J.}~\bibnamefont {Zhou}}, \bibinfo {author} {\bibfnamefont {Z.}~\bibnamefont {Hua}}, \bibinfo {author} {\bibfnamefont {Y.}~\bibnamefont {Xu}}, \bibinfo {author} {\bibfnamefont {W.}~\bibnamefont {Wang}}, \bibinfo {author} {\bibfnamefont {G.-C.}\ \bibnamefont {Guo}}, \bibinfo {author} {\bibfnamefont {C.-L.}\ \bibnamefont {Zou}}, \bibinfo {author} {\bibfnamefont {L.}~\bibnamefont {Sun}},\ and\ \bibinfo {author} {\bibfnamefont {X.-B.}\ \bibnamefont {Zou}},\ }\bibfield  {title} {\bibinfo {title} {Robust and optimal control of open quantum systems},\ }\href {https://doi.org/10.1126/sciadv.adr0875} {\bibfield  {journal} {\bibinfo  {journal} {Sci. Adv.}\ }\textbf {\bibinfo {volume} {11}},\ \bibinfo {pages} {eadr0875}
  (\bibinfo {year} {2025})}\BibitemShut {NoStop}%
\end{thebibliography}%


\begin{thebibliography}{23}%
\makeatletter
\providecommand \@ifxundefined [1]{%
 \@ifx{#1\undefined}
}%
\providecommand \@ifnum [1]{%
 \ifnum #1\expandafter \@firstoftwo
 \else \expandafter \@secondoftwo
 \fi
}%
\providecommand \@ifx [1]{%
 \ifx #1\expandafter \@firstoftwo
 \else \expandafter \@secondoftwo
 \fi
}%
\providecommand \natexlab [1]{#1}%
\providecommand \enquote  [1]{``#1''}%
\providecommand \bibnamefont  [1]{#1}%
\providecommand \bibfnamefont [1]{#1}%
\providecommand \citenamefont [1]{#1}%
\providecommand \href@noop [0]{\@secondoftwo}%
\providecommand \href [0]{\begingroup \@sanitize@url \@href}%
\providecommand \@href[1]{\@@startlink{#1}\@@href}%
\providecommand \@@href[1]{\endgroup#1\@@endlink}%
\providecommand \@sanitize@url [0]{\catcode `\\12\catcode `\$12\catcode `\&12\catcode `\#12\catcode `\^12\catcode `\_12\catcode `\%12\relax}%
\providecommand \@@startlink[1]{}%
\providecommand \@@endlink[0]{}%
\providecommand \url  [0]{\begingroup\@sanitize@url \@url }%
\providecommand \@url [1]{\endgroup\@href {#1}{\urlprefix }}%
\providecommand \urlprefix  [0]{URL }%
\providecommand \Eprint [0]{\href }%
\providecommand \doibase [0]{https://doi.org/}%
\providecommand \selectlanguage [0]{\@gobble}%
\providecommand \bibinfo  [0]{\@secondoftwo}%
\providecommand \bibfield  [0]{\@secondoftwo}%
\providecommand \translation [1]{[#1]}%
\providecommand \BibitemOpen [0]{}%
\providecommand \bibitemStop [0]{}%
\providecommand \bibitemNoStop [0]{.\EOS\space}%
\providecommand \EOS [0]{\spacefactor3000\relax}%
\providecommand \BibitemShut  [1]{\csname bibitem#1\endcsname}%
\let\auto@bib@innerbib\@empty
\bibitem [{\citenamefont {Goto}(2016)}]{Goto2016}%
  \BibitemOpen
  \bibfield  {author} {\bibinfo {author} {\bibfnamefont {H.}~\bibnamefont {Goto}},\ }\bibfield  {title} {\bibinfo {title} {Universal quantum computation with a nonlinear oscillator network},\ }\href {https://doi.org/10.1103/PhysRevA.93.050301} {\bibfield  {journal} {\bibinfo  {journal} {Phys. Rev. A}\ }\textbf {\bibinfo {volume} {93}},\ \bibinfo {pages} {050301(R)} (\bibinfo {year} {2016})}\BibitemShut {NoStop}%
\bibitem [{\citenamefont {Minganti}\ \emph {et~al.}(2016)\citenamefont {Minganti}, \citenamefont {Bartolo}, \citenamefont {Lolli}, \citenamefont {Casteels},\ and\ \citenamefont {Ciuti}}]{Minganti2016}%
  \BibitemOpen
  \bibfield  {author} {\bibinfo {author} {\bibfnamefont {F.}~\bibnamefont {Minganti}}, \bibinfo {author} {\bibfnamefont {N.}~\bibnamefont {Bartolo}}, \bibinfo {author} {\bibfnamefont {J.}~\bibnamefont {Lolli}}, \bibinfo {author} {\bibfnamefont {W.}~\bibnamefont {Casteels}},\ and\ \bibinfo {author} {\bibfnamefont {C.}~\bibnamefont {Ciuti}},\ }\bibfield  {title} {\bibinfo {title} {Exact results for schr{\"o}dinger cats in driven-dissipative systems and their feedback control},\ }\href {https://doi.org/10.1038/srep26987} {\bibfield  {journal} {\bibinfo  {journal} {Sci. Rep.}\ }\textbf {\bibinfo {volume} {6}},\ \bibinfo {pages} {26987} (\bibinfo {year} {2016})}\BibitemShut {NoStop}%
\bibitem [{\citenamefont {Puri}\ \emph {et~al.}(2017)\citenamefont {Puri}, \citenamefont {Boutin},\ and\ \citenamefont {Blais}}]{Puri2017}%
  \BibitemOpen
  \bibfield  {author} {\bibinfo {author} {\bibfnamefont {S.}~\bibnamefont {Puri}}, \bibinfo {author} {\bibfnamefont {S.}~\bibnamefont {Boutin}},\ and\ \bibinfo {author} {\bibfnamefont {A.}~\bibnamefont {Blais}},\ }\bibfield  {title} {\bibinfo {title} {Engineering the quantum states of light in a kerr-nonlinear resonator by two-photon driving},\ }\href {https://doi.org/10.1038/s41534-017-0019-1} {\bibfield  {journal} {\bibinfo  {journal} {npj Quantum Inf.}\ }\textbf {\bibinfo {volume} {3}},\ \bibinfo {pages} {18} (\bibinfo {year} {2017})}\BibitemShut {NoStop}%
\bibitem [{\citenamefont {Puri}\ \emph {et~al.}(2019)\citenamefont {Puri}, \citenamefont {Grimm}, \citenamefont {Campagne-Ibarcq}, \citenamefont {Eickbusch}, \citenamefont {Noh}, \citenamefont {Roberts}, \citenamefont {Jiang}, \citenamefont {Mirrahimi}, \citenamefont {Devoret},\ and\ \citenamefont {Girvin}}]{Puri2019}%
  \BibitemOpen
  \bibfield  {author} {\bibinfo {author} {\bibfnamefont {S.}~\bibnamefont {Puri}}, \bibinfo {author} {\bibfnamefont {A.}~\bibnamefont {Grimm}}, \bibinfo {author} {\bibfnamefont {P.}~\bibnamefont {Campagne-Ibarcq}}, \bibinfo {author} {\bibfnamefont {A.}~\bibnamefont {Eickbusch}}, \bibinfo {author} {\bibfnamefont {K.}~\bibnamefont {Noh}}, \bibinfo {author} {\bibfnamefont {G.}~\bibnamefont {Roberts}}, \bibinfo {author} {\bibfnamefont {L.}~\bibnamefont {Jiang}}, \bibinfo {author} {\bibfnamefont {M.}~\bibnamefont {Mirrahimi}}, \bibinfo {author} {\bibfnamefont {M.~H.}\ \bibnamefont {Devoret}},\ and\ \bibinfo {author} {\bibfnamefont {S.~M.}\ \bibnamefont {Girvin}},\ }\bibfield  {title} {\bibinfo {title} {Stabilized cat in a driven nonlinear cavity: A fault-tolerant error syndrome detector},\ }\href {https://doi.org/10.1103/PhysRevX.9.041009} {\bibfield  {journal} {\bibinfo  {journal} {Phys. Rev. X}\ }\textbf {\bibinfo {volume} {9}},\ \bibinfo {pages} {041009} (\bibinfo {year} {2019})}\BibitemShut {NoStop}%
\bibitem [{\citenamefont {Grimm}\ \emph {et~al.}(2020)\citenamefont {Grimm}, \citenamefont {Frattini}, \citenamefont {Puri}, \citenamefont {Mundhada}, \citenamefont {Touzard}, \citenamefont {Mirrahimi}, \citenamefont {Girvin}, \citenamefont {Shankar},\ and\ \citenamefont {Devoret}}]{Grimm2020}%
  \BibitemOpen
  \bibfield  {author} {\bibinfo {author} {\bibfnamefont {A.}~\bibnamefont {Grimm}}, \bibinfo {author} {\bibfnamefont {N.~E.}\ \bibnamefont {Frattini}}, \bibinfo {author} {\bibfnamefont {S.}~\bibnamefont {Puri}}, \bibinfo {author} {\bibfnamefont {S.~O.}\ \bibnamefont {Mundhada}}, \bibinfo {author} {\bibfnamefont {S.}~\bibnamefont {Touzard}}, \bibinfo {author} {\bibfnamefont {M.}~\bibnamefont {Mirrahimi}}, \bibinfo {author} {\bibfnamefont {S.~M.}\ \bibnamefont {Girvin}}, \bibinfo {author} {\bibfnamefont {S.}~\bibnamefont {Shankar}},\ and\ \bibinfo {author} {\bibfnamefont {M.~H.}\ \bibnamefont {Devoret}},\ }\bibfield  {title} {\bibinfo {title} {Stabilization and operation of a kerr-cat qubit},\ }\href {https://doi.org/10.1038/s41586-020-2587-z} {\bibfield  {journal} {\bibinfo  {journal} {Nature (London)}\ }\textbf {\bibinfo {volume} {584}},\ \bibinfo {pages} {205} (\bibinfo {year} {2020})}\BibitemShut {NoStop}%
\bibitem [{\citenamefont {Han}\ \emph {et~al.}(2026)\citenamefont {Han}, \citenamefont {Yang}, \citenamefont {Ning}, \citenamefont {Zhang}, \citenamefont {Kang}, \citenamefont {Qiu},\ and\ \citenamefont {Yang}}]{Han2026}%
  \BibitemOpen
  \bibfield  {author} {\bibinfo {author} {\bibfnamefont {P.-R.}\ \bibnamefont {Han}}, \bibinfo {author} {\bibfnamefont {T.-L.}\ \bibnamefont {Yang}}, \bibinfo {author} {\bibfnamefont {W.}~\bibnamefont {Ning}}, \bibinfo {author} {\bibfnamefont {H.-L.}\ \bibnamefont {Zhang}}, \bibinfo {author} {\bibfnamefont {H.}~\bibnamefont {Kang}}, \bibinfo {author} {\bibfnamefont {H.}~\bibnamefont {Qiu}},\ and\ \bibinfo {author} {\bibfnamefont {Z.-B.}\ \bibnamefont {Yang}},\ }\bibfield  {title} {\bibinfo {title} {Liouvillian exceptional-point-induced phase transition in a single kerr-cat qubit},\ }\bibfield  {journal} {\bibinfo  {journal} {Commun. Phys.}\ }\href {https://doi.org/10.1038/s42005-026-02919-7} {10.1038/s42005-026-02919-7} (\bibinfo {year} {2026})\BibitemShut {NoStop}%
\bibitem [{\citenamefont {Wong}\ \emph {et~al.}(2026)\citenamefont {Wong}, \citenamefont {Zeng},\ and\ \citenamefont {Li}}]{Wong2026}%
  \BibitemOpen
  \bibfield  {author} {\bibinfo {author} {\bibfnamefont {W.~C.}\ \bibnamefont {Wong}}, \bibinfo {author} {\bibfnamefont {B.}~\bibnamefont {Zeng}},\ and\ \bibinfo {author} {\bibfnamefont {J.}~\bibnamefont {Li}},\ }\bibfield  {title} {\bibinfo {title} {Non-markovian exceptional points by interpolating quantum channels},\ }\href {https://doi.org/10.1038/s41534-026-01205-2} {\bibfield  {journal} {\bibinfo  {journal} {npj Quantum Inf.}\ }\textbf {\bibinfo {volume} {12}},\ \bibinfo {pages} {63} (\bibinfo {year} {2026})}\BibitemShut {NoStop}%
\bibitem [{\citenamefont {Wolf}\ \emph {et~al.}(2008)\citenamefont {Wolf}, \citenamefont {Eisert}, \citenamefont {Cubitt},\ and\ \citenamefont {Cirac}}]{Wolf2008}%
  \BibitemOpen
  \bibfield  {author} {\bibinfo {author} {\bibfnamefont {M.~M.}\ \bibnamefont {Wolf}}, \bibinfo {author} {\bibfnamefont {J.}~\bibnamefont {Eisert}}, \bibinfo {author} {\bibfnamefont {T.~S.}\ \bibnamefont {Cubitt}},\ and\ \bibinfo {author} {\bibfnamefont {J.~I.}\ \bibnamefont {Cirac}},\ }\bibfield  {title} {\bibinfo {title} {Assessing non-markovian quantum dynamics},\ }\href {https://doi.org/10.1103/PhysRevLett.101.150402} {\bibfield  {journal} {\bibinfo  {journal} {Phys. Rev. Lett.}\ }\textbf {\bibinfo {volume} {101}},\ \bibinfo {pages} {150402} (\bibinfo {year} {2008})}\BibitemShut {NoStop}%
\bibitem [{\citenamefont {Davalos}\ \emph {et~al.}(2019)\citenamefont {Davalos}, \citenamefont {Ziman},\ and\ \citenamefont {Pineda}}]{Davalos2019}%
  \BibitemOpen
  \bibfield  {author} {\bibinfo {author} {\bibfnamefont {D.}~\bibnamefont {Davalos}}, \bibinfo {author} {\bibfnamefont {M.}~\bibnamefont {Ziman}},\ and\ \bibinfo {author} {\bibfnamefont {C.}~\bibnamefont {Pineda}},\ }\bibfield  {title} {\bibinfo {title} {Divisibility of qubit channels and dynamical maps},\ }\href {https://doi.org/10.22331/q-2019-05-20-144} {\bibfield  {journal} {\bibinfo  {journal} {{Quantum}}\ }\textbf {\bibinfo {volume} {3}},\ \bibinfo {pages} {144} (\bibinfo {year} {2019})}\BibitemShut {NoStop}%
\bibitem [{\citenamefont {Culver}(1966)}]{Culver1966}%
  \BibitemOpen
  \bibfield  {author} {\bibinfo {author} {\bibfnamefont {W.~J.}\ \bibnamefont {Culver}},\ }\bibfield  {title} {\bibinfo {title} {On the existence and uniqueness of the real logarithm of a matrix},\ }\href {https://doi.org/10.1090/S0002-9939-1966-0202740-6} {\bibfield  {journal} {\bibinfo  {journal} {Proc. Am. Math. Soc.}\ }\textbf {\bibinfo {volume} {17}},\ \bibinfo {pages} {1146} (\bibinfo {year} {1966})}\BibitemShut {NoStop}%
\bibitem [{\citenamefont {Minganti}\ \emph {et~al.}(2018)\citenamefont {Minganti}, \citenamefont {Biella}, \citenamefont {Bartolo},\ and\ \citenamefont {Ciuti}}]{Minganti2018}%
  \BibitemOpen
  \bibfield  {author} {\bibinfo {author} {\bibfnamefont {F.}~\bibnamefont {Minganti}}, \bibinfo {author} {\bibfnamefont {A.}~\bibnamefont {Biella}}, \bibinfo {author} {\bibfnamefont {N.}~\bibnamefont {Bartolo}},\ and\ \bibinfo {author} {\bibfnamefont {C.}~\bibnamefont {Ciuti}},\ }\bibfield  {title} {\bibinfo {title} {Spectral theory of liouvillians for dissipative phase transitions},\ }\href {https://doi.org/10.1103/PhysRevA.98.042118} {\bibfield  {journal} {\bibinfo  {journal} {Phys. Rev. A}\ }\textbf {\bibinfo {volume} {98}},\ \bibinfo {pages} {042118} (\bibinfo {year} {2018})}\BibitemShut {NoStop}%
\bibitem [{\citenamefont {Brandão}(2026)}]{Brandao2026}%
  \BibitemOpen
  \bibfield  {author} {\bibinfo {author} {\bibfnamefont {P.~A.}\ \bibnamefont {Brandão}},\ }\bibfield  {title} {\bibinfo {title} {Operational signatures of exceptional quantum channels},\ }\href {https://doi.org/10.48550/arXiv.2610.13622} {\bibfield  {journal} {\bibinfo  {journal} {arXiv}\ }\textbf {\bibinfo {volume} {2610.13622}} (\bibinfo {year} {2026})}\BibitemShut {NoStop}%
\bibitem [{\citenamefont {Gorini}\ \emph {et~al.}(1976)\citenamefont {Gorini}, \citenamefont {Kossakowski},\ and\ \citenamefont {Sudarshan}}]{Gorini1976}%
  \BibitemOpen
  \bibfield  {author} {\bibinfo {author} {\bibfnamefont {V.}~\bibnamefont {Gorini}}, \bibinfo {author} {\bibfnamefont {A.}~\bibnamefont {Kossakowski}},\ and\ \bibinfo {author} {\bibfnamefont {E.~C.~G.}\ \bibnamefont {Sudarshan}},\ }\bibfield  {title} {\bibinfo {title} {Completely positive dynamical semigroups of n‐level systems},\ }\href {https://doi.org/10.1063/1.522979} {\bibfield  {journal} {\bibinfo  {journal} {J. Math. Phys.}\ }\textbf {\bibinfo {volume} {17}},\ \bibinfo {pages} {821} (\bibinfo {year} {1976})}\BibitemShut {NoStop}%
\bibitem [{\citenamefont {Lindblad}(1976)}]{Lindblad1976}%
  \BibitemOpen
  \bibfield  {author} {\bibinfo {author} {\bibfnamefont {G.}~\bibnamefont {Lindblad}},\ }\bibfield  {title} {\bibinfo {title} {On the generators of quantum dynamical semigroups},\ }\href {https://doi.org/10.1007/BF01608499} {\bibfield  {journal} {\bibinfo  {journal} {Commun. Math. Phys.}\ }\textbf {\bibinfo {volume} {48}},\ \bibinfo {pages} {119} (\bibinfo {year} {1976})}\BibitemShut {NoStop}%
\bibitem [{\citenamefont {Wolf}\ and\ \citenamefont {Cirac}(2008)}]{Wolf2008CMP}%
  \BibitemOpen
  \bibfield  {author} {\bibinfo {author} {\bibfnamefont {M.~M.}\ \bibnamefont {Wolf}}\ and\ \bibinfo {author} {\bibfnamefont {J.~I.}\ \bibnamefont {Cirac}},\ }\bibfield  {title} {\bibinfo {title} {Dividing quantum channels},\ }\href {https://doi.org/10.1007/s00220-008-0411-y} {\bibfield  {journal} {\bibinfo  {journal} {Commun. Math. Phys.}\ }\textbf {\bibinfo {volume} {279}},\ \bibinfo {pages} {147} (\bibinfo {year} {2008})}\BibitemShut {NoStop}%
\bibitem [{\citenamefont {Choi}(1975)}]{Choi1975}%
  \BibitemOpen
  \bibfield  {author} {\bibinfo {author} {\bibfnamefont {M.-D.}\ \bibnamefont {Choi}},\ }\bibfield  {title} {\bibinfo {title} {Completely positive linear maps on complex matrices},\ }\href {https://doi.org/https://doi.org/10.1016/0024-3795(75)90075-0} {\bibfield  {journal} {\bibinfo  {journal} {Lin. Algebra Appl.}\ }\textbf {\bibinfo {volume} {10}},\ \bibinfo {pages} {285} (\bibinfo {year} {1975})}\BibitemShut {NoStop}%
\bibitem [{\citenamefont {Breuer}\ \emph {et~al.}(2016)\citenamefont {Breuer}, \citenamefont {Laine}, \citenamefont {Piilo},\ and\ \citenamefont {Vacchini}}]{Breuer2016}%
  \BibitemOpen
  \bibfield  {author} {\bibinfo {author} {\bibfnamefont {H.-P.}\ \bibnamefont {Breuer}}, \bibinfo {author} {\bibfnamefont {E.-M.}\ \bibnamefont {Laine}}, \bibinfo {author} {\bibfnamefont {J.}~\bibnamefont {Piilo}},\ and\ \bibinfo {author} {\bibfnamefont {B.}~\bibnamefont {Vacchini}},\ }\bibfield  {title} {\bibinfo {title} {Colloquium: Non-markovian dynamics in open quantum systems},\ }\href {https://doi.org/10.1103/RevModPhys.88.021002} {\bibfield  {journal} {\bibinfo  {journal} {Rev. Mod. Phys.}\ }\textbf {\bibinfo {volume} {88}},\ \bibinfo {pages} {021002} (\bibinfo {year} {2016})}\BibitemShut {NoStop}%
\bibitem [{\citenamefont {Rivas}\ \emph {et~al.}(2014)\citenamefont {Rivas}, \citenamefont {Huelga},\ and\ \citenamefont {Plenio}}]{Rivas2014}%
  \BibitemOpen
  \bibfield  {author} {\bibinfo {author} {\bibfnamefont {{\'A}.}~\bibnamefont {Rivas}}, \bibinfo {author} {\bibfnamefont {S.~F.}\ \bibnamefont {Huelga}},\ and\ \bibinfo {author} {\bibfnamefont {M.~B.}\ \bibnamefont {Plenio}},\ }\bibfield  {title} {\bibinfo {title} {Quantum non-markovianity: characterization, quantification and detection},\ }\href {https://doi.org/10.1088/0034-4885/77/9/094001} {\bibfield  {journal} {\bibinfo  {journal} {Rep. Prog. Phys.}\ }\textbf {\bibinfo {volume} {77}},\ \bibinfo {pages} {094001} (\bibinfo {year} {2014})}\BibitemShut {NoStop}%
\bibitem [{\citenamefont {Chru\ifmmode \acute{s}\else \'{s}\fi{}ci\ifmmode~\acute{n}\else \'{n}\fi{}ski}\ and\ \citenamefont {Kossakowski}(2010)}]{Chruscinski2010}%
  \BibitemOpen
  \bibfield  {author} {\bibinfo {author} {\bibfnamefont {D.}~\bibnamefont {Chru\ifmmode \acute{s}\else \'{s}\fi{}ci\ifmmode~\acute{n}\else \'{n}\fi{}ski}}\ and\ \bibinfo {author} {\bibfnamefont {A.}~\bibnamefont {Kossakowski}},\ }\bibfield  {title} {\bibinfo {title} {Non-markovian quantum dynamics: Local versus nonlocal},\ }\href {https://doi.org/10.1103/PhysRevLett.104.070406} {\bibfield  {journal} {\bibinfo  {journal} {Phys. Rev. Lett.}\ }\textbf {\bibinfo {volume} {104}},\ \bibinfo {pages} {070406} (\bibinfo {year} {2010})}\BibitemShut {NoStop}%
\bibitem [{\citenamefont {Hall}\ \emph {et~al.}(2014)\citenamefont {Hall}, \citenamefont {Cresser}, \citenamefont {Li},\ and\ \citenamefont {Andersson}}]{Hall2014}%
  \BibitemOpen
  \bibfield  {author} {\bibinfo {author} {\bibfnamefont {M.~J.~W.}\ \bibnamefont {Hall}}, \bibinfo {author} {\bibfnamefont {J.~D.}\ \bibnamefont {Cresser}}, \bibinfo {author} {\bibfnamefont {L.}~\bibnamefont {Li}},\ and\ \bibinfo {author} {\bibfnamefont {E.}~\bibnamefont {Andersson}},\ }\bibfield  {title} {\bibinfo {title} {Canonical form of master equations and characterization of non-markovianity},\ }\href {https://doi.org/10.1103/PhysRevA.89.042120} {\bibfield  {journal} {\bibinfo  {journal} {Phys. Rev. A}\ }\textbf {\bibinfo {volume} {89}},\ \bibinfo {pages} {042120} (\bibinfo {year} {2014})}\BibitemShut {NoStop}%
\bibitem [{\citenamefont {Breuer}\ \emph {et~al.}(2009)\citenamefont {Breuer}, \citenamefont {Laine},\ and\ \citenamefont {Piilo}}]{Breuer2009}%
  \BibitemOpen
  \bibfield  {author} {\bibinfo {author} {\bibfnamefont {H.-P.}\ \bibnamefont {Breuer}}, \bibinfo {author} {\bibfnamefont {E.-M.}\ \bibnamefont {Laine}},\ and\ \bibinfo {author} {\bibfnamefont {J.}~\bibnamefont {Piilo}},\ }\bibfield  {title} {\bibinfo {title} {Measure for the degree of non-markovian behavior of quantum processes in open systems},\ }\href {https://doi.org/10.1103/PhysRevLett.103.210401} {\bibfield  {journal} {\bibinfo  {journal} {Phys. Rev. Lett.}\ }\textbf {\bibinfo {volume} {103}},\ \bibinfo {pages} {210401} (\bibinfo {year} {2009})}\BibitemShut {NoStop}%
\bibitem [{\citenamefont {Rivas}\ \emph {et~al.}(2010)\citenamefont {Rivas}, \citenamefont {Huelga},\ and\ \citenamefont {Plenio}}]{Rivas2010}%
  \BibitemOpen
  \bibfield  {author} {\bibinfo {author} {\bibfnamefont {{\'A}.}~\bibnamefont {Rivas}}, \bibinfo {author} {\bibfnamefont {S.~F.}\ \bibnamefont {Huelga}},\ and\ \bibinfo {author} {\bibfnamefont {M.~B.}\ \bibnamefont {Plenio}},\ }\bibfield  {title} {\bibinfo {title} {Entanglement and non-markovianity of quantum evolutions},\ }\href {https://doi.org/10.1103/PhysRevLett.105.050403} {\bibfield  {journal} {\bibinfo  {journal} {Phys. Rev. Lett.}\ }\textbf {\bibinfo {volume} {105}},\ \bibinfo {pages} {050403} (\bibinfo {year} {2010})}\BibitemShut {NoStop}%
\bibitem [{\citenamefont {Fanchini}\ \emph {et~al.}(2014)\citenamefont {Fanchini}, \citenamefont {Karpat}, \citenamefont {\ifmmode~\mbox{\c{C}}\else \c{C}\fi{}akmak}, \citenamefont {Castelano}, \citenamefont {Aguilar}, \citenamefont {Far\'{\i}as}, \citenamefont {Walborn}, \citenamefont {Ribeiro},\ and\ \citenamefont {de~Oliveira}}]{Fanchini2014}%
  \BibitemOpen
  \bibfield  {author} {\bibinfo {author} {\bibfnamefont {F.~F.}\ \bibnamefont {Fanchini}}, \bibinfo {author} {\bibfnamefont {G.}~\bibnamefont {Karpat}}, \bibinfo {author} {\bibfnamefont {B.}~\bibnamefont {\ifmmode~\mbox{\c{C}}\else \c{C}\fi{}akmak}}, \bibinfo {author} {\bibfnamefont {L.~K.}\ \bibnamefont {Castelano}}, \bibinfo {author} {\bibfnamefont {G.~H.}\ \bibnamefont {Aguilar}}, \bibinfo {author} {\bibfnamefont {O.~J.}\ \bibnamefont {Far\'{\i}as}}, \bibinfo {author} {\bibfnamefont {S.~P.}\ \bibnamefont {Walborn}}, \bibinfo {author} {\bibfnamefont {P.~H.~S.}\ \bibnamefont {Ribeiro}},\ and\ \bibinfo {author} {\bibfnamefont {M.~C.}\ \bibnamefont {de~Oliveira}},\ }\bibfield  {title} {\bibinfo {title} {Non-markovianity through accessible information},\ }\href {https://doi.org/10.1103/PhysRevLett.112.210402} {\bibfield  {journal} {\bibinfo  {journal} {Phys. Rev. Lett.}\ }\textbf {\bibinfo {volume} {112}},\ \bibinfo {pages} {210402} (\bibinfo {year} {2014})}\BibitemShut {NoStop}%
\end{thebibliography}%

\end{document}


\title{Supplementary Material for ``Control of Non-Markovianity via Channel Exceptional Points in a Driven-Dissipative Kerr-Cat Qubit"}

\author{Tian-Le Yang}
\affiliation{Fujian Key Laboratory of Quantum Information and Quantum Optics, Fuzhou University, Fuzhou 350108, China}
\affiliation{Department of Physics, Fuzhou University, Fuzhou 350108, China}
\author{Hao-Long Zhang}
\affiliation{Fujian Key Laboratory of Quantum Information and Quantum Optics, Fuzhou University, Fuzhou 350108, China}
\affiliation{Department of Physics, Fuzhou University, Fuzhou 350108, China}
\author{Pei-Rong Han}
\affiliation{School of Physics and Mechanical and Electrical Engineering, Longyan University, Longyan 364012, China}
\author{Zhen-Biao Yang} \email{zbyang@fzu.edu.cn}
\affiliation{Fujian Key Laboratory of Quantum Information and Quantum Optics, Fuzhou University, Fuzhou 350108, China}
\affiliation{Department of Physics, Fuzhou University, Fuzhou 350108, China}
\affiliation{Hefei National Laboratory, Hefei 230088, China}
\author{Shi-Biao Zheng} \email{t96034@fzu.edu.cn}
\affiliation{Fujian Key Laboratory of Quantum Information and Quantum Optics, Fuzhou University, Fuzhou 350108, China}
\affiliation{Department of Physics, Fuzhou University, Fuzhou 350108, China}
\affiliation{Hefei National Laboratory, Hefei 230088, China}
\date{\today}

\maketitle
\tableofcontents

\section{Calculation of Quantum Channels and Exceptional Points}
\subsection{Liouvillian matrix}
We consider a Kerr-nonlinear resonator subject to a two-photon drive (TPD)~\cite{Goto2016,Minganti2016,Puri2017,Puri2019,Grimm2020}.
In the frame rotating at the TPD frequency and assuming a zero-temperature Markovian thermal bath, the dynamics is governed by the Lindblad master equation ($\hbar=1$)
\begin{equation}
\dot{\rho} = \mathcal{L}[\rho]= -i[\Delta a^\dagger a -Ka^{\dagger2}a^{2} + (\varepsilon_2 a^{\dagger 2} + \varepsilon_2^* a^2),\,\rho] + \kappa a\rho a^\dagger - \frac{\kappa}{2}\{a^\dagger a, \rho\}.
\label{eq:master}
\end{equation}
To analyze the Liouvillian spectrum and locate Liouvillian exceptional points (EPs), we use the vectorized representation $\rho \mapsto \boldsymbol{V} = \mathrm{vec}(\rho)$, under which Eq. ~\eqref{eq:master} becomes the linear ordinary differential equation $\dot{\boldsymbol{V}}(t) = \mathcal{L}_{\mathrm{M}} \boldsymbol{V}(t)$.
The matrix representation of $\mathcal{L}$ is
\begin{align}
\mathcal{L}_{\mathrm{M}} = -i\left(H \otimes \mathbb{I} - \mathbb{I} \otimes H^{\mathsf{T}}\right) + \Big[\Gamma_{a} \otimes \Gamma_{a}^{*}
- \frac{1}{2}\,\Gamma_{a}^\dagger \Gamma_{a} \otimes \mathbb{I}
	- \frac{1}{2}\,\mathbb{I} \otimes (\Gamma_{a}^{\mathsf{T}} \Gamma_{a}^{*})\Big],
\end{align}
with $\Gamma_a = \sqrt{\kappa}\,a$.

\begin{figure*}
\includegraphics[width =0.7\textwidth]{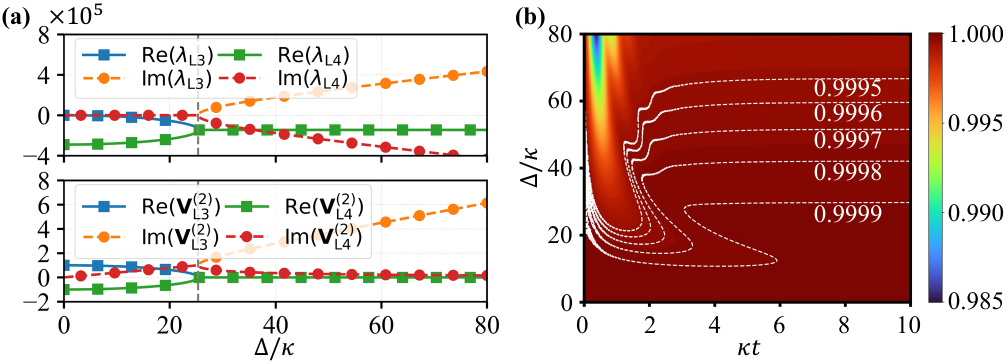}
\caption{\label{FigS1}
(a) Spectrum of $\mathcal{L}_{\text{M}}$. Top: third and fourth eigenvalues of $\mathcal{L}_{\text{M}}$ at $\kappa t=1$.
Bottom: second components $\{\mathbf{V}^{(2)}_{\text{L}3}, \mathbf{V}^{(2)}_{\text{L}4}\}$ of the eigenvectors of $\mathcal{L}_{\text{M}}$.
At the Liouvillian EP, both eigenvalues and eigenvectors coalesce (gray dashed line).
(b) Validation of the projected Liouvillian dynamics.
Fidelity $F(\rho_{\mathrm{H}}, \rho_{\mathrm{L}})$ between the full Lindblad evolution $\rho_{\mathrm{H}}$ and the projected Liouvillian dynamics $\rho_{\mathrm{L}}$ versus $\kappa t$ and detuning $\Delta / \kappa$, for initial state $|\alpha\rangle$. Parameters: $K/\kappa=0.65\times10^3$ and $\varepsilon_{2}/\kappa=1.50\times10^3$.
}
\end{figure*}

Projecting $\mathcal{L}$ onto the logical subspace $\{ |\mathcal{C}^+_\alpha\rangle, |\mathcal{C}^-_\alpha\rangle \}$ and using the identities
\begin{align}
	a |\mathcal{C}^\pm_\alpha\rangle = \alpha p^{\pm 1} |\mathcal{C}^\mp_\alpha\rangle,\, a^\dagger |\mathcal{C}^\pm_\alpha\rangle = \alpha^{*} p^{\mp 1} |\mathcal{C}^\mp_\alpha\rangle,
\end{align}
the projected Liouvillian reduces to a $4\times4$ matrix
\begin{align}
\mathcal{L}_{\mathrm{M}} =
\begin{pmatrix}
-\kappa\alpha^{2}p^{2} & 
0 & 
0 & 
\kappa\alpha^{2}p^{-2} \\
0 & 
i\Delta\alpha^{2}p_{2}^{-} - \dfrac{\kappa}{2}\alpha^{2}p_{2}^{+} & 
\kappa\alpha^{2} & 
0 \\
0 & 
\kappa\alpha^{2} & 
-i\Delta\alpha^{2}p_{2}^{-} - \dfrac{\kappa}{2}\alpha^{2}p_{2}^{+} & 
0 \\
\kappa\alpha^{2}p^{2} & 
0 & 
0 & 
-\kappa\alpha^{2}p^{-2}
\end{pmatrix}.
\end{align}
Assuming $\mathcal{L}_{\mathrm{M}}$ is diagonalizable, the time evolution of $\boldsymbol{V}(t)$ can be expressed as
\begin{equation}
	\mathbf{V}(t) = \sum_{k=1}^4 c_k e^{\lambda_{\text{L}k} t} \mathbf{V}_{\text{L}k},
	\label{eq:V(t)}
\end{equation}
where $\lambda_{\text{L}k}$ are the eigenvalues of $\mathcal{L}_{\mathrm{M}}$, $\mathbf{V}_{\text{L}k}$ are the corresponding eigenvectors, and $c_k$ are expansion coefficients fixed by the initial condition.
The eigenvalue equation $\mathcal{L}_{\mathrm{M}}\mathbf{V}_{\text{L}k}=\lambda_{\text{L}k}\mathbf{V}_{\text{L}k}$ yields 
\begin{align}
\lambda_{\text{L}1} &= 0,\\
\lambda_{\text{L}2} &= -\alpha^2\kappa p^{+}_{2},\\
\lambda_{\text{L}3} &= -\alpha^2(\kappa p^{+}_2/2 - \sqrt{\Lambda}),\\
\lambda_{\text{L}4} &= -\alpha^2(\kappa p^{+}_2/2 + \sqrt{\Lambda}),
\end{align}
with the corresponding eigenvectors
\begin{align}
\mathbf{V}_{\text{L1}}&=\{1/p^4,\,0,\,0,\,1\}^{\mathsf{T}},\\
\mathbf{V}_{\text{L2}}&=\{-1,\,0,\,0,\,1\}^{\mathsf{T}},\\
\mathbf{V}_{\text{L3}}&=\{0,\,\mathbf{V}^{(2)}_{\text{L}3},\,1,\,0\}^{\mathsf{T}},\\
\mathbf{V}_{\text{L4}}&=\{0,\,\mathbf{V}^{(2)}_{\text{L}4},\,1,\,0\}^{\mathsf{T}},
\end{align}where 
\begin{align}
\mathbf{V}^{(2)}_{\text{L}3}=(i\Delta p^{-}_{2}+\sqrt{\Lambda})/\kappa, \mathbf{V}^{(2)}_{\text{L}4}=(i\Delta p^{-}_{2}-\sqrt{\Lambda})/\kappa,  
\end{align}
with $p^{\pm}_j = p^{-j} \pm p^{j}$ and $\Lambda=\kappa^2-(\Delta p^{-}_{2})^2$.
A Liouvillian EP occurs when $\lambda_{\text{L3}}=\lambda_{\text{L4}}$ and $\mathbf{V}^{(2)}_{\text{L3}}=\mathbf{V}^{(2)}_{\text{L4}}$, as shown in Fig.~\ref{FigS1}(a), where the gray dashed line marks the Liouvillian EP condition $\Delta/\kappa=1/p^{-}_{2}$~\cite{Han2026}.
The spectrum exhibits two phases: it is fully real for $|\Delta|/\kappa < 1/p^{-}_{2}$, $\{0,\,\lambda_{\text{L}2},\,\lambda_{\text{L}3},\,\lambda_{\text{L}4}\}\in\mathbb{R}$, and contains one complex-conjugate pair for $|\Delta|/\kappa > 1/p^{-}_{2}$ i.e., $\{0,\,\lambda_{\text{L}2}\}\in\mathbb{R}$, $\{\lambda_{\text{L}3},\,\lambda_{\text{L}4}=\lambda^{*}_{\text{L}3}\}\in\mathbb{C}$.

To assess how well the system dynamics remains confined to the Kerr-cat-qubit subspace, we numerically simulate the full master equation~\eqref{eq:master}, yielding the exact density matrix $\rho_{\mathrm{H}}$, and compare it with the reduced dynamics obtained from the projected Liouvillian in Eq.~\eqref{eq:V(t)}, denoted by $\rho_{\mathrm{L}}$.
The validity of the approximation is quantified by the fidelity~\cite{Han2026} 
\begin{align}
F(\rho_{\text{H}},\,\rho_{\text{L}})=(\text{Tr}\sqrt{ \sqrt{\rho_{\text{H}}} \rho_{\text{L}} \sqrt{\rho_{\text{H}}} })^{2}.
\end{align}
Figure~\ref{FigS1}(b) shows $F(\rho_{\text{H}},\,\rho_{\text{L}})$ as a function of $\kappa t$ and $\Delta/\kappa$. The initial state is chosen as $|\alpha\rangle=|\sqrt{\varepsilon_{2}/K}\rangle$.
Across the parameter range, the projected dynamics closely tracks the full dynamics, with fidelity remaining above 0.985 for the parameters used in this work.

\subsection{Markovian quantum channel}
The quantum channel $\mathcal{E}=e^{\mathcal{L}_{\mathrm{M}}t}$ is
\begin{align}
\mathcal{E} =
\begin{pmatrix}
\frac{1+e^{-p^{+}_{2}\alpha^{2}\kappa t}p^{4}}{1+p^{4}} & 
0 & 
0 & 
\frac{1-e^{-p^{+}_{2}\alpha^{2}\kappa t}}{1+p^{4}} \\
0 & 
\mathcal{E}_{22} & 
\frac{ e^{-\frac{p^{+}_{2}\alpha^{2}\kappa t}{2}} \kappa
\sinh{(\frac{\Lambda_{t}}{2})}}{\sqrt{\Lambda}}&
0 \\
0 & 
\frac{ e^{-\frac{p^{+}_{2}\alpha^{2}\kappa t}{2}} \kappa
\sinh{(\frac{\Lambda_{t}}{2})}}{\sqrt{\Lambda}} & 
\mathcal{E}_{33} & 
0 \\
\frac{ (1-e^{-p^{+}_{2}\alpha^{2}\kappa t})p^{4} }{ 1+p^{4} } & 
0 & 
0 & 
\frac{ e^{-p^{+}_{2}\alpha^{2}\kappa t} + p^{4} }{ 1+p^{4} }
\end{pmatrix},
\label{eq:channel}
\end{align}
with
\begin{align}
\mathcal{E}_{22}&=\frac{e^{\frac{-(p^{+}_{2}\alpha^{2}\kappa t+\Lambda_{t})}{2}}[-i(e^{\Lambda_{t}}-1)(p^{4}-1)\Delta
+(e^{\Lambda_{t}}+1)p^{2}\sqrt{\Lambda}]}{2p^{2}\sqrt{\Lambda}},\\
\mathcal{E}_{33}&=\mathcal{E}^{*}_{22},
\end{align}
where $\Lambda_{t}=2\alpha^{2}t\sqrt{\Lambda}$.
The eigenvalues and corresponding eigenvectors are:
\begin{align}
\lambda_{\text{C}1} &= e^{\lambda_{\text{L}1}t}=1,\\
\lambda_{\text{C}2} &= e^{\lambda_{\text{L}2}t}=e^{-p^{+}_{2}\alpha^2\kappa t},\\
\lambda_{\text{C}3} &= e^{\lambda_{\text{L}3}t}=e^{-(p^{+}_2\alpha^2\kappa t - \Lambda_{t}) /2},\\
\lambda_{\text{C}4} &= e^{\lambda_{\text{L}4}t}=e^{-(p^{+}_2\alpha^2\kappa t + \Lambda_{t}) /2},
\end{align}
with eigenvectors
\begin{align}
\mathbf{V}_{\text{C}1}&=\{1/p^4,\,0,\,0,\,1\}^{\mathsf{T}},\\
\mathbf{V}_{\text{C}2}&=\{-1,\,0,\,0,\,1\}^{\mathsf{T}},\\
\mathbf{V}_{\text{C}3}&=\{0,\,\mathbf{V}^{(2)}_{\text{L}3},\,1,\,0\}^{\mathsf{T}},\\
\mathbf{V}_{\text{C}4}&=\{0,\,\mathbf{V}^{(2)}_{\text{L}4},\,1,\,0\}^{\mathsf{T}}.
\end{align}
Because $\mathcal{E}$ and $\mathcal{L}$ share the same eigenvectors, the channel exceptional point (CEP) of $\mathcal{E}$ inherits the time-independent Liouvillian EP structure of $\mathcal{L}_M$.

\subsection{Interpolated quantum channel}
The matrix form of the interpolated quantum channel 
\begin{align}
\mathcal{E}(m)=(1-m)\mathcal{E}_{1}+m\mathcal{E}_{2}
\end{align}
is
\begin{align}
\mathcal{E}(m) = 
\begin{pmatrix}
\frac{1+e^{-p^{+}_{2}\alpha^{2}\kappa t}p^{4}}{1+p^{4}} & 
0 & 
0 & 
\frac{1-e^{-p^{+}_{2}\alpha^{2}\kappa t}}{1+p^{4}} \\
0 & 
(1-m)E_{1}+mE_{2} & 
[\mathcal{E}(m)]_{23} & 
0 \\
0 & 
[\mathcal{E}(m)]_{32} & 
(1-m)E^{*}_{1}+mE^{*}_{2} & 
0 \\
\frac{ (1-e^{-p^{+}_{2}\alpha^{2}\kappa t})p^{4} }{ 1+p^{4} } & 
0 & 
0 & 
\frac{ e^{-p^{+}_{2}\alpha^{2}\kappa t} + p^{4} }{ 1+p^{4} }
\end{pmatrix},
\label{eq:Ichannel}
\end{align}
with
\begin{align}
E_{j}&=e^{\frac{-\alpha^{2}t(p^{+}_{2}\kappa +2\sqrt{\Lambda_{j}})}{2}}\frac{-i(e^{2\alpha^{2}t\sqrt{\Lambda_{j}}}-1)(p^{4}-1)\Delta_{j}
+(e^{2\alpha^{2}t\sqrt{\Lambda_{j}}}+1)p^{2}\sqrt{\Lambda_{j}}}{2p^{2}\sqrt{\Lambda_{j}}},\\
[\mathcal{E}(m)]_{23}&=[\mathcal{E}(m)]_{32}=e^{-\frac{p^{+}_{2}\alpha^{2}\kappa t}{2}} \kappa\bigg[(1-m)\frac{ 
\sinh{(\alpha^{2}t\sqrt{\Lambda_{1}})}}{\sqrt{\Lambda_{1}}} + m\frac{ 
\sinh{(\alpha^{2}t\sqrt{\Lambda_{2}})}}{\sqrt{\Lambda_{2}}}\bigg],\\
\Lambda_{j}&=\kappa^2-(\Delta_{j} p^{-}_{2})^2,\,(j=1,\,2).
\end{align}
The eigenvalues and corresponding eigenvectors are
\begin{align}
\lambda_{\text{IC}1} =& 1,\\
\lambda_{\text{IC}2} =& e^{-p^{+}_{2}\alpha^2\kappa t},\\
\lambda_{\text{IC}3} =& e^{-\frac{p^{+}_{2}\alpha^2\kappa t}{2}}\bigg[(1-m)\cosh(\alpha^{2}t\sqrt{\Lambda_{1}})+m\cosh(\alpha^{2}t\sqrt{\Lambda_{2}})\nonumber\\
&+\Bigg( \frac{
\Lambda_{1}\Lambda_{2}
[ (1-m)^{2}\cosh(2\alpha^{2}t\sqrt{\Lambda_{1}}) - 1+2(1-m)m] 
}
{ 2\Lambda_{1}\Lambda_{2} } \nonumber\\
&+\frac{
m[ m\Lambda_{1}\Lambda_{2}\cosh(2\alpha^{2}t\sqrt{\Lambda_{2}}) + 4(1-m)\Lambda_{12}\sqrt{\Lambda_{1}}\sqrt{\Lambda_{2}}\sinh(\alpha^{2}t\sqrt{\Lambda_{1}})\sinh(\alpha^{2}t\sqrt{\Lambda_{2}}) ] 
}
{ 2\Lambda_{1}\Lambda_{2} }
\Bigg)^{1/2} \Bigg],\\
\lambda_{\text{IC}4} =& e^{-\frac{p^{+}_{2}\alpha^2\kappa t}{2}}\bigg[(1-m)\cosh(\alpha^{2}t\sqrt{\Lambda_{1}})+m\cosh(\alpha^{2}t\sqrt{\Lambda_{2}})\nonumber\\
&-\Bigg( \frac{
\Lambda_{1}\Lambda_{2}
[ (1-m)^{2}\cosh(2\alpha^{2}t\sqrt{\Lambda_{1}}) - 1+2(1-m)m] 
}
{ 2\Lambda_{1}\Lambda_{2} } \nonumber\\
&+\frac{
m[ m\Lambda_{1}\Lambda_{2}\cosh(2\alpha^{2}t\sqrt{\Lambda_{2}}) + 4(1-m)\Lambda_{12}\sqrt{\Lambda_{1}}\sqrt{\Lambda_{2}}\sinh(\alpha^{2}t\sqrt{\Lambda_{1}})\sinh(\alpha^{2}t\sqrt{\Lambda_{2}}) ] 
}
{ 2\Lambda_{1}\Lambda_{2} }
\Bigg)^{1/2} \Bigg],
\end{align}
with
\begin{align}
\mathbf{V}_{\text{IC}1}&=\{1/p^4,\,0,\,0,\,1\}^{\mathsf{T}},\\
\mathbf{V}_{\text{IC}2}&=\{-1,\,0,\,0,\,1\}^{\mathsf{T}},\\
\mathbf{V}_{\text{IC}3}&=\{0,\,\mathbf{V}^{(2)}_{\text{IC}3},\,1,\,0\}^{\mathsf{T}},\\
\mathbf{V}_{\text{IC}4}&=\{0,\,\mathbf{V}^{(2)}_{\text{IC}4},\,1,\,0\}^{\mathsf{T}},
\end{align}
where
\begin{align}
\mathbf{V}^{(2)}_{\text{IC}3} = &
\Bigg[-2ip^{-}_{2}[(1-m)\Delta_{1}\sqrt{\Lambda_{2}}\sinh(\alpha^{2}t\sqrt{\Lambda_{1}})+m\Delta_{2}\sqrt{\Lambda_{1}}\sinh(\alpha^{2}t\sqrt{\Lambda_{2}})]\nonumber\\
&-\sqrt{2}\Bigg( \Lambda_{1}\Lambda_{2}
[ (1-m)^{2}\cosh(2\alpha^{2}t\sqrt{\Lambda_{1}}) - 1+2(1-m)m] 
\nonumber\\
&+
m[ m\Lambda_{1}\Lambda_{2}\cosh(2\alpha^{2}t\sqrt{\Lambda_{2}}) + 4(1-m)\Lambda_{12}\sqrt{\Lambda_{1}}\sqrt{\Lambda_{2}}\sinh(\alpha^{2}t\sqrt{\Lambda_{1}})\sinh(\alpha^{2}t\sqrt{\Lambda_{2}}) ] 
\Bigg)^{1/2} \Bigg]\nonumber\\
&\bigg/2\kappa\bigg[(m-1)\sqrt{\Lambda_{2}}\sinh(\alpha^{2}t\sqrt{\Lambda_{1}})-m\sqrt{\Lambda_{1}}\sinh(\alpha^{2}t\sqrt{\Lambda_{2}})\bigg],
\end{align}

\begin{figure*}
\includegraphics[width =\textwidth]{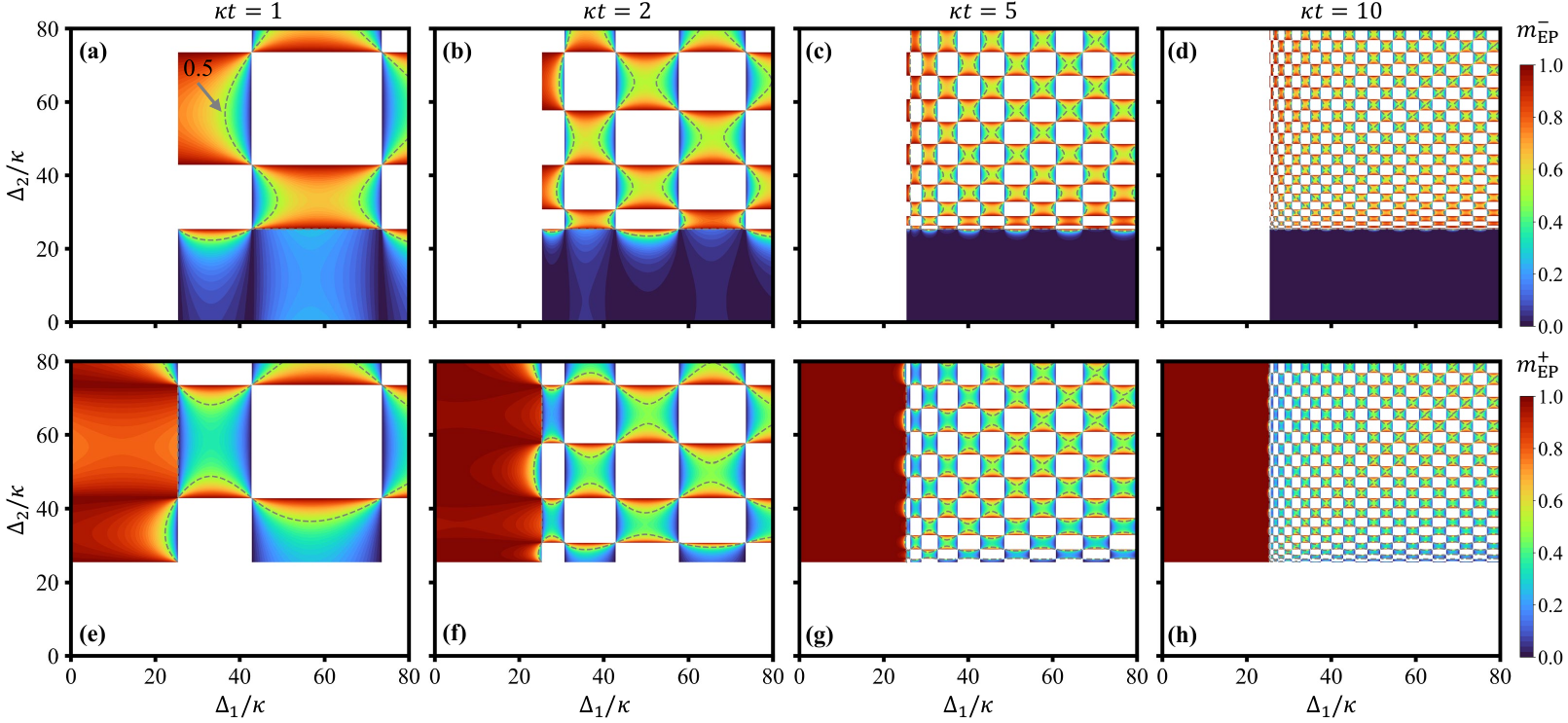}
\caption{\label{FigS2} Second-order ICEP solutions $m^{-}_{\text{EP}}$ [(a)--(d)] and $m^{+}_{\text{EP}}$ [(e)--(h)] versus $\Delta_1/\kappa$ and $\Delta_2/\kappa$ at different evolution times. White regions denote parameter domains where no valid ICEP exists for $m \in [0,1]$.
Parameters: $K/\kappa=0.65\times10^3$ and $\varepsilon_{2}/\kappa=1.50\times10^3$.
}
\end{figure*}

\begin{align}
\mathbf{V}^{(2)}_{\text{IC}4} = &
\Bigg[-2ip^{-}_{2}[(1-m)\Delta_{1}\sqrt{\Lambda_{2}}\sinh(\alpha^{2}t\sqrt{\Lambda_{1}})+m\Delta_{2}\sqrt{\Lambda_{1}}\sinh(\alpha^{2}t\sqrt{\Lambda_{2}})]\nonumber\\
&+\sqrt{2}\Bigg( \Lambda_{1}\Lambda_{2}
[ (1-m)^{2}\cosh(2\alpha^{2}t\sqrt{\Lambda_{1}}) - 1+2(1-m)m] 
\nonumber\\
&+
m[ m\Lambda_{1}\Lambda_{2}\cosh(2\alpha^{2}t\sqrt{\Lambda_{2}}) + 4(1-m)\Lambda_{12}\sqrt{\Lambda_{1}}\sqrt{\Lambda_{2}}\sinh(\alpha^{2}t\sqrt{\Lambda_{1}})\sinh(\alpha^{2}t\sqrt{\Lambda_{2}}) ] 
\Bigg)^{1/2} \Bigg]\nonumber\\
&\bigg/2\kappa\bigg[(m-1)\sqrt{\Lambda_{2}}\sinh(\alpha^{2}t\sqrt{\Lambda_{1}})-m\sqrt{\Lambda_{1}}\sinh(\alpha^{2}t\sqrt{\Lambda_{2}})\bigg].
\end{align}

Solving $\lambda_{\text{IC}3}=\lambda_{\text{IC}4}$, or equivalently $\mathbf{V}^{(2)}_{\text{IC}3}=\mathbf{V}^{(2)}_{\text{IC}4}$, gives the ICEP conditions
\begin{align}
m^{\pm}_{\text{EP}} =&\frac{\Lambda_1\Lambda_2 \mathcal{C}_1 -2\sqrt{\Lambda_1}\sqrt{\Lambda_2}\Lambda_{12}\mathcal{S}_{12} \pm\sqrt{4\Lambda_1\Lambda_2\Lambda^2_{12}\mathcal{S}^2_{12}-\Lambda^2_1\Lambda^2_2\mathcal{C}_1\mathcal{C}_2}}
{\Lambda_1\Lambda_2(\mathcal{C}_1+\mathcal{C}_2)-4\sqrt{\Lambda_1}\sqrt{\Lambda_2}\Lambda_{12}\mathcal{S}_{12}},
\end{align}
where 
\begin{align}
\Lambda_j=\kappa^2-(\Delta_{j}p^{-}_{2})^2, \Lambda_{12}=\kappa^2-\Delta_1\Delta_2 (p^{-}_{2})^2, \mathcal{C}_{j}=\cosh(2x_{j})-1, \mathcal{S}_{12}=\prod_{j=1}^{2} \sinh(x_{j}),
\end{align}
with $x_{j}=t\alpha^2\sqrt{\Lambda_j}$.
At an ICEP, the coalesced eigenvalue is
\begin{equation}
\lambda_{\mathrm{EP}}
=e^{-\frac{p^{+}_{2}}{2}\alpha^{2}\kappa t}
\Big[(1-m^{\pm}_{\mathrm{EP}})\cosh x_{1}+m^{\pm}_{\mathrm{EP}}\cosh x_{2}\Big].
\label{eq:lamEP-sm}
\end{equation}

Any solution satisfying $0<m^{\pm}_{\text{EP}}<1$ signals the emergence of an ICEP.
Conversely, the boundary solutions $m^{-}_{\text{EP}}=1$ ($m^{+}_{\text{EP}}=0$) or $m^{+}_{\text{EP}}=1$ ($m^{-}_{\text{EP}}=0$) correspond to recovering the original Markovian channel $\mathcal{E}_1(\Delta_1)$ or $\mathcal{E}_2(\Delta_2)$.

Figure~\ref{FigS2} shows $m^{-}_{\text{EP}}$ and $m^{+}_{\text{EP}}$ as functions of $\Delta_1/\kappa$ and $\Delta_2/\kappa$ at various times $\kappa t$.
White regions denote the absence of valid ICEPs.
At each time, the lower panel is the mirror image of the upper panel about $\Delta_2=\Delta_1$: the $m^{+}_{\mathrm{EP}}$ landscape equals the $m^{-}_{\mathrm{EP}}$ landscape under exchange of $\Delta_1$ and $\Delta_2$ between rows (not within a panel).
Two regimes emerge in parameter space.

(i) \textit{Single-EP phase regime}: A single ICEP appears when $\Delta_1$ and $\Delta_2$ lie on opposite sides of the Liouvillian EP (or equivalently the CEP) boundary.
This follows the conventional phase-boundary picture: when the two endpoint channels are drawn from distinct spectral phases---one with a fully real spectrum and the other with a complex-conjugate pair---their interpolation yields a single EP accompanying the phase boundary, as generally expected in Ref.~\cite{Wong2026}.
Specifically,  $m^{-}_{\mathrm{EP}}$ is valid for $\Delta_2/\kappa<1/p^{-}_2<\Delta_1/\kappa$, whereas $m^{+}_{\mathrm{EP}}$ applies to $\Delta_1/\kappa<1/p^{-}_2<\Delta_2/\kappa$.
At short times (e.g., $\kappa t=1$), the physical solution mainly lies within $(0,1)$, with $m^{-}_{\mathrm{EP}}$ predominantly below $0.5$ and $m^{+}_{\mathrm{EP}}$ above $0.5$, confirming that the ICEPs lie in the interior of the interpolation.
As time increases, the valid parameter rectangles remain fixed, but $m^{-}_{\text{EP}}\rightarrow0$ and $m^{+}_{\text{EP}}\rightarrow1$.
The interpolation effect thus vanishes asymptotically, and the system recovers Markovian dynamics asymptotically.

(ii) \textit{Double-EP phase regime}: Dual ICEPs appear when both detunings exceed the Liouvillian EP (CEP) boundary, i.e., $(\Delta_1,\,\Delta_2)/\kappa>1/p^{-}_2$.
No phase boundary separates the endpoints, so the emergence of not one but two ICEPs within a single phase goes beyond the conventional phase-boundary picture.
At short time (e.g., $\kappa t=1$), dual ICEPs populate four rectangular regions within $(0,1)$.
With increasing time, the valid ranges of $m^{\pm}_{\text{EP}}$ persist, but the distribution fragments into smaller subregions.
This fragmentation renders the ICEPs increasingly sensitive to variations in $\Delta_1/\kappa$, $\Delta_2/\kappa$, and $\kappa t$.

\begin{figure*}
\includegraphics[width =\textwidth]{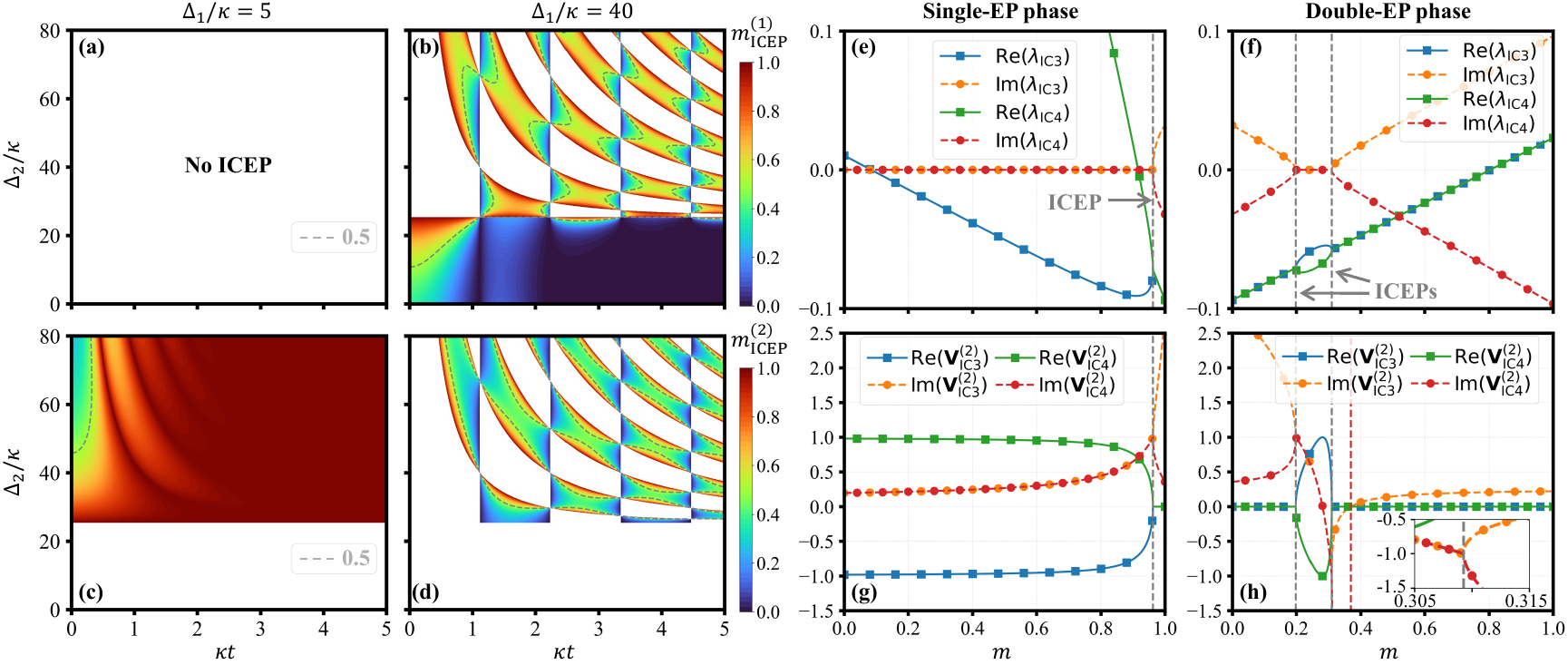}
\caption{\label{FigS3} (a)--(d) Temporal persistence of the ICEPs. ICEP locations $m^{-}_{\mathrm{EP}}$ [(a), (b)] and $m^{+}_{\mathrm{EP}}$ [(c), (d)] versus $\kappa t$ and $\Delta_2/\kappa$ at fixed $\Delta_1/\kappa=5$ [(a), (c)] and $\Delta_1/\kappa=35$ [(b), (d)].
Color encodes $m$; white denotes the absence of a valid ICEP with $m\in[0,1]$. Within each colored domain the ICEP persists and drifts continuously with time at fixed $(\Delta_1/\kappa,\Delta_2/\kappa)$. Coincident colored regions above the boundary in (b) and (d) signals dual ICEPs; part of this region loses its ICEP $\kappa t$ increases.
(e)--(h)
Spectrum of the interpolated quantum channel versus $m$ in different phase regimes: the single-EP phase regime with $(\Delta_1/\kappa,\,\Delta_2/\kappa)=(5,\,40)$ in (e) and (g), and the double-EP phase regime with  $(\Delta_1/\kappa,\,\Delta_2/\kappa)=(40,\,60)$ is shown in (f) and (h).
In all panels, $K/\kappa=0.65\times10^3$ and $\varepsilon_{2}/\kappa=1.50\times10^3$.
}
\end{figure*}

To examine how the ICEPs evolve in time, we show $m^{\pm}_{\mathrm{EP}}$ as functions of $\kappa t$ and $\Delta_2/\kappa$ at fixed $\Delta_1/\kappa$; the color scale encodes $m^{\pm}_{\mathrm{EP}}\in(0,1)$, and white regions denote the absence of a valid ICEP. 
Figures~\ref{FigS3}(a) and \ref{FigS3}(c) show the slice $\Delta_1/\kappa=5<1/p^{-}_{2}$ (single-EP phase throughout).
Only $m^{+}_{\mathrm{EP}}$ is physical: it exists for every $\Delta_2/\kappa$ above the Liouvillian EP boundary $\Delta/\kappa=1/p^{-}_{2}$ and at every evolution time, i.e., for each fixed $(\Delta_1,\Delta_2)$ the ICEP location $m^{\pm}_{\mathrm{EP}}$ persists and drifts continuously along the whole time axis.
Figures~\ref{FigS3}(b) and \ref{FigS3}(d) show the slice $\Delta_1/\kappa=35>1/p^{-}_{2}$, which spans both regimes.
Below the boundary (single-EP phase part), only $m^{-}_{\mathrm{EP}}$ exists and persists in time; above the boundary (double-EP phase part), the valid regions of $m^{-}_{\mathrm{EP}}$ and $m^{+}_{\mathrm{EP}}$ coincide, so the interpolated channel carries two distinct ICEPs (dual ICEPs) at the same parameters. These ICEPs persist throughout the shown time window, while part of the region loses its ICEP with increasing $\kappa t$.
The persistence in time seen in all colored domains is a structural consequence of defining the EP at the channel level: the ICEP conditions involve the evolution time $t$ only through the endpoint channels $\mathcal{E}_{j}$, so wherever a solution $m^{\pm}_{\mathrm{EP}}$ exists at one instant, it persists, drifting continuously, over finite time intervals.

To verify the ICEPs in both regimes, Fig.~\ref{FigS3} shows the channel spectrum at $\kappa t=1$.
In the single-EP phase regime, $(\Delta_1/\kappa,\,\Delta_2/\kappa)=(5,\,40)$ [Figs.~\ref{FigS3}(e) and \ref{FigS3}(g)], both eigenvalues ($\lambda_{\text{IC}3},\,\lambda_{\text{IC}4}$) and eigenvectors ($\mathbf{V}_{{\text{IC}}3},\,\mathbf{V}_{{\text{IC}}4}$) coalesce at a single ICEP, $m \approx 0.9233$.
In the double-EP phase regime, $(\Delta_1/\kappa,\,\Delta_2/\kappa)=(40,\,60)$ [Figs.~\ref{FigS3}(f) and \ref{FigS3}(h)], dual ICEPs occur at $m \approx 0.3093$ and $m \approx 0.1979$, where both eigenvalues and eigenvectors coalesce simultaneously.
These numerical results agree fully with the analytical predictions in Fig.~\ref{FigS2}.

\section{Proof of the Markovian and non-Markovian criteria}
\label{sm:criterion}

Following Ref.~\cite{Wolf2008}, a snapshot channel is Markovian if it is belong to a time-independent Markovian semigroup, $\mathcal{E}=e^{\mathcal{L}t}$. This holds if and only if some branch of $\log\mathcal{E}$ is Hermiticity preserving (HP), trace annihilating, and conditionally completely positive (cCP). The spectral limitation established in Ref.~\cite{Wolf2008}---that is a negative real eigenvalue admits no HP logarithm---was derived for a nondefective, nondegenerate spectrum, in which every negative eigenvalue is unpaired~\cite{Wolf2008}.
Davalos et al.~\cite{Davalos2019} recast the existence of a real
logarithm as $\mathcal{L}$-divisibility and lifted this limitation for diagonalizable qubit channels: a negative eigenvalue $\lambda_{\mathrm{DP}}<0$ of even multiplicity, carried by two $1\times1$ Jordan blocks $\mathrm{diag}(\lambda_{\mathrm{DP}},\lambda_{\mathrm{DP}})$---a
diabolical point (DP) where the degenerate eigenvectors do not coalesce, unlike at an EP---admits real logarithms in which the pair coalesces into a rotation by an odd multiple of $\pi$ (Theorm~3 therein); the defective case was referred to Ref.~\cite{Culver1966}.

An EP lies precisely in this gap: eigenvector coalescence forces the merged eigenvalue into a single Jordan block, so neither the unpaired shortcut of Ref.~\cite{Wolf2008} nor the rotation escape of Ref.~\cite{Davalos2019} applies.
Here we settle this case by writing down all possible logarithms of the snapshot and keeping only those whose spectrum is conjugate-symmetric, as required for Hermiticity preservation (Step~1).
The fate of the candidates is decided by the sign of the coalesced eigenvalue $\lambda_{\mathrm{EP}}$, and only two possibilities can arise.
If no candidate survives--which occurs for $\lambda_{\mathrm{EP}}<0$--then no HP logarithm exists and the snapshot is NMC: this is the spectral criterion derived in Case~(iii) below.
If exactly one candidate survives--which happens for $\lambda_{\mathrm{EP}}>0$--it is the principal logarithm $\mathcal{L}_{0}=\log\mathcal{E}_{\mathrm{C}}$; the remaining question is whether it generates a valid evolution, i.e., whether it is cCP~\cite{Wolf2008}.
This is the second certification route, developed in Case~(i) below. The singular case $\lambda_{\mathrm{EP}}=0$, where no logarithm exists at all, is treated as Case~(ii).

Hereafter, $\mathcal{E}_{\mathrm{C}}$ denotes a generic snapshot channel
carrying an EP---either the Markovian channel $\mathcal{E}$ or an
interpolated snapshot $\mathcal{E}(m)$---and $\lambda_{\mathrm{EP}}$ its
coalesced eigenvalue, i.e.,
$\lambda_{\mathrm{EP}}=\lambda_{\mathrm{C}3}=\lambda_{\mathrm{C}4}$ for $\mathcal{E}_{\mathrm{C}}=\mathcal{E}$ and $\lambda_{\mathrm{EP}}=\lambda_{\mathrm{IC}3}=\lambda_{\mathrm{IC}4}$ for $\mathcal{E}_{\mathrm{C}}=\mathcal{E}(m)$; the argument below is identical for both.

\textit{Step 1: conjugate symmetry.} A GKSL Liouvillian preserves Hermiticity, $\mathcal{L}(\mathcal{X}^\dagger)=\mathcal{L}(\mathcal{X})^\dagger$, so $\mathcal{L}(\mathbf{V})=\nu\mathbf{V}$ implies
$\mathcal{L}(\mathbf{V}^\dagger)=\nu^{*}\mathbf{V}^\dagger$~\cite{Wolf2008,Minganti2018}:
the spectrum of $\mathcal{L}$ is conjugate-symmetric,
$\{\nu_k\}=\{\nu_k^*\}$, including multiplicities~\cite{Davalos2019}.
This holds for any HP map and gives a necessary condition used throughout: a candidate logarithm whose spectrum is not conjugate  symmetric cannot be HP.

\textit{Step 2: one branch on the merged eigenmodes.}
Suppose $\mathcal{E}_{\mathrm{C}}=e^{\mathcal{L}t}$; then $[\mathcal{L},\mathcal{E}_{\mathrm{C}}]=0$.
Unlike the singular case $\lambda_{\mathrm{EP}}=0$ of Ref.~\cite{Wong2026}, the channels considered here are nonsingular ($\lambda_{\mathrm{EP}}\neq0$), so logarithms of $\mathcal{E}_{\mathrm{C}}$ exist; the question is whether any of them is HP.
At an EP, eigenvector coalescence gives a Jordan chain~\cite{Wong2026,Brandao2026}
\begin{equation}
\mathcal{E}_{\mathrm{C}}\mathbf{V}_{\mathrm{C}}
=\lambda_{\mathrm{EP}}\mathbf{V}_{\mathrm{C}},
\qquad
(\mathcal{E}_{\mathrm{C}}-\lambda_{\mathrm{EP}}I)\mathbf{W}_{\mathrm{C}}
=\mathbf{V}_{\mathrm{C}},
\label{eq:chain}
\end{equation}
with $\mathbf{W}_{\mathrm{C}}$ unique up to adding multiples of $\mathbf{V}_{\mathrm{C}}$.
The two-dimensional space $\mathcal{G}=\mathrm{span}\{\mathbf{V}_{\mathrm{C}},\mathbf{W}_{\mathrm{C}}\} =\{\mathbf{x}:(\mathcal{E}_{\mathrm{C}}-\lambda_{\mathrm{EP}}I)^2\mathbf{x}=0\}$ is $\mathcal{L}$-invariant, i.e.,
$\mathcal{L}(\mathcal{G})=\{\mathcal{L}\mathbf{x}:\mathbf{x}\in\mathcal{G}\}\subseteq\mathcal{G}$: for $\mathbf{x}=c_1\mathbf{V}_{\mathrm{C}}+c_2\mathbf{W}_{\mathrm{C}}$, $(\mathcal{E}_{\mathrm{C}}-\lambda_{\mathrm{EP}}I)^2(\mathcal{L}\mathbf{x}) =\mathcal{L}(\mathcal{E}_{\mathrm{C}}-\lambda_{\mathrm{EP}}I)^2\mathbf{x}=0$, because $\mathcal{L}$ commutes with $(\mathcal{E}_{\mathrm{C}}-\lambda_{\mathrm{EP}}I)^2$.
Since $\mathcal{E}_{\mathrm{C}}$ has only one linearly independent eigenvector at $\lambda_{\mathrm{EP}}$, namely $\mathbf{V}_{\mathrm{C}}$, commutation gives $\mathcal{E}_{\mathrm{C}}(\mathcal{L}\mathbf{V}_{\mathrm{C}})
=\mathcal{L}(\mathcal{E}_{\mathrm{C}}\mathbf{V}_{\mathrm{C}})
=\lambda_{\mathrm{EP}}(\mathcal{L}\mathbf{V}_{\mathrm{C}})$:
$\mathcal{L}\mathbf{V}_{\mathrm{C}}$ is an eigenvector of $\mathcal{E}_{\mathrm{C}}$ with eigenvalue $\lambda_{\mathrm{EP}}$, and all such eigenvectors are multiples of $\mathbf{V}_{\mathrm{C}}$.
Hence $\mathcal{L}$ has an eigenvalue---one of the $\nu_k$, call it $\nu$---with eigenvector $\mathbf{V}_{\mathrm{C}}$,
\begin{equation}
\mathcal{L}\mathbf{V}_{\mathrm{C}}=\nu\mathbf{V}_{\mathrm{C}}.
\label{eq:LV}
\end{equation}
Applying $e^{\mathcal{L}t}=\mathcal{E}_{\mathrm{C}}$ to $\mathbf{V}_{\mathrm{C}}$ gives $e^{\nu t}\mathbf{V}_{\mathrm{C}}=e^{\mathcal{L}t}\mathbf{V}_{\mathrm{C}}
=\mathcal{E}_{\mathrm{C}}\mathbf{V}_{\mathrm{C}}
=\lambda_{\mathrm{EP}}\mathbf{V}_{\mathrm{C}}$, i.e.\
$e^{\nu t}=\lambda_{\mathrm{EP}}$: $\nu$ is a branch of the logarithm of $\lambda_{\mathrm{EP}}$, henceforth denoted $\nu_{\mathrm{EP}}$.
To see that the same $\nu_{\text{EP}}$ governs $\mathbf{W}_{\mathrm{C}}$, write $\mathcal{L}\mathbf{W}_{\mathrm{C}}=a\mathbf{V}_{\mathrm{C}}+b\mathbf{W}_{\mathrm{C}}$ (the $\mathcal{L}$-invariance of $\mathcal{G}$ guarantees this form) and apply $\mathcal{E}_{\mathrm{C}}-\lambda_{\mathrm{EP}}I$ on both sides:
\begin{align}
(\mathcal{E}_{\mathrm{C}}-\lambda_{\mathrm{EP}}I) (\mathcal{L}\mathbf{W}_{\mathrm{C}}) &=\mathcal{L}\big((\mathcal{E}_{\mathrm{C}}-\lambda_{\mathrm{EP}}I) \mathbf{W}_{\mathrm{C}}\big) =\mathcal{L}\mathbf{V}_{\mathrm{C}} =\nu_{\text{EP}}\mathbf{V}_{\mathrm{C}},\nonumber\\
(\mathcal{E}_{\mathrm{C}}-\lambda_{\mathrm{EP}}I)
\big(a\mathbf{V}_{\mathrm{C}}
+b\mathbf{W}_{\mathrm{C}}\big)
&=b\,\mathbf{V}_{\mathrm{C}},
\label{eq:lock}
\end{align}
where the first line uses $[\mathcal{L},\mathcal{E}_{\mathrm{C}}]=0$ and Eq.~\eqref{eq:chain}, and the second uses $(\mathcal{E}_{\mathrm{C}}-\lambda_{\mathrm{EP}}I)\mathbf{V}_{\mathrm{C}}=0$.
Equality of the two lines forces $b=\nu_{\text{EP}}$, i.e.,
\begin{equation}
\mathcal{L}\mathbf{W}_{\mathrm{C}}
=a\mathbf{V}_{\mathrm{C}}+\nu_{\text{EP}}\mathbf{W}_{\mathrm{C}},
\end{equation}
so that, in the Jordan basis $\{\mathbf{V}_{\mathrm{C}},\mathbf{W}_{\mathrm{C}}\}$, the restriction of $\mathcal{L}$ to $\mathcal{G}$ is the $2\times2$ matrix
\begin{equation}
\mathcal{L}|_{\mathcal{G}}=
\begin{pmatrix}
\nu_{\text{EP}} & a\\
0 & \nu_{\text{EP}}
\end{pmatrix},
\qquad
e^{\nu_{\text{EP}} t}=\lambda_{\mathrm{EP}}.
\label{eq:Lblock}
\end{equation}
Writing $\lambda_{\mathrm{EP}}=|\lambda_{\mathrm{EP}}|e^{i\vartheta}$, the eigenvalue $\nu_{\text{EP}}$ is a branch of the complex logarithm of $\lambda_{\mathrm{EP}}$,
\begin{equation}
\nu_{\text{EP}}=t^{-1}\big[\ln|\lambda_{\mathrm{EP}}|+i\vartheta+2\pi i k\big],
\qquad
\vartheta=
\begin{cases}
\pi, & \lambda_{\mathrm{EP}}<0,\\
0, & \lambda_{\mathrm{EP}}>0,
\end{cases}
\qquad k\in\mathbb{Z},
\label{eq:branch}
\end{equation}
where the integer $k$ labels the countably many logarithm branches.
Both diagonal entries of $\mathcal{L}|_{\mathcal{G}}$ in Eq.~\eqref{eq:Lblock} share the same $(\vartheta,k)$: the two merged eigenmodes cannot host different logarithm branches.
Thus every candidate generator is forced to use a single logarithm branch on the merged pair; the entire freedom is one integer $k$.

\textit{Step 3: all logarithm branches and the HP requirement.}
We now list all the logarithm branches and keep only those whose spectrum is conjugate-symmetric, as HP requires.
The snapshot is block diagonal,
$\mathcal{E}_{\mathrm{C}}=\mathcal{E}_{\{1,4\}}\oplus
J_{2}(\lambda_{\mathrm{EP}})$,
with a $2\times2$ block $\mathcal{E}_{\{1,4\}}$ carrying the two distinct
eigenvalues $1$ and $e^{-\alpha^{2}\kappa p^{+}_{2}t}$.
This block is diagonalizable, so its logarithm is unique and real, denoted $\mathcal{L}_{\{1,4\}}$ (the two eigenvalues map to $0$ and $-\alpha^{2}\kappa p^{+}_{2}t$); the interpolation acts only on the $\{2,3\}$ block, the $\{1,4\}$ block is identical for $\mathcal{E}$ and $\mathcal{E}(m)$ [see Eqs.~\eqref{eq:channel} and~\eqref{eq:Ichannel}] and require no further attention.
For the $2\times2$ Jordan block, let $\mathcal{L}_{\mathrm{EP}}$ be any
matrix with $e^{\mathcal{L}_{\mathrm{EP}}}=J_{2}(\lambda)$.
Every eigenvalue $\nu_{\text{EP}}$ of $\mathcal{L}_{\mathrm{EP}}$ must obey $e^{\nu_{\text{EP}}}=\lambda_{\text{EP}}$, i.e.\ $\nu_{\text{EP}}=\ln\lambda_{\text{EP}}+2\pi i k$ with one and the same $k\in\mathbb{Z}$ on both Jordan seats; and since $e^{\mathcal{L}_{\mathrm{EP}}}$ is a single Jordan block, $\mathcal{L}_{\mathrm{EP}}$ is itself a single Jordan block.
Writing $J_{2}(\lambda_{\text{EP}})=\lambda_{\text{EP}}(\mathbb{I}+\breve{N})$ with $\breve{N}=(J_{2}(\lambda_{\text{EP}})-\lambda_{\text{EP}}\mathbb{I})/\lambda_{\text{EP}}$ (a $2\times2$ matrix with $\breve{N}^{2}=0$), one readily verifies that the unique solution is
\begin{equation}
\mathcal{L}_{\mathrm{EP}}=(\ln\lambda_{\text{EP}}+2\pi i k)\,\mathbb{I}+\breve{N},
\label{eq:Jlog}
\end{equation}
because $e^{\mathcal{L}_{\mathrm{EP}}}=\lambda\,e^{\breve{N}}
=\lambda(\mathbb{I}+\breve{N})=J_{2}(\lambda_{EP})$: the exponential series of
$\breve{N}$ stops at first order since $\breve{N}^{2}=0$. Hence every
logarithm of the full snapshot is the $4\times4$ matrix
\begin{equation}
\mathcal{L}_{k}=\mathcal{L}_{\{1,4\}}\oplus
\big[(\ln\lambda_{\mathrm{EP}}+2\pi i k)\mathbb{I}+\breve{N}\big],
\label{eq:Lk}
\end{equation}
whose eigenvalues are
\begin{equation}
\{0,\,\alpha^{2}\kappa p^{+}_{2}t,\,\ln\lambda_{\mathrm{EP}}+2\pi i k,\,\ln\lambda_{\mathrm{EP}}+2\pi i k\}.
\label{eq:specLk}
\end{equation}
Now apply the test:
\begin{itemize}
\item $k\neq0$ is rejected in all cases.
Whatever the sign of $\lambda_{\mathrm{EP}}$, the eigenvalue $\ln\lambda_{\mathrm{EP}}+2\pi i k$ is non-real and appears twice, while its conjugate is absent from Eq.~\eqref{eq:specLk}: $\mathcal{L}_{k}$ violates the conjugate symmetry required of any HP map (Step~1).
\item $k=0$ depends on the sign.
For $\lambda_{\mathrm{EP}}>0$ the eigenvalue $\ln\lambda_{\mathrm{EP}}$ is real and poses no obstacle; for $\lambda_{\mathrm{EP}}<0$ it reads $\ln|\lambda_{\mathrm{EP}}|+i\pi$---non-real already at $k=0$, appearing twice with no conjugate partner, so it violates the conjugate symmetry.
\end{itemize}
Three cases therefore remain, distinguished by the sign of $\lambda_{\mathrm{EP}}$, leading to three criteria: the non-HP criterion (a negative coalesced eigenvalue leaves no HP logarithm), the non-cCP criterion (a positive coalesced eigenvalue with a unique HP logarithm fails cCP), and the HP-cCP criterion (a positive coalesced eigenvalue with a unique HP logarithm satisfies cCP).

\begin{figure*}
\includegraphics[width =\textwidth]{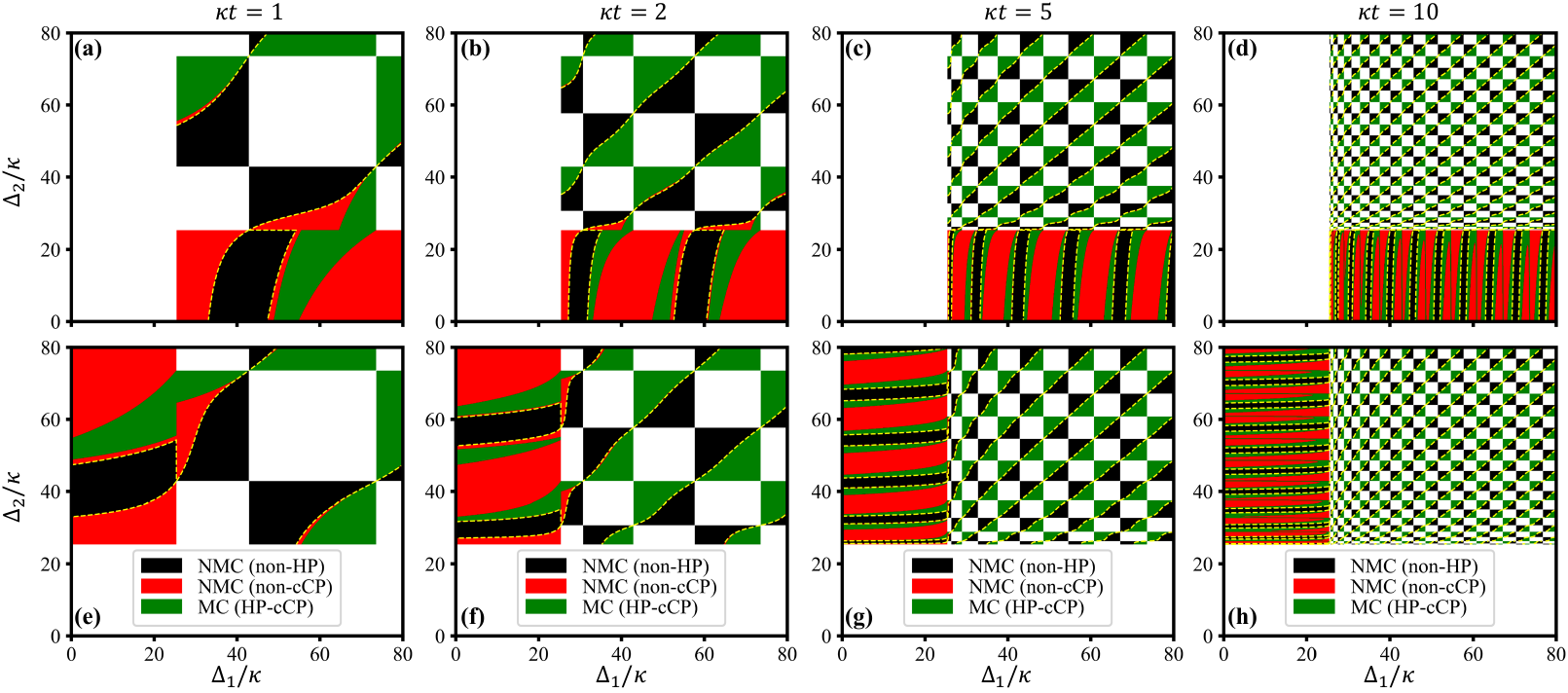}
\caption{\label{FigS4} Certified Markovian--non-Markovian phase diagram
of the ICEP snapshots.
(a)--(d): Markovian--non-Markovian phase diagrams in the $(\Delta_1/\kappa,\Delta_2/\kappa)$ plane at $\kappa t=1,2,5,10$ (left to right), for the ICEP solutions $m^{-}_{\mathrm{EP}}(\Delta_{1},\Delta_{2})$ (upper row) and $m^{+}_{\mathrm{EP}}(\Delta_{1},\Delta_{2})$ (lower row); $\lambda_{\mathrm{EP}}$ is real throughout the colored regions, while white indicates no valid ICEP exists.
Black: $\lambda_{\mathrm{EP}}<0$, certified as an NMC by the non-HP criterion [Eq.~\eqref{eq:criterion-non-HP-sm}].
Red: $\lambda_{\mathrm{EP}}>0$ with $\mu<0$, certified as an NMC by the non-cCP criterion [Eq.~\eqref{eq:criterion-non-cCP-sm}].
Green: $\lambda_{\mathrm{EP}}>0$ with $\mu\geq0$, certified as an MC by passing the HP-cCP test.
The yellow dashed line marks $\lambda_{\mathrm{EP}}=0$.
The lower row mirrors the upper row under $\Delta_1\leftrightarrow\Delta_2$.
Parameters: $K/\kappa=0.65\times10^{3}$ and $\varepsilon_{2}/\kappa=1.50\times10^{3}$.
}
\end{figure*}

\textit{Case (i): $\lambda_{\mathrm{EP}}>0$---one candidate left, to be
tested for cCP.}
Of all branches $\mathcal{L}_{k}$, only the $k=0$ branch meets the
conjugate symmetry condition of Step~1, and it is indeed HP: all four
eigenvalues of $\mathcal{E}_{\mathrm{C}}$ are positive, so the existence
condition of Culver's theorem~\cite{Culver1966} (an even number of Jordan
blocks for every negative eigenvalue) is automatically satisfied---there are
no negative eigenvalues---and a real logarithm exists; in a basis of
Hermitian operators, real means HP, and it is precisely the $k=0$ branch.
Hence $\mathcal{L}_{0}$ is the unique HP logarithm of the snapshot.

HP, however, is only the first of Wolf's three conditions.
It remains to verify cCP.
By the GKSL theorem~\cite{Gorini1976,Lindblad1976} and its channel-divisibility formulation~\cite{Wolf2008,Wolf2008CMP}, a Liouvillian $\mathcal{L}$ is cCP if and only if $e^{s\mathcal{L}}$ is CP for all $s>0$; by Choi's criterion~\cite{Choi1975}, this is equivalent to the Choi matrix of $e^{s\mathcal{L}}$ being positive semidefinite for all $s>0$.
We therefore examine the smallest eigenvalue of the Choi matrix of
$e^{s\mathcal{L}_{0}}$ as $s\to0^{+}$.
Expanding the exponential and using linearity of the Choi map
$J(\mathcal{A})=\sum_{ij}\mathcal{A}(|i\rangle\langle j|)
\otimes|i\rangle\langle j|$~\cite{Wolf2008,Wolf2008CMP},
\begin{equation}
J\big[e^{s\mathcal{L}_{0}}\big]
=J[\mathbb{I}]+s\,J[\mathcal{L}_{0}]+O(s^{2})
=|\Omega\rangle\langle\Omega|+s\,J(\mathcal{L}_{0})+O(s^{2}),
\label{eq:expand}
\end{equation}
where $|\Omega\rangle=\sum_{i}|ii\rangle$ and
$J[\mathbb{I}]=|\Omega\rangle\langle\Omega|$ follows directly from the
definition ($\mathbb{I}(|i\rangle\langle j|)=|i\rangle\langle j|$).
In the orthonormal basis consisting of $|\Omega\rangle/\sqrt{\langle\Omega|\Omega\rangle}$ and the three vectors
\begin{equation}
|\mathbf{e}_{1}\rangle=\frac{|00\rangle-|11\rangle}{\sqrt{2}},
\qquad
|\mathbf{e}_{2}\rangle=|01\rangle,
\qquad
|\mathbf{e}_{3}\rangle=|10\rangle,
\label{eq:kerbasis}
\end{equation}
which are orthogonal to $|\Omega\rangle$, Eq.~\eqref{eq:expand} takes the $1\oplus3$ block form
\begin{equation}
J\big[e^{s\mathcal{L}_{0}}\big]
=\begin{pmatrix}
2+s\,c & s\,\mathbf{g}^{\dagger}\\
s\,\mathbf{g} & s\,M
\end{pmatrix}+O(s^{2}),
\qquad
M_{lr}=\langle\mathbf{e}_{l}|J(\mathcal{L}_{0})|\mathbf{e}_{r}\rangle,
\label{eq:block}
\end{equation}
where $c=\tfrac12\langle\Omega|J(\mathcal{L}_{0})|\Omega\rangle$ and $g_l=\tfrac{1}{\sqrt2}\langle\mathbf{e}_l|J(\mathcal{L}_{0})|\Omega\rangle$ ($l=1,2,3$) complete the matrix elements of $J(\mathcal{L}_{0})$ in this basis: $M$ is its $3\times3$ block within $\{|\mathbf{e}_{l}\rangle\}$, $\mathbf{g}$ its coupling to the $|\Omega\rangle$ direction, and $J(\mathcal{L}_{0})$ is Hermitian because $\mathcal{L}_{0}$ is HP.
Eliminating the $|\Omega\rangle$ component via the Schur complement of the
upper block gives the three small eigenvalues, i.e., those of the matrix $sM-(s\mathbf{g})(2+sc)^{-1}(s\mathbf{g})^{\dagger}=sM+O(s^{2})$, since the
correction carries two powers of $s$.
Therefore
\begin{equation}
\lambda_{\min}(s)=\lambda_{\mathrm{min}}[J[e^{s\mathcal{L}_{0}}]]=s\,\mu+O(s^{2}),
\qquad
\mu=\lambda_{\mathrm{min}}[M],
\label{eq:mu}
\end{equation}
where $\lambda_{\min}[\mathcal{M}]$ is the smallest eigenvalue of $\mathcal{M}$.
Hence $e^{s\mathcal{L}_{0}}$ stays CP down to $s\to0^{+}$ if and only if $\mu\geq0$.
Summarizing case (i), the \textit{non-cCP criterion} reads
\begin{align}
\lambda_{\mathrm{EP}}>0,\
\mu<0
\ \Rightarrow\
\exists\, s>0:\ \lambda_{\min}\!\big[J(e^{s\mathcal{L}_{0}})\big]<0
\ \Rightarrow\
\text{no cCP logarithm}
\ \Rightarrow\
\mathcal{E}\ \text{is NMC},
\label{eq:criterion-non-cCP-sm}
\end{align}
with $\mu$ from Eq.~\eqref{eq:mu}; conversely, the \textit{HP-cCP criterion} reads
\begin{equation}
\lambda_{\mathrm{EP}}>0,\
\mu\geq0
\ \Rightarrow\
\forall\, s>0: \ \lambda_{\min}\!\big[J(e^{s\mathcal{L}_{0}})\big]\geq0
\ \Rightarrow\
\text{a cCP logarithm exists}
\ \Rightarrow\
\mathcal{E}\ \text{is MC}.
\label{eq:criterion-cCP-sm}
\end{equation}

\textit{Case (ii): $\lambda_{\mathrm{EP}}=0$.}
$\mathcal{E}_{\mathrm{C}}$ is singular, no logarithm exists, and the channel
is trivially nonembeddable (the situation of Ref.~\cite{Wong2026}).

\textit{Case (iii): $\lambda_{\mathrm{EP}}<0$---no survivor at all.}
Now $\vartheta=\pi$, and $\nu_{\text{EP}}=t^{-1}[\ln|\lambda_{\mathrm{EP}}|+i(2k{+}1)\pi]$ is non-real for every $k$, including $k=0$.
Consequently, even the only candidate branch that would otherwise survive the HP requirement corresponds to a non-real eigenvalue of multiplicity two and lacks its complex-conjugate counterpart; it is therefore excluded by the HP condition.
The remaining eigenvalues provide no solution either: their logarithms possess mismatched real parts and imaginary parts that are even multiples of $\pi/t$ (contrary to the required odd multiples), so they cannot generate $\nu^{*}$.
Consequently, no candidate has a conjugate-symmetric spectrum, implying that no HP logarithm exists:
\begin{equation}
\lambda_{\mathrm{EP}}<0
\ \Rightarrow\
\text{no $\mathcal{L}$ with}\ \{\nu_k\}=\{\nu_k^*\} \text{ exists}
\ \Rightarrow\
\mathcal{E}\ \text{is NMC},
\label{eq:criterion-non-HP-sm}
\end{equation}
which is the \textit{non-HP criterion}.
Together with Eqs.~\eqref{eq:criterion-non-cCP-sm} and~\eqref{eq:criterion-cCP-sm}, the three criteria identify whether a channel at an EP is Markovian or non-Markovian.
To our knowledge, these criteria provide the first general spectral certification of intrinsic non-Markovianity at channel-level EPs, and the first certified Markovian--Markovian phase diagram of such EPs in a physical system.

The three cases---negative, vanishing, and positive $\lambda_{\mathrm{EP}}$---cover all possibilities.
Figure~\ref{FigS4} shows the certified classification in the $(\Delta_1/\kappa,\Delta_2/\kappa)$ plane at $\kappa t=1$, $2$, $5$, and $10$, illustrating its drift, splitting, and shrinking with time, as discussed in the main text.
Away from the ICEPs, the coalesced pair is nondegenerate, and each snapshot falls under the criteria of Refs.~\cite{Wolf2008,Davalos2019} instead.

\textit{Remark.}
Equation~\eqref{eq:criterion-non-HP-sm} is the defective instance of Theorem 1 of Ref.~\cite{Culver1966}: a nonsingular real matrix has a real logarithm if and only if every negative eigenvalue occupies an even number of Jordan blocks of each size.
In the present setting this count takes two concrete forms: an EP carries the negative eigenvalue in a single $2\times2$ Jordan block (one block of size two---odd, hence forbidden), whereas a diabolic point, where the eigenvalues coincide but the eigenvectors remain distinct, distributes it over two $1\times1$ blocks (two blocks of size one---even, hence allowed).
The criterion of Eq.~\eqref{eq:criterion-non-HP-sm} is exactly the odd case excluded by the theorem.

\textit{Scope.}
Two notions of Markovianity should be distinguished.
In the snapshot sense of Ref.~\cite{Wolf2008}, a single channel
$\mathcal{E}_{\mathrm{C}}$ is Markovian if it is an element of a
time-independent semigroup, $\mathcal{E}_{\mathrm{C}}=e^{\mathcal{L}t}$;
our criteria determine for each ICEP whether the snapshot is of this form: the HP and cCP criteria show that snapshots with $\lambda_{\mathrm{EP}}<0$ and with $\lambda_{\mathrm{EP}}>0$, $\mu<0$, respectively, are not of this form, while snapshots with $\lambda_{\mathrm{EP}}>0$, $\mu\geq0$ are MCs.
More generally, a dynamics $\mathcal{E}(t,t_0)$ is called Markovian if it is CP-divisible---a larger class that includes evolutions generated by time-dependent Liouvillians $\mathcal{L}(t)$ with non-negative rates~\cite{Breuer2016,Rivas2014,Chruscinski2010,Hall2014}, whose snapshots need not be of the form $e^{\mathcal{L}t}$.
Note that the CP tests here and in the cCP criterion concern different maps: the former checks the intermediate maps $\mathcal{E}(t,t_{0})$ of the actual dynamics, whereas the latter checks the maps $e^{s\mathcal{L}_{0}}$ generated by the snapshot's own logarithm.
The HP criterion alone therefore does not exclude such a time-dependent Markovian origin of a snapshot, and settling that question requires an analysis of the whole dynamics, in particular of all intermediate maps $\mathcal{E}(t,t_{0})$, $t\ge t_{0}$ (the map taking states from time $t_{0}$ to time $t$), e.g., whether each of them is CP~\cite{Breuer2009,Rivas2010,Fanchini2014}; this lies beyond the present work---no Liouvillian of the interpolated channel is constructed.
Instead, we follow how the ICEP regions, i.e., the sets of parameters $(\Delta_1/\kappa,\Delta_2/\kappa)$ where a valid ICEP exists, move and deform as the evolution time $\kappa t$ increases (see Fig.~\ref{FigS2}).

\bibliography{Ref}